\documentclass{pasj00}  \draft 

\begin{document}

\def\be{\begin{equation}} 	\def\ee{\end{equation}}
\def\bc{\begin{center}} 	\def\ec{\end{center}}
\def\bfig{\begin{figure}} 	\def\bff{\begin{figure*}}
\def\ef{\end{figure}} 	\def\eff{\end{figure*}}
\def\ig{includegraphics}  
\def\bt{\begin{table}} 	\def\btf{\begin{table*}}
\def\et{\end{table}} 	\def\etf{\end{table*}}
\def\v{\vskip 2mm}  	\def\cen{\centerline}  
\def\dv{de Vaucouleurs }
\def\gc{Galactic Center }
\def\kms{km s$^{-1}$}  	\def\Msun{M_\odot} 
\def\Vsun{V_0} 		\def\Rsun{R_0}
\def\rzero{$R_0$} 	\def\vzero{$V_0$} 
\def\sin{{\rm sin}} 	\def\cos{{\rm cos}}
\def\w{\omega} 		\def\wsun{\omega_0}
\def\kmsmpc{ km s$^{-1}$ Mpc$^{-1}$ }
\def\vrot{ $V_{\rm rot}$ } \def\Vrot{ V_{\rm rot} } 
\def\Msun{ M_{\odot \hskip-5.2pt \bullet} }
\def\msun{ $M_{\odot \hskip-5.2pt \bullet}$ } 
\def\ha{ H$\alpha$ } 	\def\Ha{ H$\alpha$ } 
\def\deg{$^\circ$} 	\def\Deg{^\circ}
\def\meleven{\times 10^{11}\Msun} \def\mtwelve{\times 10^{12}\Msun}  
\def\vr{v_{\rm r}} 	\def\vp{v_{\rm p}}
\def\msqpc{\Msun {\rm pc}^{-2}} \def\mcupc{\Msun {\rm pc}^{-3}}

\title{{\it \small Planets, Stars and Stellar Systems, Springer, Berlin 2013, Vol. 5, ed. G. Gilmore., Chap 19} \\
The Mass Distribution and Rotation Curve in the Galaxy}
\author{Yoshiaki {\sc Sofue} }  
\affil{Institute of Astronomy, The University of Tokyo, Mitaka, 181-0015 Tokyo, \& \\
Department of Physics, Meisei University, Hino, 191-8506 Tokyo, Japan \\  
Email:{\it sofue@ioa.s.u-tokyo.ac.jp} } 


\maketitle 

\begin{abstract}

The mass distribution in the Galaxy is determined by dynamical and photometric methods. The dynamical method is based on the Virial theorem, and calculates the mass from kinematical data such as rotation velocities, velocity dispersions, and motions of satellite galaxies. Rotation curves are the major tool for determining the dynamical mass distribution in the Milky Way and spiral galaxies. The photometric (statistical) method utilizes luminosity profiles from optical and infrared observations, and assumes empirical values of the mass-to-luminosity (M/L) ratio to convert the luminosity to mass. This method is convenient to separate the mass components such as bulge and disk, while the uncertainty is large due to ambiguous M/L ratio arising from the variety of stellar populations. Also, the methods cannot detect the dark matter that dominates in the outer regions and central black holes. 

In this chapter the dynamical method is described in detail, and rotation curves and mass distribution in the Milky Way and nearby spiral galaxies are presented. The dynamical method is further categorized into two methods: the decomposition method and direct method. The former fits the rotation curve by calculated curve assuming several mass components such as a bulge, disk and halo, and adjust the dynamical parameters of each component. Explanations are given of the mass profiles as the de Vaucouleurs law, exponential disk, and dark halo profiles inferred from numerical simulations. Another method is the direct method, with which the mass distribution can be directly calculated from the data of rotation velocities without employing any mass models. Some results from both methods are presented, and the Galactic structure is discussed in terms of the mass. Rotation curves and mass distributions in external galaxies are also discussed, and the fundamental mass structures are shown to be universal.
\\

\\

\\

\noindent{\bf Note: Preprint with full figures is available from\\
 http://www.ioa.s.u-tokyo.ac.jp/${\sim}$sofue/htdocs/2013psss/ \\}

\end{abstract}


\section{INTRODUCTION}

The mass of a galaxy and its distribution are obtained by two ways: one is to use photometric data such as luminosity profiles assuming the mass-to-luminosity ratio, and the other is to apply dynamics to kinematical data based on the Virial theorem. 
The photometric method has been used to estimate the mass of the Galaxy early in the beginning of the 20th century by counting stellar density in the solar neighborhood. The density was multiplied by the total volume of the Galaxy to calculate the number of solar-mass stars. This estimation already gave an approximate mass on the order of $10^{11}\Msun$ of the Galaxy (Fich and Tremain 1991). 

Most of the luminous mass is occupied by stars, and the rest $\sim 10$ percent is by interstellar gases. Hence, the stellar luminosity distribution roughly represents the luminous mass distribution. The distribution of stellar mass can be obtained from optical and infrared surface photometry of spiral galaxies, by assuming the mass-to-luminosity (M/L) ratio. Particularly, near-infrared K band (2.2 $\mu$m) images are useful to obtain an approximate distribution of stars occupying most of the luminous mass. Infrared photometry along the Galactic plane has been used to derive the mass distribution in the Galaxy. Although the photometric method is convenient to approximately map the luminous mass, it varies with the employed M/L ratio. In order to determine the M/L ratio, measurement of mass independent of luminosity observation is in any way necessary beforehand. Besides the ambiguity of the assumed M/L ratio, it cannot give information about the dark matter.

Nevertheless, the luminous matter distribution is helpful to overview the mass components in the Galaxy. Figure \ref{Gillust} shows a schematic view of a spiral galaxy, which is composed of the bulge, disk, and halo. The luminous halo is dominated by globular clusters and high-temperature gas, whereas the mass is dominated by dark matter. In addition to these luminous components, a galaxy nests a central massive object, or a black hole, and a massive core in the bulge. The entire galaxy is surrounded by a huge dark halo composed of dark matter.

\bfig
\bc
\includegraphics[width=7cm]{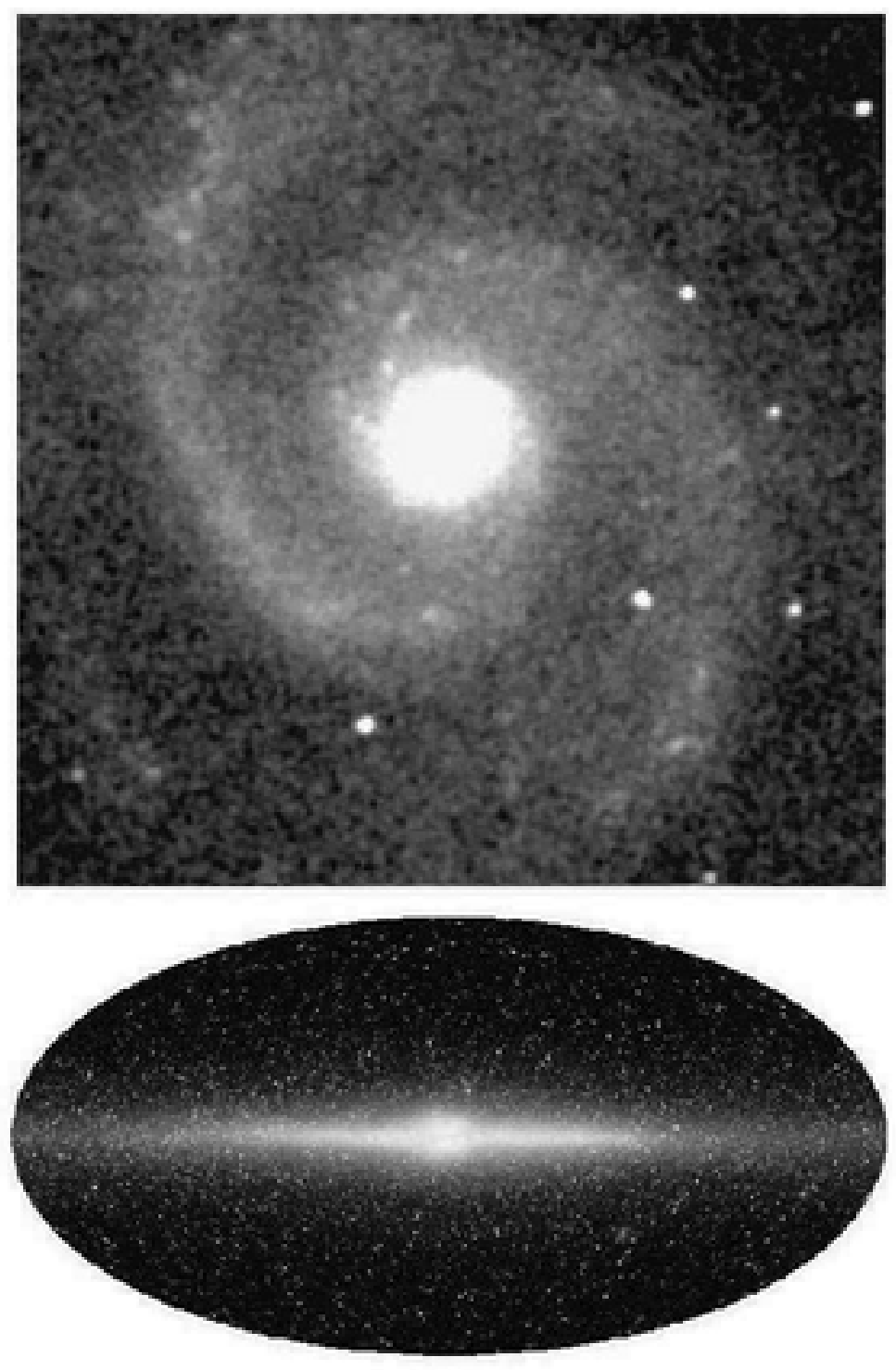}  \\
\includegraphics[width=8cm]{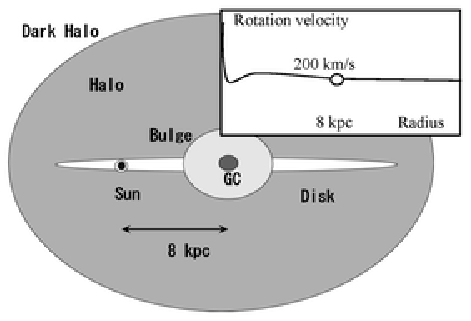}  
\ec
\caption{[Top] Near-infrared (K band, 2.2 $\mu$m) image of a face-on spiral galaxy M51 from 2MASS survey. (The Milky Way is rotating clockwise as seen from the North Galactic Pole, showing trailing arms in opposite sense to this image.) [Middle] Whole sky near-infrared view of the Milky Way from COBE, showing the Galaxy's edge-on view. These images approximately represent the distributions of stars, therefore, the luminous mass. They are embedded in a massive dark halo of comparable mass. [Bottom] Schematic galactic structure. The Galaxy consists of the nucleus nesting a massive black hole, high-density massive core, bulge, disk, halo objects, and a dark matter halo. Inserted is a schematic rotation curve.}
\label{m51} 
\label{Gillust}
\ef

For the mass including the dark matter and invisible masses like black holes, it is necessary to apply the dynamical method using kinematical data. The dynamical mass is calculated based on the Virial theorem that the kinetic and gravitational energies are in equilibrium for a dynamically relaxed system. For a rotating object such as the galactic disk, the balance between the gravitational and centrifugal forces is applied. 

The dynamical mass of the Galaxy on the order of $\sim 10^{11} \Msun$ inside the solar circle was already calculated early in 1950's (Oort 1958), when the circular orbit of the local standard of rest (LSR) was obtained in terms of the galactic-centric distance \rzero  and rotation velocity \vzero (see Fich and Tremaine 1991 for a review). For a standard set of the parameters of \rzero = 8 kpc and \vzero = 200 \kms, they yield the most fundamental quantity, the mass inside the solar circle on an assumption of spherical distribution, to be on the order of
\be
M_0=R_0 V_0^2/G = 7.44 \times 10^{10} \Msun \sim 10^{11}\Msun,
\label{eq-masssolcirc}
\ee 
where $G$ is the gravitational constant. 
Although this approximate estimation is not far from the true value, the mass distribution in the Galaxy, as well as those in any galaxies, is not simply spherical, and it is principally derived by analyzing the rotation curves on the assumption that the centrifugal force of the circular motion is balancing with the gravitational force in a spheroid or a disk. Hence, the first step to derive the mass and mass distributions in galaxies is to obtain the rotation curves. Given a rotation curve, it is deconvolved to several components representing the mass distribution using various methods.
 
 The inner rotation curve is simply measured by the terminal (tangential)-velocity method applied to radio line observations such as the HI and CO lines. The mass profiles of the galactic disk and bulge were thus known since 1970's. The central mass condensation and the nuclear massive black hole have been measured since 1980's when kinematics of interstellar gas and stars close to Sgr A$^*$, our Galaxy's nucleus, were  measured by infrared observations. For the total mass of the Galaxy including the outermost regions, they had to wait until an outer rotation curve and detailed analyses of motions of member galaxies in the Local Group were obtained. It was only recent when the total mass was estimated to be $\sim 3 \times 10^{11}\Msun$ including the dark halo up to $\sim 150$ kpc by considering the outer rotation curve and motions of satellite galaxies.

The galactic constants, which are the galactocentric distance of the Sun $R_0$ and the rotation velocity  $V_0$ of the Local Standard of Rest (LSR) around the Galactic Center (GC), namely the three-dimensional position and motion of the LSR, are the most fundamental parameters to derive the mass and its distribution in the Galaxy. The LSR is defined as the coordinates with its origin at the Sun and rotating on a circular orbit around the GC after correcting for the solar motion. The galactocentric distance \rzero  has been determined by various methods, which lies in the range of 7 and 9 kpc (Reid 1993).  Given the value of \rzero, the rotation velocity \vzero   is determined by using the Oort's constants $A$ and $B$ as
\be
V_0=(A-B)R_0,
\ee
where
\be
A={1 \over 2} \left({V \over R} -{dV \over dR} \right)_{R_0}
\ee
and
\be
B=-{1 \over 2} \left({V \over R} + {dV \over dR}\right)_{R_0},
\ee
with $R$ and $V$ being galactocentric distance and rotation velocity of stars in the solar neighborhood. Here the values $R_0=8$ kpc and \vzero=200 \kms are adopted, which yield approximate mass of the Galaxy inside the solar circle of $M_0=7.44 \times 10^{10}\Msun$ as in Eq. (\ref{eq-masssolcirc}). 

In this chapter the dynamical methods are described and are applied to determination of the mass distribution in the Galaxy using the rotation curve. The description will be made of two parts: one for rotation curve, and another for deconvolution into mass components and density profiles. Mass distributions in external galaxies and their rotation curves are also touched upon. More detailed description of individual methods and analyses may be referenced to various papers in the cited literature. Section 2 is partly based on the review by Sofue and Rubin (2001). Sections 3 and 4 are based on the studies of galactic mass models by Sofue et al. (2009) and Sofue (2009).
 
\section{ROTATION CURVES}

\vskip 5mm \subsection{\bf Measurements of Galactic Rotation}
 
The rotation of the Milky Way is clearly seen in longitude-radial velocity (LV) diagrams along the galactic plane, where spectral line intensities are plotted on the $(l,V_{\rm lsr})$ plane as shown in Fig.  \ref{pvdHI}, which shows the observed LV diagrams in the $\lambda$21-cm HI and $\lambda$2.6-mm CO emission lines.

The positive- and negative-velocity envelopes (terminal velocities) at $0<l<90\Deg$ and $270<l<360\Deg$, respectively, are used for determining the inner rotation curve inside the solar circle at $R\le R_0$. This terminal (tangential)-velocity method is applied to HI and CO line observations for the inner Galaxy (Burton and Gordon 1978; Clemens 1985; Fich et al. 1989). In order to derive outer rotation curve beyond the solar circle, optical distances and velocities of OB stars are combined with CO-line velocities (Blitz et al. 1982; Demers and Battinelli 2007). HI thickness method is useful to obtain rotation curve of the entire disk (Merrifield 1992; Honma and Sofue 1997). High accuracy measurements of parallax and proper motions of maser sources and Mira variable stars using VLBI technique are providing an advanced tool to derive a more accurate rotation curve (Honma et al. 2007). 

\bfig
\bc
\includegraphics[width=9cm]{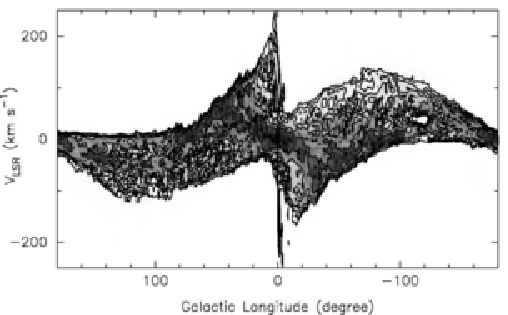}   \\
\includegraphics[width=9cm]{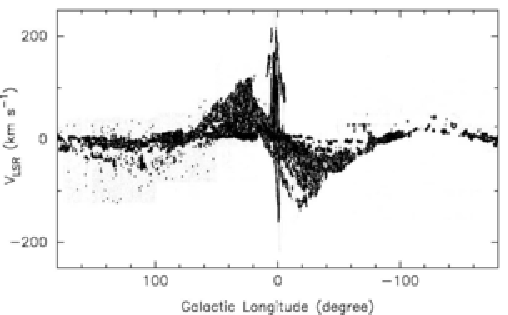} 
\ec
\caption{Longitude-radial velocity ($l-V_{\rm lsr}$) diagram of the $\lambda$21-cm HI line emission (top: Nakanishi 2007) and $\lambda$2.6-mm CO (bottom: Dame et al. 1987) lines along the galactic plane.} 
\label{pvdHI}
\label{pvdCO}
\end{figure} 

\vskip 5mm \subsubsection{\bf Rotation velocities from distance, radial velocity, and proper motion}

\begin{figure} 
\begin{center} 
\includegraphics[width=10cm]{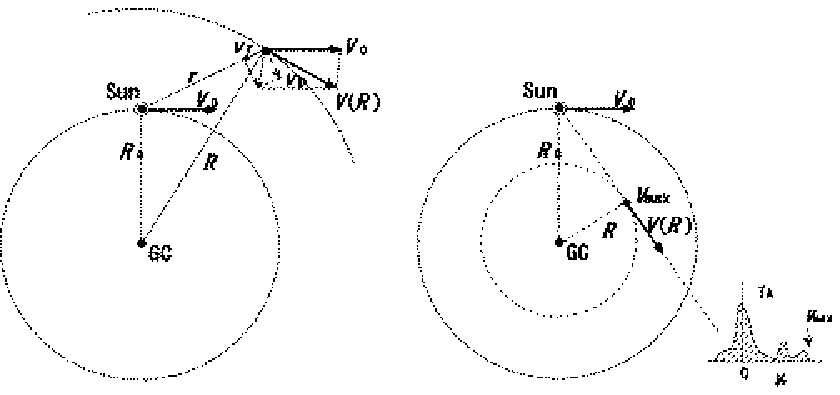} 
\end{center}
\caption{[Left] Rotation velocity at any point in the galactic plane is obtained by measuring the distance $r$ and either radial velocity $\vr$ or perpendicular velocity $\vp=\mu r$, where $\mu$ is the proper motion. [Right] Rotation curve inside the solar circle ($R<R_0$; dashed circle) is obtained by measuring the terminal radial velocity  $v_{\rm max}$ at the tangent point, where the GC distance is given by $R=R_0 \sin l $.}
\label{rc-det}
\end{figure}

Given the galactic constants  $\Rsun$ and $\Vsun$, rotation velocity $V(R)$ in the galactic disk can be obtained as a function of galacto-centric distance $R$ by measuring the distance $r$ radial velocity $\vr$ and/or perpendicular velocity $\vp=\mu r$ of an object, where $\mu$ is the proper motion (Fig. \ref{rc-det}). The velocity vector of a star at any position in the Galaxy is determined by observing its three dimensional position $(r,l,b)$ and its motion $(\vr, \vp)$, where $\vr$ is the radial velocity and $\vp$ the perpendicular velocity with $\vp=\mu r$ with $\mu$ being the proper motion on the ski.

The galacto-centric distance $R$ is calculated from the position of the object $(l, b, r)$ and $\Rsun$ as
\be
R=(r^2+R_0^2 - 2 r R_0 \cos l)^{1/2}.  
\ee
Here, the distance $r$ to the object must be measured directly by trigonometric (parallax) method, or indirectly by spectroscopic measurements.

If the orbit of the star is assumed to be circular in the galactic plane, the rotation velocity $V(R)$ may be obtained by measuring one of the radial velocity or proper motion.  The rotation velocity $V(R)$ is related to the radial velocity $v_{\rm r}$ as 
\be
 V(R)= {R \over R_0} \left({v_{\rm r} \over \sin l} +V_0 \right). 
\label{vfromvr}
\ee 
Alternatively, the rotation velocity is determined by measuring the proper motion $\mu$ as
\be
V(R) = - {R \over s} (v_{\rm p} + V_0 \cos l),
\ee 
where
\be
s=r - R_0 \cos l.
\ee  

The method using  radial velocity has been traditionally applied to various stellar objects. Star forming regions are most frequently used to determine the rotation curve beyond the solar circle. In this method, distances $r$ of OB stars are measured from their distance modulus from the apparent magnitude after correction for extinction and absolute luminosity by the star's color and spectral type. Then, the star's distance $r$ is assumed to be the same as that of its associated molecular cloud and/or HII region whose radial velocity is obtained by observing the Doppler velocity of molecular lines and/or recombination lines. In this method, the error in the distance is large, which results in the large scatter in the obtained outer rotation curve, as seen in Fig. \ref{fig-obs}. 

Accurate measurements of rotation velocities have been obtained, combining the spectroscopic and VLBI techniques by observing both the proper motion and radial velocity. Trigonometric (parallax) measurements of maser-line radio sources are used to determine the distance $r$ and the proper motion $\vp$ ($=r\mu$), and the radial velocity of the same source $\vr$ is measured by radio spectroscopic observations of the maser line. By this method, an accurate velocity has been obtained on the outer rotation curve as plotted in Fig. \ref{fig-obs} by a big dot (Honma et al. 2007).

\vskip 5mm \subsubsection{\bf  Terminal-velocity method: Rotation curve inside Solar circle}
  
From spectral profiles of interstellar gases as observed in the HI 21-cm and/or CO 2.6-mm emission lines in the first quadrant of the galactic plane ($0<l<90\Deg$), it is known that the gases within the solar circle show positive radial velocities, and those outside the solar circle have negative velocities because of the motion of the Sun (figures \ref{pvdHI}, \ref{rc-det}). Maximum positive velocity is observed at the tangent point, at which the line of sight is tangential to the radius as indicated by the dashed lines in Fig. \ref{rc-det}. This maximum radial velocity  ${\vr}_{\rm~ max}$ is called the terminal velocity or the tangent-point velocity. Using this terminal velocity, the rotation velocity $V(R)$ is simply calculated by 
\begin{equation}
 V(R) = {\vr}_{\rm ~max} + \Vsun~\sin~ l,
\end{equation}
and the galacto-centric distance is given by 
\be
R=\Rsun \sin~ l.
\ee 
The rotation curve $V(R)$ in the first and fourth quadrants of the disk is thus determined by observing terminal velocities at various longitudes at  $0<l<90\Deg$ and  $270<l<360\Deg$, and therefore, at various $R$. In the fourth quadrant at $270<l<360\Deg$ the terminal velocities have negative values.

\vskip 5mm \subsubsection{\bf  Ring thickness method}

The HI-disk thickness method utilizes apparent width of an annulus ring of HI disk in the whole Galaxy (Merrifield 1992; Honma and Sofue 1997). This method yields annulus-averaged rotation velocity in the entire galactic disk. The method is illustrated in Fig. \ref{hithickness}: the apparent latitudinal angle $\Delta b$ of the HI disk along an annulus ring of radius $R$ varies with longitude as 
\be
\Delta b = {\rm arctan} \left({ z_0 \over {R_0 \cos l + \sqrt{R^2 - R_0^2 \sin^2 l}}}\right).
\ee
with
\be
\vr = W(R) \sin l,
\ee
where 
\be
W(R)=\left[ V(R){R_0 \over R} - V_0\right].
\ee
The amplitude of $\Delta b$ normalized by its value at $l=180\Deg$ plotted against longitude $l$ is uniquely related to the galacto-centric distance $R$, which is as a function of $V(R)$ and is related to $\vr$ as above equations. This method utilizes the entire HI disk, so that they obtained rotation curve manifests an averaged kinematics of the Galaxy. Therefore, it is informative for more global rotation curve compared to the measurements of individual stars or the tangent-point method. 

\bfig
\bc
\includegraphics[width=6cm]{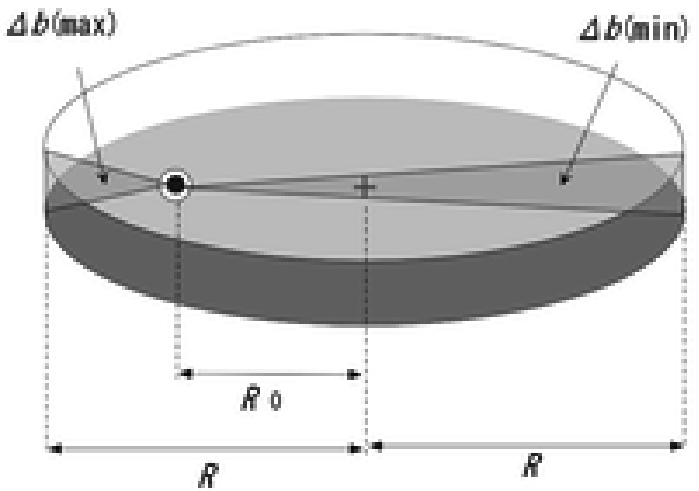} \\
\vskip 5mm
\includegraphics[width=7cm]{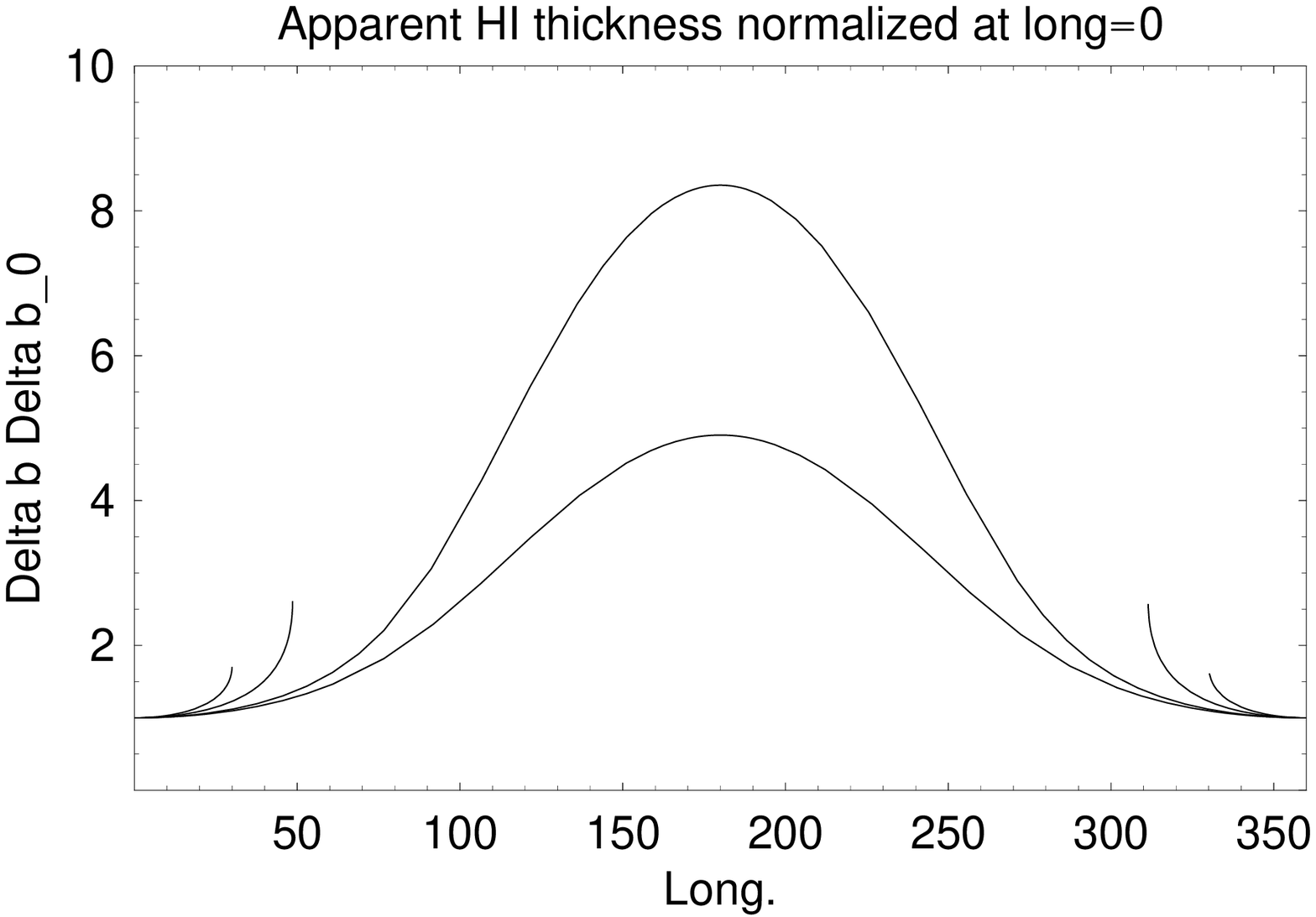}
\ec
\caption{Apparent thickness $\Delta b$ of an annulus ring at $R$ with the rotation velocity $V(R)$ varies with the longitude with the shape and relative amplitude being depending on the ring radius $R$. Indicated lines correspond to $R=4,~ 6, ~10$ and 12 kpc.}
\label{hithickness}
\ef

\vskip 5mm \subsection{\bf  Rotation curve of the Galaxy}

\begin{figure}
\begin{center} 
\includegraphics[width=11cm]{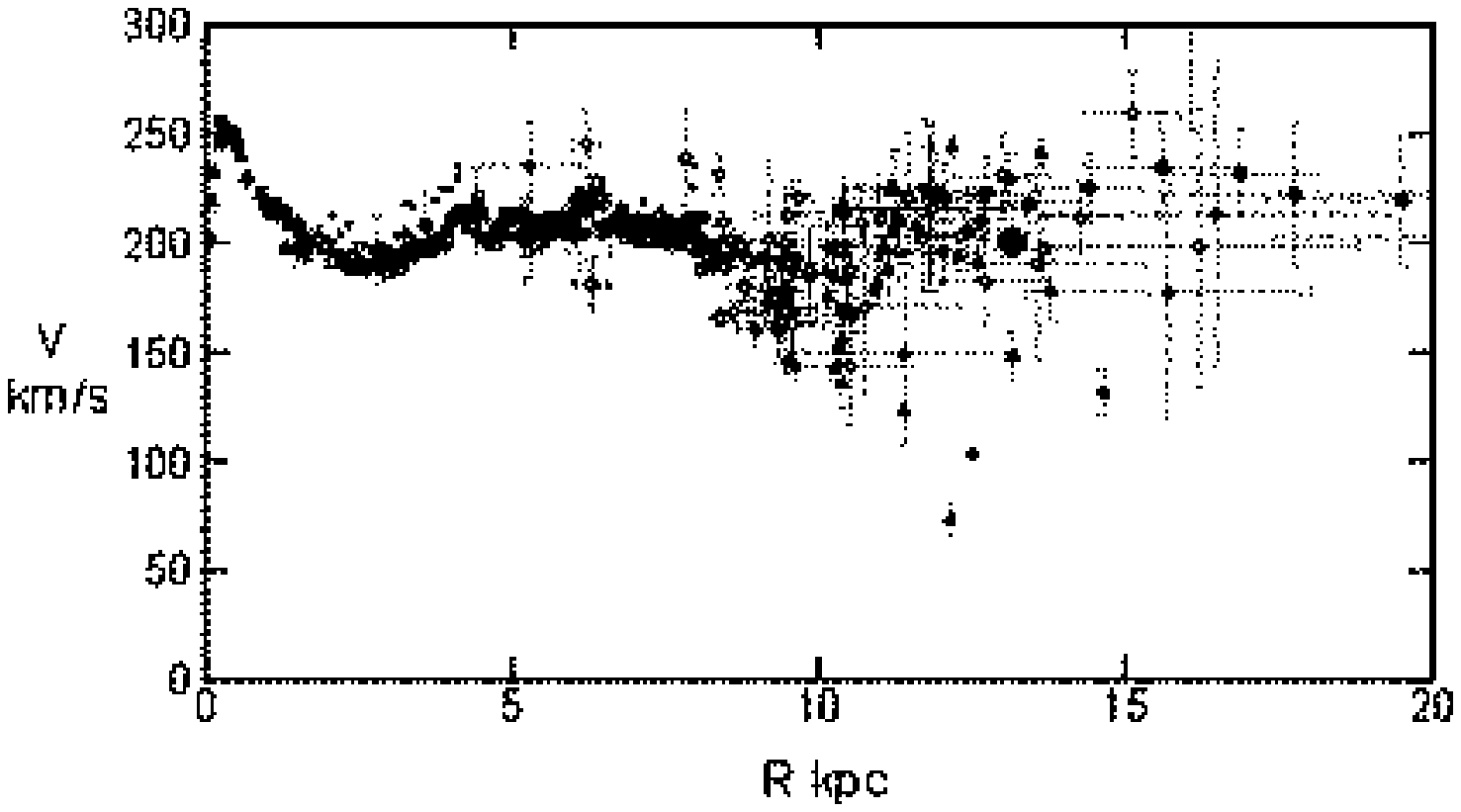}   
\end{center}
\caption{Observed circular velocities representing the rotation curve of the Galaxy (Sofue et al. 2009).. Plot was obtained using HI tangent velocity data (Burton and Gordon 1978; Fich et al. 1989); CO tangent-velocity data (Clemens 1985); CO cloud + HII region + OB stars (Fich et al.1989, Blitz et al. 1982); Late type stars (Demers and Battinelli 2007); HI thickness method (Honma and Sofue 1997). Big circle at 13.1 kpc is from VERA by parallax, radial and proper motions (Honma et al. 2007).  All data have been converted to $(R_0,V_0)=(8.0$, 200.0 \kms).}
\label{fig-obs}  
\bc \includegraphics[width=9cm]{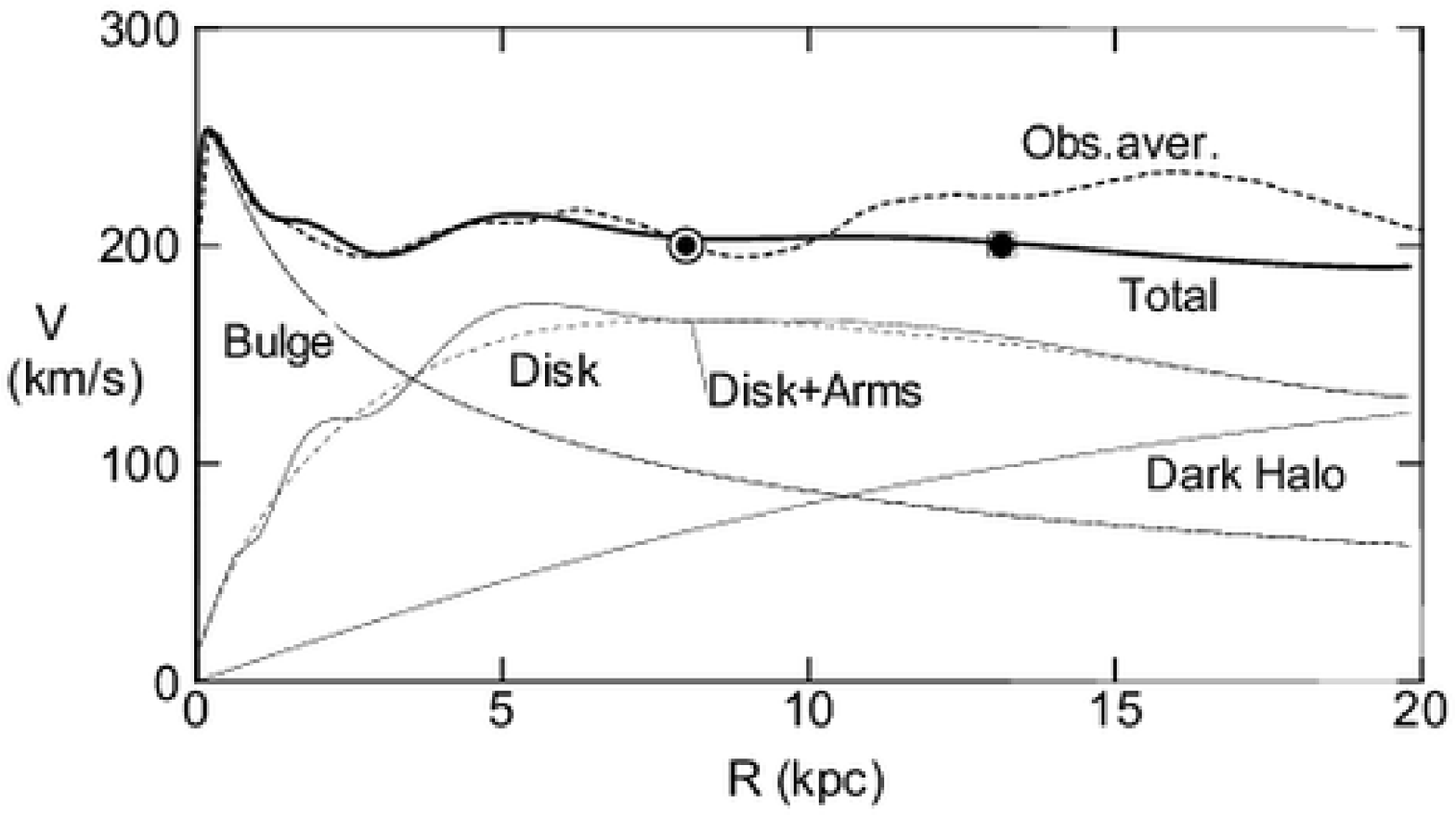} \ec 
\caption{Calculated rotation curve including the bulge, disk, spiral arms, and  dark halo. The big dot denotes the observed result from VERA (Honma et al. 2007). The pure disk component is also indicated by the thin dashed line. The thick dashed line indicates a simply averaged observed rotation curve taken from Sofue et al. (1999). } 
\label{fig-rc-arm}    
\end{figure}

The entire rotation curve of the Galaxy is obtained by combining the observed rotation velocities from the various methods. However, when they calculate the rotation velocities from the data, the observers often adopt different parameters. This had led to rotation curves of the Galaxy in different scaling both in $R$ and $V(R)$. In order to avoid this inconvenience, a unified rotation curve was obtained by integrating the existing data by re-calculating the distances and velocities for a nominal set of the galactocentric distance and the circular velocity of the Sun as  $(R_0, V_0)$=(8.0 kpc, 200 \kms) .
Figure \ref{fig-obs} show a plot of compiled data points for the rotation velocities (Sofue et al. 2009), and Fig. \ref{fig-rc-arm} is a rotation curve fitted to the observed  points, as described in the next subsections. 
Recent high accuracy measurement by Honma et al. (2007) using VERA, an exact data point was given on the outer rotation curve at  $(R, V(R))=(13.15 \pm 0.22 {\rm kpc}, 200\pm 6 {\rm km~s^{-1}}$).

\vskip 5mm \subsubsection{\bf  Rotation in the Galactic Center}

\begin{figure}
\begin{center} 
\includegraphics[width=9cm]{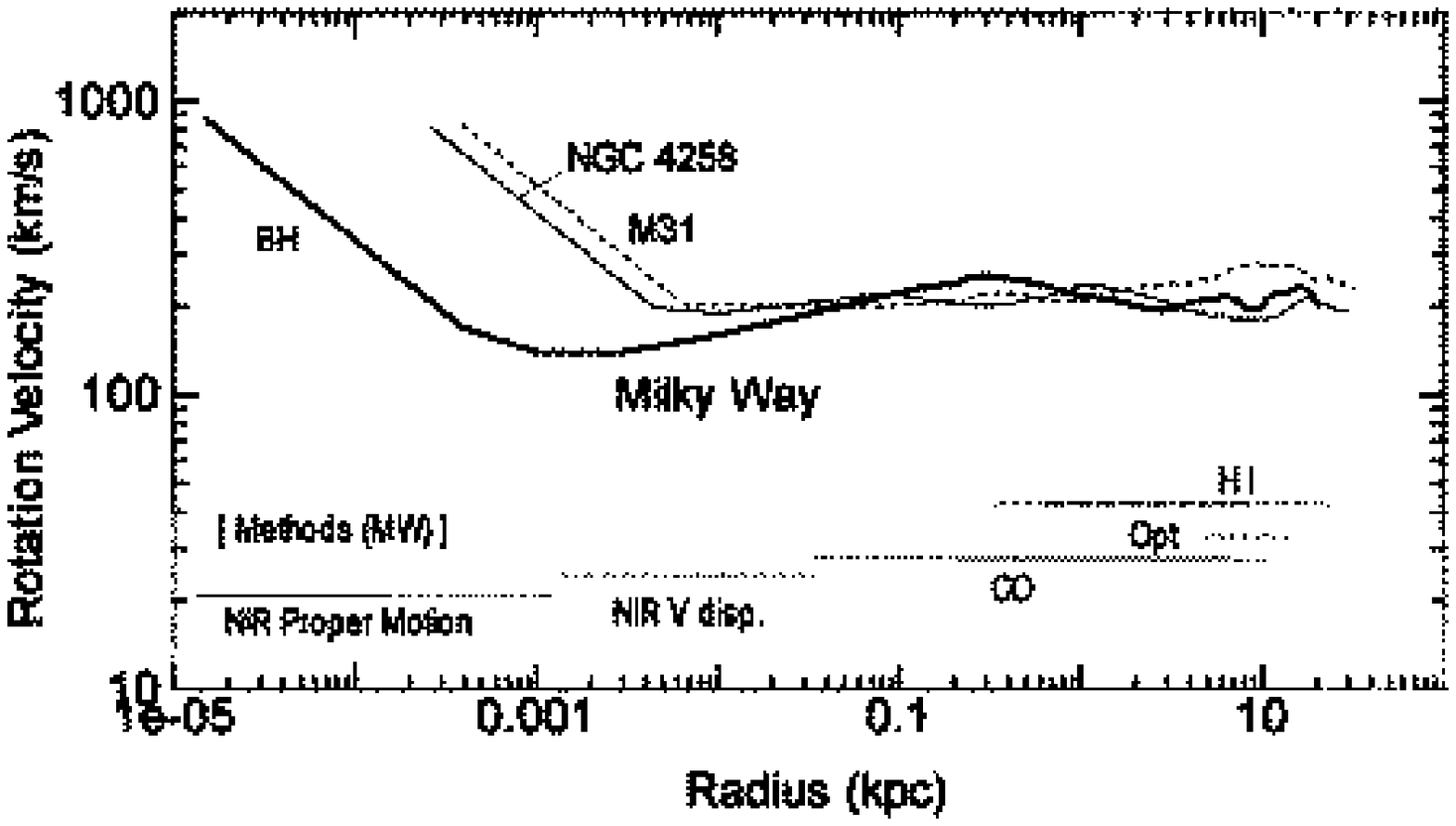}   
\end{center}
\caption{ Logarithmic rotation curve of the Galaxy (Sofue and Rubin 2001). For comparison those for two spiral galaxies are indicated. Innermost rotation velocities are Keplerian velocities calculated for the massive black holes. Observational methods for the Milky Way are shown by horizontal lines.}
\label{rc-mw-log}
\end{figure}

The Galaxy provides a unique opportunity to derive a high resolution central rotation curve (Gilmore et al. 1990). Proper-motion studies in the near infrared have revealed individual orbits of stars within the central 0.1 pc, and the velocity dispersion increases toward the center, indicating the existence of a massive black hole of mass $3\times 10^6\Msun$ (Genzel et al. 1997, 2000; Ghez et al. 1998). Outside the very nuclear region, radial velocities of OH and SiO maser lines from IR stars in the Galactic Center region are used to derive the velocity dispersion and  mean rotation (Lindqvist et al.1992). 
SiO masers from IRAS sources in the central bulge have been used to study the kinematics, and the mean rotation of the bulge was found to be in solid body rotation of the order of 100 \kms (Izumiura et al. 1999; Deguchi et al. 2000). The very central rotation curve controlled by the black hole is Keplerian. In order to overview the entire rotation characteristics from the nucleus to the outer edge, the rotation curve is shown in a logarithmic scale in Fig. \ref{rc-mw-log}.  

\vskip 5mm \subsection{\bf  Measurements of Rotation Velocities in External Galaxies} 

Rotation curves are also the major tool to derive the mass distribution in external galaxies, and are usually derived from Doppler velocities of the emission lines such as H$\alpha$, [NII], HI and CO emission lines, which are particularly useful to calculate the mass distribution because of the small velocity dispersions of $\sim 10$ \kms compared to rotation velocities of $\sim 200$ \kms. This allows us to neglect the pressure term in the equation of motion for calculating the mass distribution. Reviews on rotation curves of disk galaxies have been given by various authors (Trimble 1987; Ashman 1992; Persic \& Salucci 1996, 1997; Sofue and Rubin 2001). On the other hand, spheroidal galaxies have higher velocity dispersions that are measured from stellar absorption lines (Faber \& Gallagher 1979; Binney 1982; de Zeeuw \& Franx 1991).  

\vskip 5mm \subsubsection{\bf  \ha\ and Optical Measurements}  

In optical observations, long slit spectra are most often used to deduce the rotation curve of a galaxy from emission lines (Rubin et al. 1982, 1985; Mathewson et al. 1992, 1996; Amram et al.  1995). The H$\alpha$, [NII], and [SII] emission lines are most commonly used. For a limited number of nearby galaxies, rotation curves can be produced from velocities of individual HII regions in galactic disks (Rubin \& Ford, 1970, 1983; Zaritsky et al. 1997). In early galaxies, planetary nebulae are valuable tracers of the velocity fields in outer disks,  where the optical light is faint.  Fabry-Perot spectrometers are used to derive the H$\alpha$ velocity fields.  Velocity fields cover the entire disk and yield more accurate rotation curve than the slit observations, while it requires more sophisticated analyses.

\vskip 5mm \subsubsection{\bf  Radio lines: HI, CO, and masers}

Radio wave emission lines are useful for studying the whole galactic disk, because the extinction is negligible even in the central regions, where the optical lines are often subject to strong absorption by the interstellar dusts.
The $\lambda$ 21-cm HI line is a powerful tool to obtain kinematics of the outer rotation curves, in part because its radial extent is greater than that of the visible disk (Bosma 1981a, b; van der Kruit \& Allen 1978), while it is often weak or absent in the central regions.  
 
\begin{figure}
\bc
\includegraphics[width=6cm]{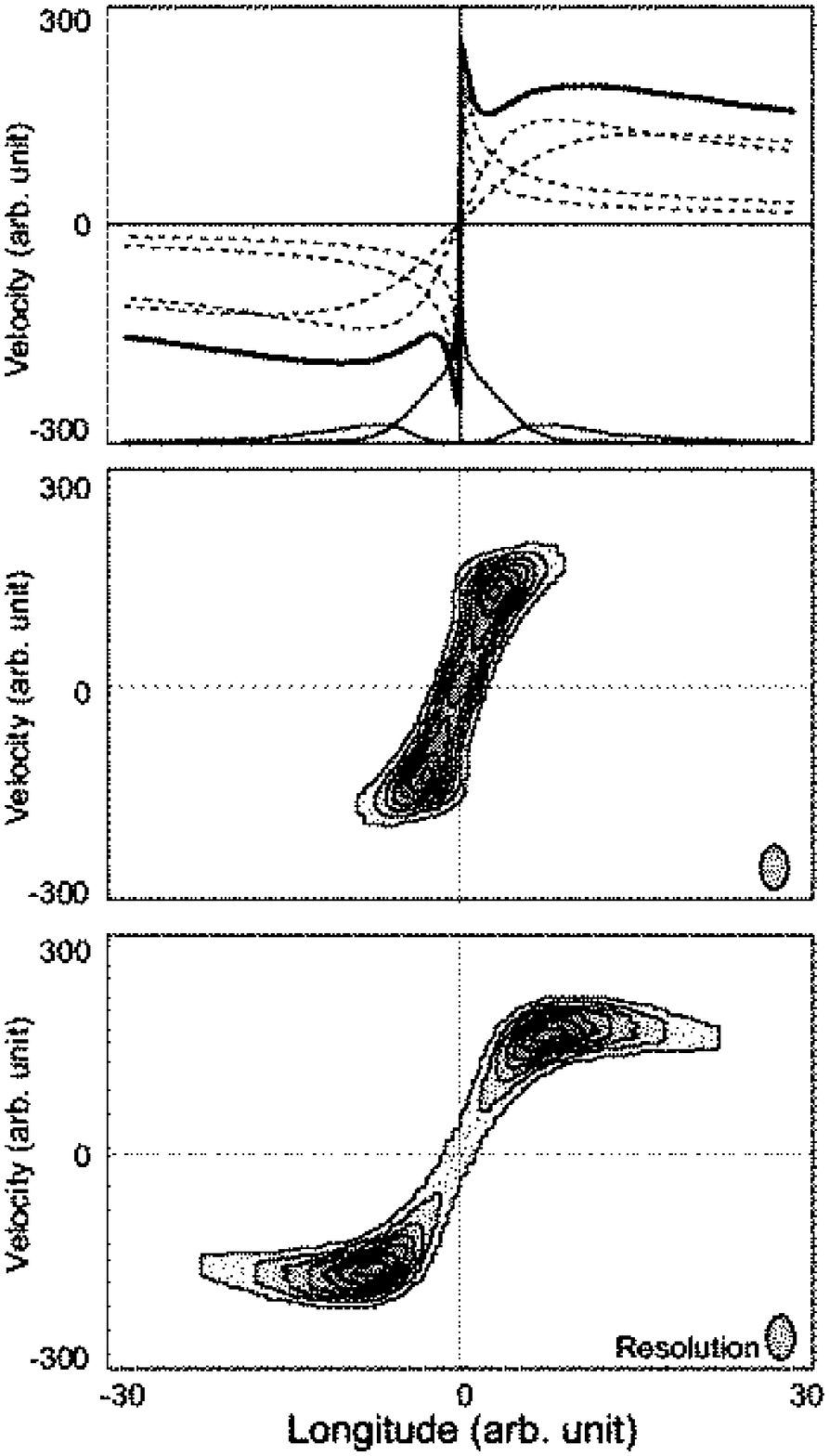}    
\ec
\caption{[Top] A model rotation curve comprising a massive core,  bulge,  disk and halo. Distributions of the molecular (CO) and HI gases are given by thin lines. [Middle] Composed position-velocity diagram in CO, and [bottom] HI.}
\label{rc-simu}
\begin{center}    
\includegraphics[width=8cm]{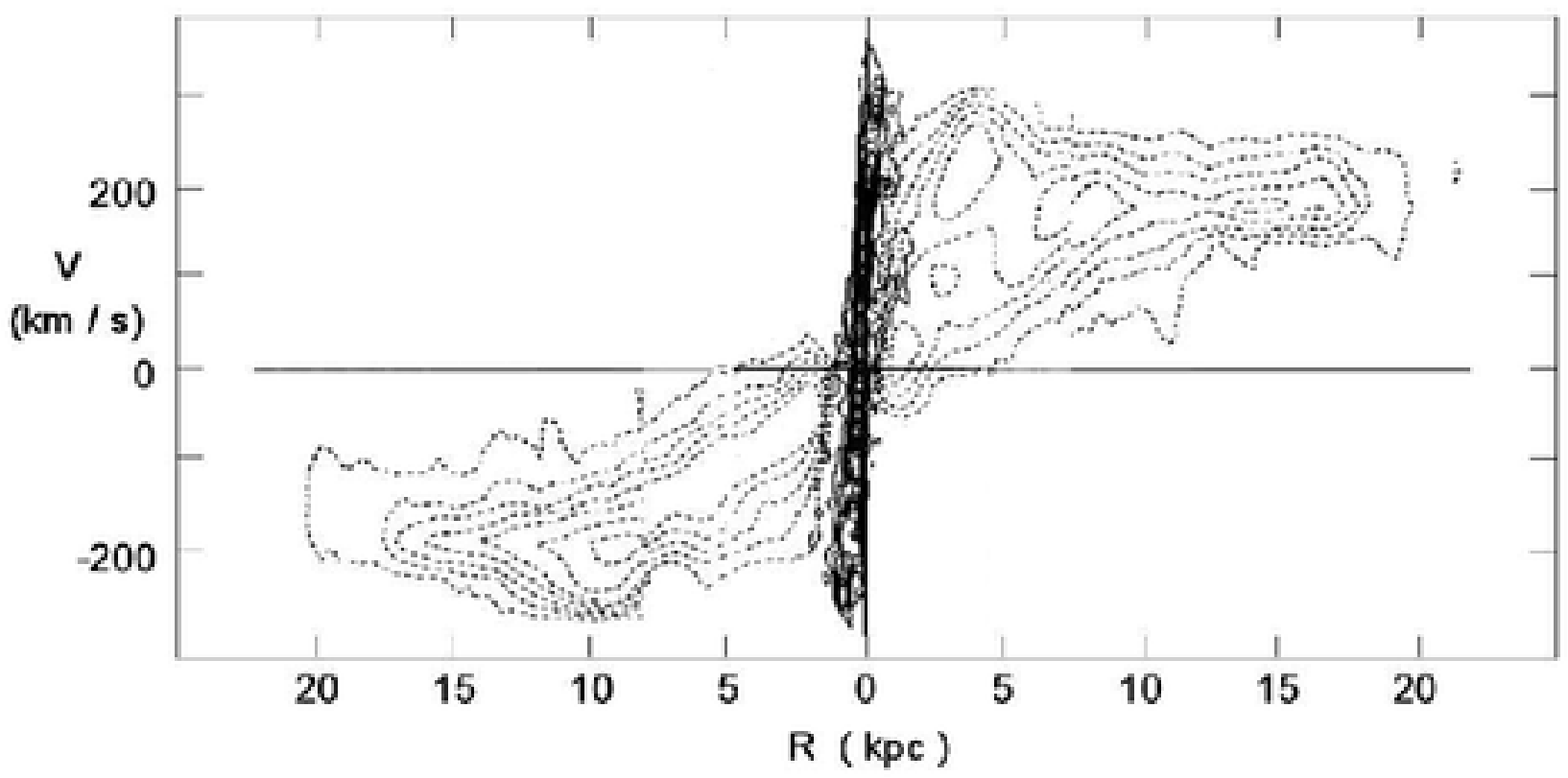}   
\end{center}
\caption{Position-velocity diagram along the major axis of the edge-on galaxy NGC 3079. Dashed and central full contours are from HI (Irwin and Seaquist 1991) and CO ($J=1-0$) (Sofue et al. 2001) line observations, respectively.  }
\label{rc-n3079} 
\end{figure}

The CO rotational transition lines in the millimeter wave
range at 2.63 and 1.32 mm are valuable in studying rotation kinematics of the inner disks not only for the concentration of the molecular gas toward the center (Sofue 1996, 1997). Edge-on and high-inclination galaxies are particularly useful for rotation curve analysis in order to minimize the uncertainty arising from inclination corrections, for which extinction-free radio measurements are crucial. Another advantage of CO spectroscopy is its high spatial and velocity resolutions by interferometric observations (Sargent and Welch 1993; Sofue and Rubin 2001). 
Figure \ref{rc-simu} illustrates by a simple simulation that the inner rotation curves is better traced by CO lines, while the outer rotation curve is well obtained from HI. Fig. \ref{rc-n3079} shows a position-velocity diagram obtained for the  edge-on galaxy NGC 3079 in the HI and CO line interferometer observations.

\vskip 5mm \subsubsection{\bf  Centroid velocity methods}

Rotation curve of a galaxy is defined as the trace of velocities on a position-velocity (PV) diagram along the major axis corrected for the inclination angle between the line-of-sight and the galaxy disk.  There are several methods to derive the rotation velocities from the PV diagrams.

A widely used method is to trace intensity-weighted velocities, which are  defined by
\be
 V_{\rm int}={1 \over {\rm sin}~l} \int I(v) v dv / \int I(v) dv,
\ee
where $I(v)$ is the intensity profile at a given radius as a function of the radial velocity $v$ corrected for the systemic velocity of the galaxy  and  $i$ is the inclination angle. For convenience, the intensity-weighted velocity is often approximated by centroid velocity or peak-intensity velocity, which is close to each other.
The centroid velocity is often obtained by tracing the values on the mean-velocity map of a disk galaxy, which is usually produced from a spectral data cube by taking the 1st moment. The mean velocities are obtained also by tracing values along the major axis of a mean-velocity map (moment 1 map) produced from a spectral data cube.

The obtained rotation velocity is reasonable for the outer disk. However, it is largely deviated from the true rotation speed in the innermost region, where the velocity structure is complicated. It should be remembered that the mean velocity near the nucleus gives always underestimated rotation velocity, because the finite resolution of observation inevitably results in zero value at the center by averaging plus and minus values in both sides of nucleus along the major axis. Hence, the derived rotation curve often starts from zero velocity in the center. But the nucleus is the place where the stars and gases are most violently moving, often nesting a black hole with the surrounding objects moving at high-velocities close to the light speed.

\vskip 5mm \subsubsection{\bf  Terminal velocity method}

This method makes use of the terminal velocity in a PV diagram along the
major axis. The rotation velocity is derived by using the terminal velocity
$V_{\rm t}$:
\be
\Vrot=V_{\rm t} / {\rm sin}~i~ -(\Sigma_{\rm obs}^2 + \Sigma_{\rm ISM}^2)^{1/2},
\ee
where $\Sigma_{\rm ISM}$ and $\Sigma_{\rm obs}$ are
the velocity dispersion of the interstellar gas and the velocity resolution
of observations, respectively.
The interstellar velocity dispersion is of the order of
$\Sigma_{\rm ISM} \sim 5$ to 10 \kms, while $\Sigma_{\rm obs}$ depends on the instruments.

Here, the terminal velocity is defined by a velocity at which the intensity becomes equal to 
$
I_{\rm t}=[(\eta I_{\rm max})^2+I_{\rm lc}^2]^{1/2}
$
on the observed PV diagram, where $I_{\rm max}$ and $I_{\rm lc}$ are the maximum intensity and intensity corresponding to the lowest contour level, respectively, and $\eta$ is usually taken to be $\sim 0.2$\% so that the 20\% level of the intensity profile is traced. If the intensity is weak, the equation gives $I_{\rm t}\simeq I_{\rm lc}$ which approximately defines the loci along the lowest contour level. 

\vskip 5mm \subsubsection{\bf  Iteration Method}

A more reliable method is to reproduce the observed position-velocity diagram by correcting the iteratively obtained rotation curves (Takamiya and Sofue 2000). The method  comprises the following procedure. An initial rotation curve, RC0, is adopted from a PV diagram (PV0), obtained by any method as above (e.g. a peak-intensity method). 
Using this rotation curve and an observed radial distribution of intensity (emissivity) of the line used in the analysis, a PV diagram, PV1, is constructed. The difference between this calculated PV diagram and the original PV0 (e.g. the difference between peak-intensity velocities) is used to correct for the initial rotation curve to obtain a corrected rotation curve RC1.
This RC is used to calculated another PV diagram PV2 using the observed intensity distribution, and to obtain the next iterated rotation curve, RC2 by correcting for the difference between PV2 and PV0.
This iteration is repeated until PV$i$ and PV0 becomes identical, such that the summation of root mean square of the differences between PV$i$ and PV0 becomes minimum and stable. RC$i$ is adopted as the most reliable rotation curve. 

\vskip 5mm \subsubsection{\bf  Three-dimensional cube method for the future}

The methods so far described utilize only a portion of the kinematical data of a galaxy. The next-generation method to deduce galactic rotation would be to utilize three-dimensional spectral data from the entire galaxy. This may be particularly useful for radio line observations, because the lines are transparent at any places in the disk. In the iteration method position-velocity diagrams along the major axis are used. In the three-dimensional method, the entire spectral data will be employed, which consist of spectra at all two-dimensional grids in the galactic disk on the sky. Such data are usually recorded as a spectral cube, or intensities $I$ at all points in the $(x, y, v)$ space, as obtained by spectral imaging observations of radio lines such as CO and HI. Here, $(x, y)$ is the position on the sky and $v$ is radial velocity. 

The reduction procedure would be similar to that for the iteration method: First, an approximate rotation curve is given, and a spectral cube is calculated from the curve based on the density distribution already derived from the projected intensity distribution in the galaxy on the sky. Next, the calculated cube (C) is compared with the observed cube (O) to find the difference. Then, the assumed rotation curve is iteratively corrected so that the difference between the O and C get minimized. The density distribution itself may be taken as another unknown profile to be measured during the iteration in addition to the rotation curve.

\vskip 5mm \subsection{\bf Rotation curves of Spiral Galaxies}

\begin{figure}
\bc
\includegraphics[width=9cm]{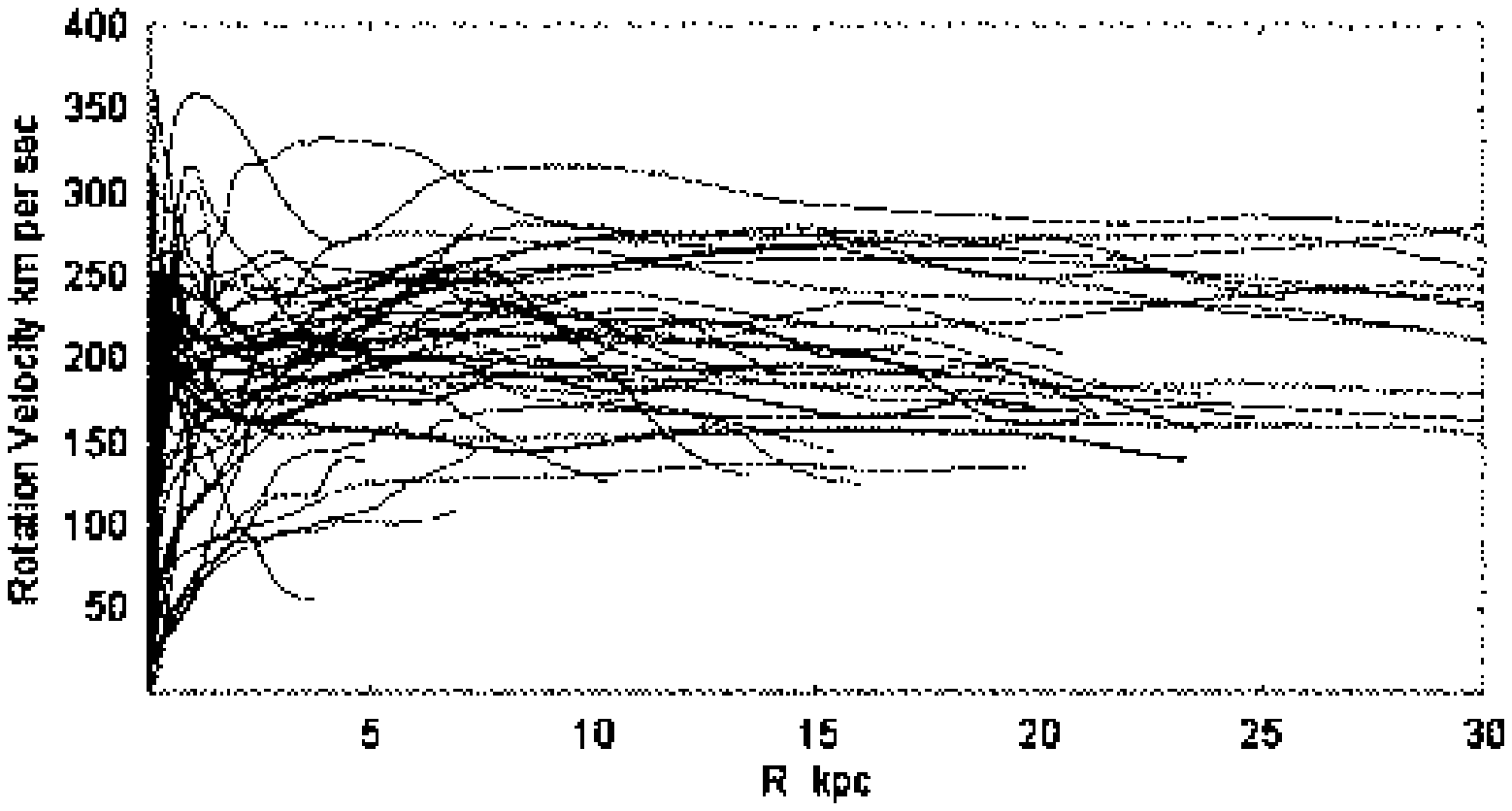}   \\ 
\ec
\caption{ Rotation curves of spiral galaxies obtained by combining CO data for the central regions, optical for disks, and HI for outer disk and halo (Sofue et al. 1999).
}
\label{rc-all} 
\end{figure}

Fig. \ref{rc-all} shows the rotation curves in nearby spiral galaxies, which have been obtained mainly by the terminal-velocity methods from optical, CO and HI line data (Sofue et al. 1999). Fig. \ref{rc-type} shows rotation curves for individual galaxies types from Sa to Sc. 
It is remarkable that the form, but not amplitude, of the disk and halo rotation curves is similar to each other for different morphologies from Sa to Sc.
This suggests that the form of the gravitational potential in the disk and halo is not strongly dependent on the galaxy type.
A moderate correlation is found between total luminosity and maximum rotation velocity, which is known as the Tully-Fisher relation (Tully and Fisher 1977; Mathewson et al. 1992, 1996).
Less luminous galaxies tend to show increasing outer rotation curve,
while most massive galaxies have flat or slightly declining rotation in the
outmost part (Persic et al. 1996).

\vskip 5mm \subsubsection{\bf  Sa, Sb, Sc Galaxies}

The form of central rotation curves depend on the total mass and galaxy types (Sofue et al. 1999). Massive galaxies of Sa and Sb types show a steeper rise and higher central velocities within a few hundred pc of the nucleus compared to less massive Sc galaxies and dwarfs (Noordmere et al. 2007). Dwarf galaxies generally show a gentle central rise. This is related to the bulge-to-disk mass ratio: the smaller is the ratio, and therefore, the later is the type, the weaker is the central rise (Rubin et al. 1985; Casertano \& van Gorkom 1991).

For massive Sb galaxies, the rotation maximum appears at a radius of 5 or 6 kpc, which is about twice the scale radius of the disk. Beyond the maximum, the rotation curve is usually flat, merging with the flat part due to the massive dark halo. Fluctuations of a few tens of \kms\ due to non-axisymmetric structures such as spiral arms and bars are superposed as velocity ripples. The fluctuations of order of 50 \kms are often superposed for barred galaxies, which indicate non circular motions in the oval potential.
  
There have been several attempts to represent the observed rotation curves by simple functions (Persic et al. 1996; Courteau 1997; Roscoe 1999). 
Persic et al. fit the curves by a formula, which is a function of total luminosity and radius, comprising both disk and halo components. Both the forms and amplitudes are functions of the luminosity, and  the outer gradient of the RC is a decreasing function of luminosity.

\bfig
\bc
\includegraphics[width=7cm]{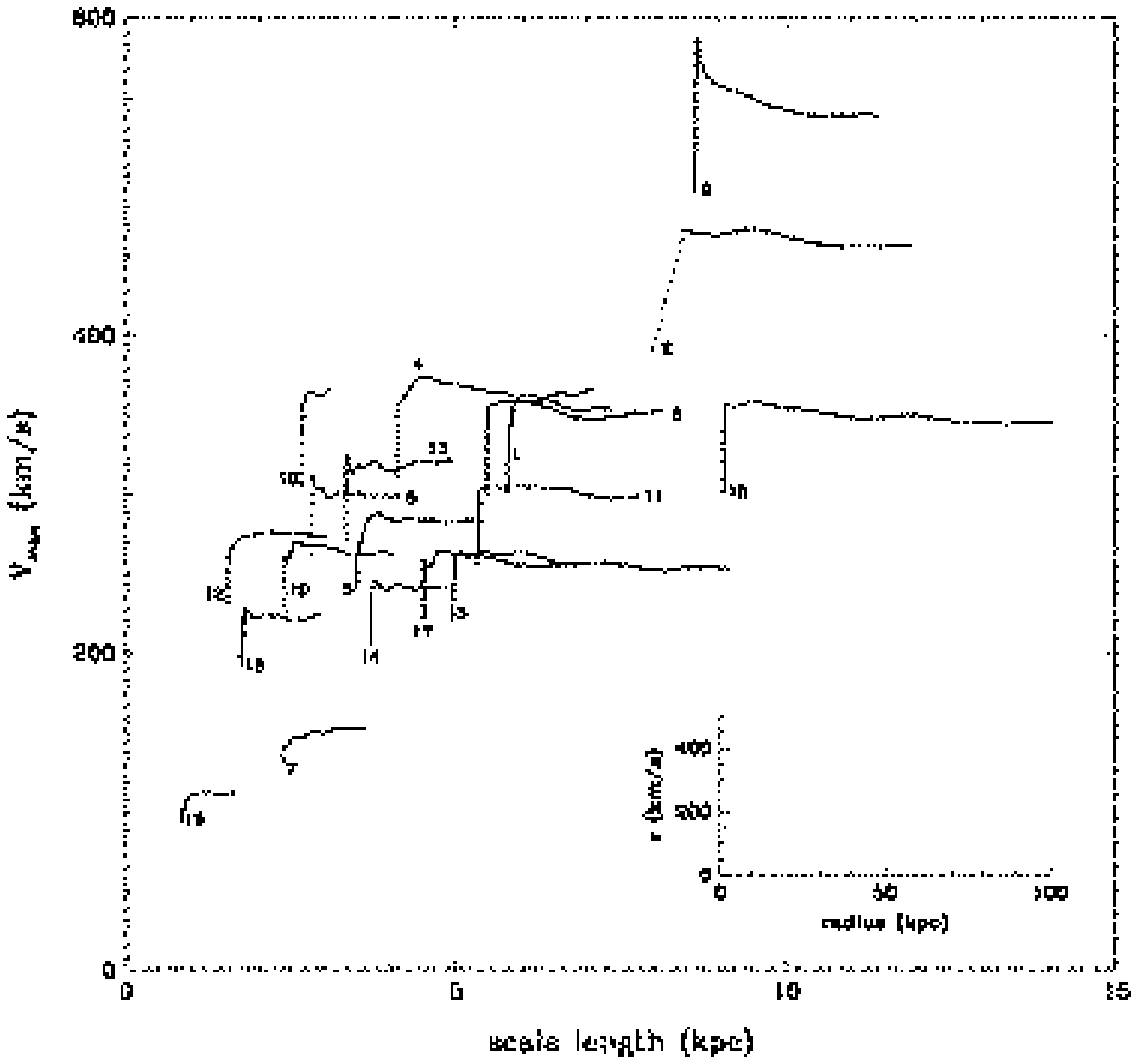}\\
\includegraphics[width=6cm]{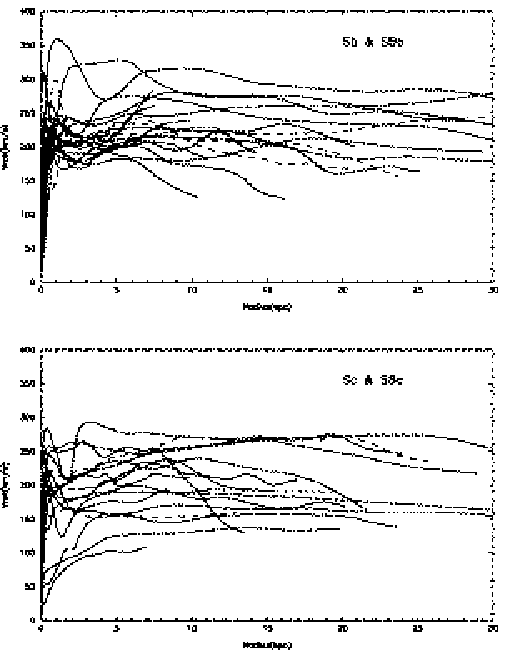}  \\ 
\ec
\caption{[Top] Rotation curves of early type spiral galaxies (Sa) listed on the maximum velocity-scale radius plane (Noordmere et al. 2007). [Middle]: Rotation curves for Sb galaxies (full lines) and barred SBb galaxies (dashed lines).  
[Lower panel]: Same, but for Sc and SBc galaxies (Sofue, et al. 1999) }
\label{rc-type} 
\ef

Rotation curves of Sa and Sb galaxies, including the Milky Way, can be summarized to have: 
(a) high-velocity rotation or dispersion in the nucleus due to a high-density core and massive black hole, with declining velocity to a minimum at $R\sim$ a few tens pc;
(b) steep rise of RC within the central 100 pc due to the bulge;
(c) maximum at radius of a few hundred pc due to the bulge, followed by a decline to a minimum at 1 to 2 kpc; then,
(d) gradual rise to the maximum at 5 to 7 kpc due to the disk; and
(e) nearly flat outer rotation due to the dark halo up to $R\sim 20-30$ kpc.
Earlier type galaxies have higher maximum rotation velocities than later type galaxies: Sa galaxies have maximum velocities of around 300 \kms and Sb 220 (Rubin et al. 1985). Figure \ref{rc-type} (top panel) shows the observed rotation curves for Sa galaxies listed on the maximum velocity-scale radius plane (Noordmere 2007). The diagram shows that the larger is the scale radius, the higher is the rotation velocity.

Sc galaxies have lower maximum velocities than Sa and Sb, ranging from $\le 100$ to $\sim 200$ \kms with the median value of 175 \kms (Rubin et al. 1985).
Massive Sc galaxies show a steep nuclear rise similar to Sb's, while less-massive Sc galaxies have gentler rise. They also have a flat rotation to their outer edges. Less luminous (lower surface brightness) Sc galaxies have a gentle central rise of rotation velocity, which monotonically increases till the outer edge. This behavior is similar to rotation curves of dwarf galaxies.

\vskip 5mm \subsubsection{\bf  Barred Galaxies}

Barred spiral galaxies constitute a considerable fraction of all disk galaxies.  Large-scale rotation properties of SBb and SBc galaxies are generally similar to those of non-barred galaxies of Sb and Sc types. However, their kinematics is more complicated due to the non-circular streaming motion by the oval potential, which results in skewed velocity fields and ripples on the rotation curves. 
Large velocity variation arises from the barred potential of several kpc length on  the order of $\pm \sim 50 - 100$ \kms  (e.g., Kuno et al.  2000). 
Simulations of PV diagrams for edge-on barred galaxies show many large velocity fluctuations superposed on the flat rotation curve (Bureau and Athanassoula 1999; Weiner \& Sellwood 1999). However, distinguishing the existence of a bar and quantifying it are not uniquely done from such limited edge-on information.
For more quantitative results, two-dimensional velocity analyses are
necessary (Wozniak \& Pfenniger 1997). 
However, in contrast to the basic mass structures as the bulge, disk and halo, determination of the mass distribution in the bar from observed data is still not straightforward because of its large number of parameters to be fixed, which are the bar's three axial lengths, major axis direction, density amplitude and mass distribution. Also, its dynamical connection to the disk and bulge is a subject for further investigation. 

\vskip 5mm \subsubsection{\bf  Dwarf galaxies}

The Large Magellanic Cloud is the nearest dwarf galaxy, which shows a flat rotation curve from the dynamical center to the outer edge at $R\sim 5$ kpc at a velocity $\sim 100$ \kms (Sofue 1999). This galaxy reveals a particular kinematical property: The dynamical center is significantly displaced from the optical bar center, indicating the existence of a massive bulge-like density component that is not visible. The Small Magellanic Cloud shows complex velocity field spitted into two velocity components. The velocity data are not appropriate to discuss the mass in this very disturbed galaxy.

Dwarf and low surface brightness galaxies show slow rotation at $\le 100$ \kms with monotonically rising rotation curves their last measured points (de Blok 2005; Swaters et al. 2009; Carignan \& Freeman 1985;  
Blais-Quellette et al. 2001). Due to the low luminosity, the mass-to-luminosity ratio is usually higher than that for normal spiral, and the dark matter fraction is much higher in dwarf galaxy than in spirals (Carignan 1985; Jobin \& Carignan 1990).  

\vskip 5mm \subsubsection{\bf  Irregular Galaxies}  

The peculiar morphology of irregular galaxies is mostly produced by gravitational interaction with the companion and/or an encountered. When galaxies gravitationally interact, they tidally distort each other, and produce the peculiar morphology that had until recently defied classification.  Toomre \& Toomre (1972) for the first time computed tidal interactions to simulate some typical distorted galaxies.

The starburst dwarf galaxy NGC 3034 (M82) shows an exceptionally peculiar rotation property (Burbidge et al. 1964; 1975). It has a normal nuclear rise and rotation velocities  which have a Keplerian decline beyond the nuclear peak. This may arise from a tidal truncation of the disk and/or halo by an encounter with M81 (Sofue 1998).
   
\section{GALACTIC MASS DISTRIBUTION}

\vskip 5mm \subsection{\bf Approximate mass distribution}

\vskip 5mm \subsubsection{\bf Flat rotation and isothermal mass distribution}

It has been shown that the rotation curve of the Galaxy as well as those of most spiral galaxies are nearly flat, indicating that the rotation velocity $V$ is approximately constant in a galaxy. Before describing the detailed mass models, an approximate mass profile in the Galaxy may be estimated from the flat characteristics of the entire rotation curve. Simply assuming a spherical distribution, the mass involved within a radius $R$ is approximated by 
\begin{equation}
 M(R)\sim {RV(R)^2 \over G}.
\end{equation} 
If the rotation velocity $V$ is constant, remembering that $M(R)=2 \pi \int _0^R \Sigma r dr$ and $M(R)=4 \pi \int_0^R \rho r^2 dr$, the surface mass density (SMD) is approximately given by 
\be
\Sigma \propto R^{-1},
\ee
 and volume mass density behaves similarly to an isothermal gas sphere as
\be
\rho \propto R^{-2}.
\ee
 These are very rough and the first-order mass distributions in a galaxy having flat rotation curve.
 
\vskip 5mm \subsubsection{\bf Vertical mass distribution near the Sun} 

\begin{figure} \bc
\includegraphics[width=8cm]{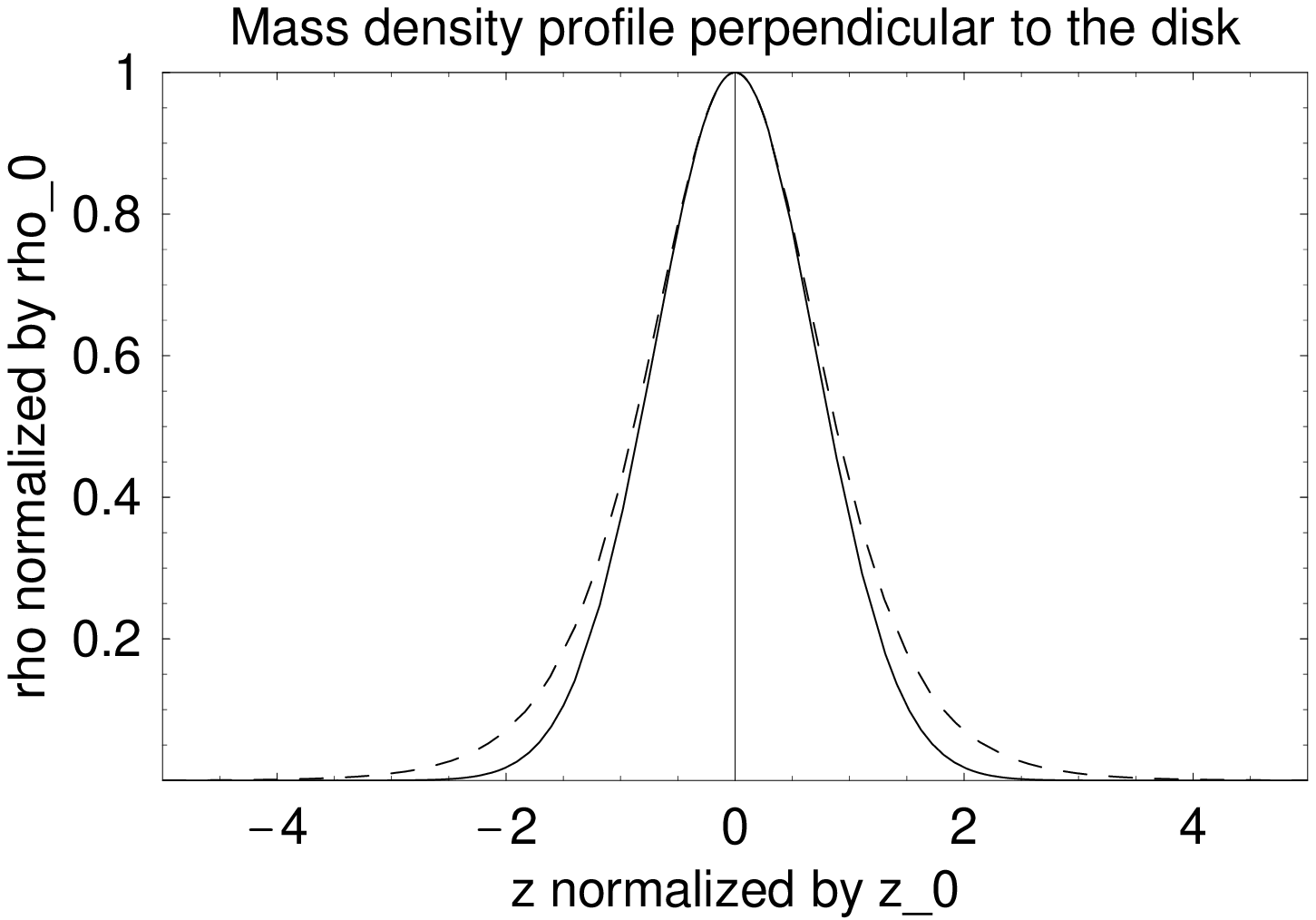} 
\ec
\caption{Mass density profile perpendicular to the local disk. The solid curve represents the Gaussian profile for a hydrostatic disk, and the dashed line is for a self-gravitating disk (Spitzer 1942). }
\label{rho-z}
\end{figure}

If the local stellar distribution is assumed to be a flat disk around the galactic plane, an approximate density profile perpendicular to the galactic plane in the solar vicinity can be calculated without considering the galactic rotation.  Since motions of stars and gas are deviating from circular orbits, the deviation acts as the pressure, which yields the disk thickness. The local distribution of stars and gas perpendicular to the galactic plane is approximated by static equilibrium between the $z$-directional gravity (perpendicular to the disk) and the velocity dispersion in the $z$ direction $v_z$. By denoting the equivalent pressure of the matter in the disk as $p=\rho v_z^2$, the equilibrium in the $z$ direction can be written as 
\begin{equation}
 {dp \over dz}={d (\rho v_z^2) \over dz} = g_z \rho,
\end{equation}
where $g_z$ and $\rho$ are the $z$-directional gravitational acceleration and the density of the matter (stars and gas). If the galactic disk is assumed to be a flat plane of infinite surface area, the gravity can be approximated as 
\begin{equation}
 g_z \sim - \alpha z,
\end{equation}
and therefore, 
\begin{equation}
 {d \rho \over dz} \sim - {\alpha z \over v_z^2} \rho.
\end{equation}
This equation can be solved if the velocity dispersion is constant with $z$: 
\be
 \rho = \rho_0 ~\exp \left( -{z^2 \over z_0^2} \right),
\label{eq-rho-gauss}
\ee
where $\rho_0$ and $z_0$ are constants depending on the species.
If the Poisson equation is solved for a self-gravitating flat disk, a slightly different profile (Spitzer 1942) is obtained as
\be
\rho=\rho_0 {\rm sech}^2 \left( {z \over z_0} \right).
\label{eq-rho-sech}
\ee
Here, $z_0$, is the $z$-directional scale height representing the typical thickness of the disk, and is related to the velocity dispersion as
\be
 z_0 = {v_z \over \sqrt{\alpha} }  \sim \sqrt{{k \over \alpha m_{\rm H}} T_{\rm gas}}.
\ee
In Fig. \ref{rho-z} the two results from Eq. (\ref{eq-rho-gauss}) and (\ref{eq-rho-sech}) are compared. 
Table \ref{tabscaleheight} summarizes the typical values of $v_z$ and the disk scale-thickness $z_0$ for various species. Population I objects are distributed in a disk of thickness $2 z_0 \sim 100$ pc, and Population II stars in a disk with thickness $2 z_0 \sim 300$ pc. See table \ref{tab_local} for more recent observed values.
 
\begin{table}[htbp]
\bc
\begin{tabular}{lll}
\hline
   Component & $v_z$ (\kms) & Thickness, $2z_0$ (pc) \\
\hline
   Molecular clouds & 5 & 60  \\
   OB stars & 5  &60 \\
   HI gas & 10 & 100 \\
   Disk stars & 20-30 & 300 \\
   Hot gas ($10^6$K) & 100 ($\sim \sqrt{{k \over m_{\rm H}}T}$) & $\sim 2 $ kpc \\
\hline
\end{tabular}
\ec
\caption{Velocity dispersion in the $z$ direction and scale thickness of
the disk.}
\label{tabscaleheight} 
\end{table}

\vskip 5mm \subsection{\bf Decomposition Methods}

The observed rotation curve is used to derive the mass distribution in the Galaxy. There are two ways to derive the mass distribution from the rotation curve. 

{\it Decomposition method:} One way is to fit the observed rotation curve by a calculated one assuming that the galaxy is composed of several components, such as a central bulge, disk, and a dark  halo. The total rotation velocity is given by summing up the squares of individual corresponding velocities:
\be
V(R)=\sqrt{\Sigma V_i^2}
\simeq \sqrt{V_{\rm b}(R)^2+V_{\rm d}(R)^2+V_{\rm h}(R)^2}.
\ee
If the resolution in the central region is sufficiently high, a more central component such as a black hole such as $V_{\rm BH}(R)^2$ may be added to the right-hand side. Here, $V_{i}$ indicates circular velocity corresponding to the $i$-th component alone, and suffices BH, b, d and h denote a black hole, bulge, disk, and halo components, respectively. For the fitting, the parameters  such as the mass and scale radius of individual components are adjusted so that the residual between the calculation and observation gets minimized. The fitting is usually started from the inner component. First, the innermost steep rise and peak of the rotation curve is fitted by a bulge component;  second, the gradual rise and flat part is fitted by the disk, and finally, the residual outskirt is fitted by a dark halo. Alternatively, these fittings may be done at the same time by giving approximate parameters, which are adjusted iteratively until the entire curve is best fitted.

{\it Direct method:} Another way is to calculate the mass distribution directly from the rotation curve. This method gives more accurate mass distributions without being affected by the assumed components and their functional forms. 

\vskip 5mm \subsection{\bf  Decomposition into bulge, disk and halo} 

The rotation curve of the Milky Way in Fig. \ref{fig-obs} shows clearly three dominant components: the galactic bulge peaking at $R\sim 300$ pc, disk broadly peaking at $R\sim 6$ kpc, and outer flat part due to the dark halo till the outer edge. The black hole does not contribute so much in the plotted scale, and may be treated as an independent object in the very central region within $\sim 1$ pc. These characteristics are commonly observed in most of external disk galaxies, and may be considered to be the universal property. Rotation curves, and therefore the mass distributions, in any galaxies including the Milky Way are, thus, similar to each other. Hence, the same functional forms used for the Galaxy may be applied to any other galaxies, and vice versa, by modifying the  parameters such as the masses and scale radii of individual   components.
Decomposition of a rotation curve into several mass components such as a bulge, disk and dark halo has been extensively applied to observed data (Bosma 1981a;  Kent 1986; Sofue 1996). 

There are various functional forms to represent the mass components.  The commonly employed mass profiles are:\\
\noindent (i) {\it de Vaucouleurs bulge + exponential disk:}  
It is well established that the luminosity profile of the spheroidal bulge component in galaxies is represented by the \dv\ ($e^{-(r/r_e)^{1/4}}$; 1953, 1958) or  S{\'e}rsic ($e^{-(r/r_e)^n}$; 1968) law.  
The disk component is represented by the exponential law ($e^{-r/r_e}$  ($n=1$); Freeman 1970). For the halo, either isothermal, Navaro-Frenk-White, or Burkert profile are adopted. \\
\noindent (ii) {\it Miyamoto-Nagai potential:} Another way is to assume a modified Plummer type potential for each component, and the most convenient profile is given by the Miyamoto-Nagai (1975) potential.

After fitting the rotation curve by the bulge, exponential disk, and a dark halo, smaller structures superposed on the rotation curves are discussed as due to more local structures such as spiral arms and/or bar.  

\vskip 5mm \subsubsection{\bf de Vaucouleurs Bulge} 
 
\begin{figure} 
\bc
\includegraphics[width=8cm]{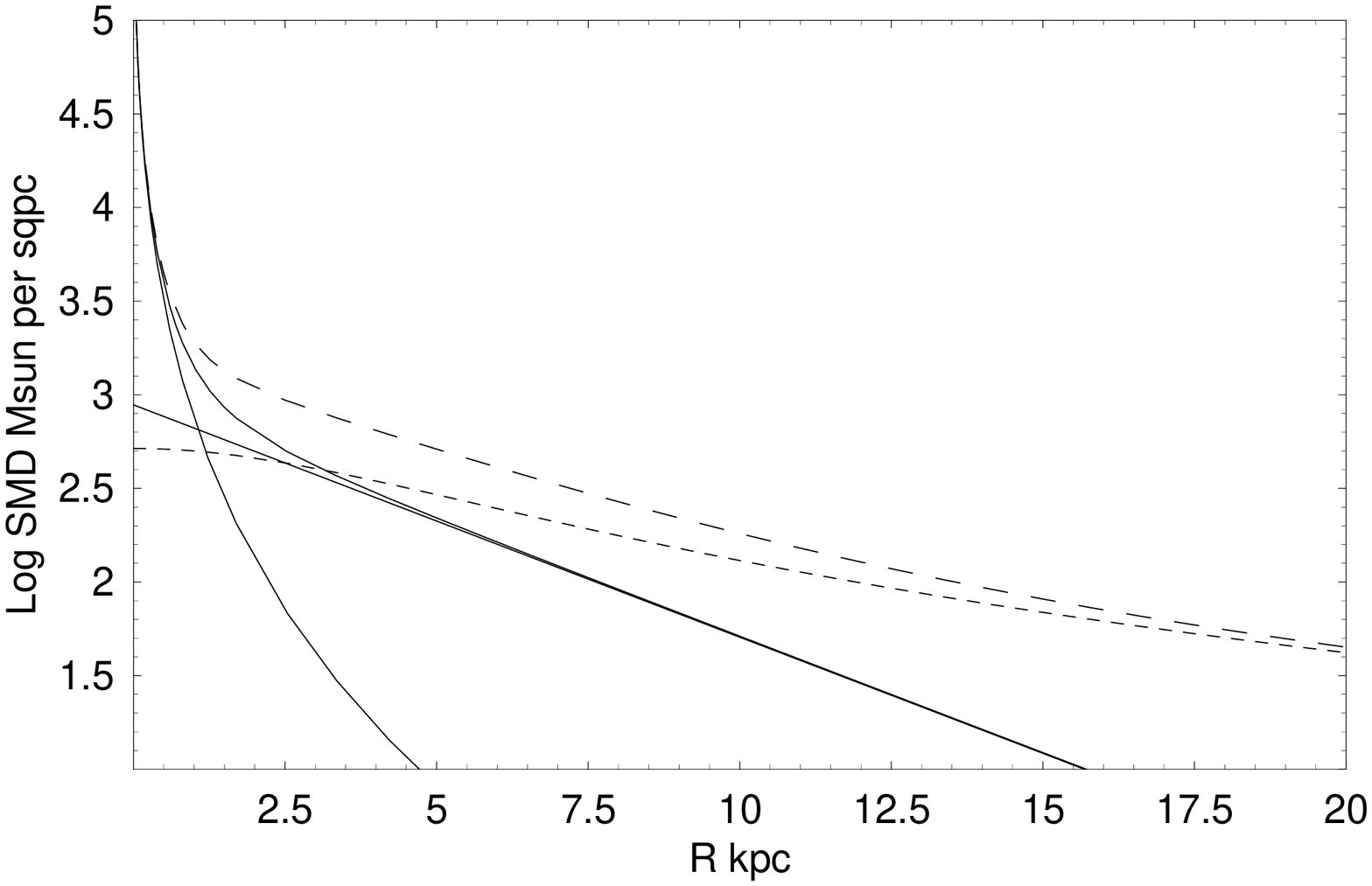}  \\ 
\ec
\caption{Surface mass density (SMD) distributions in the Galaxy. The thin curve shows the \dv  bulge, and the straight line the exponential disk. Their sum is also shown. The dashed line is the dark halo, and long-dashed line indicates the total SMD. High density at the center due to the bulge reaches $6.8\times10^6\Msun{\rm pc}^{-2}$. 
}
\label{fig-smd}  
\end{figure}

The most conventional method among the decomposition methods is to use functional forms similar to luminosity profiles. 
The inner region of the galaxy is assumed to be composed of two luminous components, which are a bulge and disk. The mass-to-luminosity ratio within each component is assumed to be constant, so that the mass density distribution has the same profile. The bulge is assumed to have a spherically symmetric mass distribution, whose surface mass density obeys the \dv\ law, as shown in Fig. \ref{fig-smd}. 

The \dv\ (1958) law for the surface brightness profile as a function of the projected radius $R$ is expressed by
\be
{\rm log} \beta = - \gamma(\alpha^{1/4}-1),
\label{eq-dv}
\ee
with  $\gamma=3.3308$. Here, $ \beta=B_{\rm b}(R)/B_{\rm be}$,  $ \alpha=R/R_{\rm b} $, and $B_{\rm b}(R)$ is the brightness distribution normalized by $B_{\rm be}$, which is the brightness at radius $R_{\rm b}$.  
The same \dv\ profile for the surface mass density is adopted as
\be \Sigma_{\rm b}(R)=\lambda_{\rm b} B_{\rm b}(R)= \Sigma_{\rm be} {\rm exp} \left[-\kappa \left(\left(R \over R_{\rm b} \right)^{1/4}-1\right)\right]
\label{eq-smdb}
\ee 
with $ \Sigma_{\rm bc} = 2142.0 \Sigma_{\rm be} $ for $\kappa=\gamma {\rm ln} 10=7.6695$.  
Here, $\lambda_{\rm b}$ is the mass-to-luminosity ratio, which is assumed to be constant within a bulge.  
Equations (\ref{eq-dv}) and (\ref{eq-smdb}) has a particular characteristics: The central value at $r=0$ is finite, and the function decreases very steeply with radius near the center. However, the gradient gets milder as radius increases, and the SMD decreases very slowly at large radius, forming an extended outskirt. The function well represents the brightness distribution in spheroidal components and elliptical galaxies that have a strong concentration towered the center with finite amplitude, while the outskirt extends widely.

The total mass is calculated by
\be M_{\rm bt}= 2 \pi \int_0^\infty r \Sigma_{\rm b}(r) dr =\eta R_{\rm b}^2 \Sigma_{\rm be},
\ee
where $\eta=22.665$ is a dimensionless constant. By definition a half of the total projected mass (luminosity) is equal to that inside a cylinder of radius $R_{\rm b}$.

In order to describe the bulge component in the Galaxy, it is often assumed that the bulge is spherical having the \dv type profile.  The volume mass density $\rho(R)$ at radius $r$ for a spherical bulge is calculated by using the surface density as 
\be
\rho(R) = {1 \over \pi} \int_R^{\infty} {d \Sigma_b(x) \over dx} {1 \over \sqrt{x^2-R^2}}dx.
\label{eq-rhob}
\ee 
Since the mass distribution   is assumed to be spherical, the circular velocity is calculated from the total mass enclosed within a sphere of radius $R$:
\be
 V_{\rm b}(R) = \sqrt{ GM_{\rm b}(R) \over R} .
 \ee
 
The velocity approaches the Keplerian-law value at radii sufficiently greater than the scale radius. The shape of the rotation curve is similar to each other for varying total mass and scale radius. For a given scale radius, the peak velocity varies proportionally to a square root of the mass. For a fixed total mass, the peak-velocity position moves inversely proportionally to the scale radius along a Keplerian line. 

Decomposition of rotation curves by the $e^{-(R/r_e)^{1/4}}$ law surface mass profiles have been extensively applied to spheroidal components of late type galaxies (Noordermeer 2007, 2008).  S\'ersic (1968) has proposed a more general form  $e^{-(R/r_e)^n}$ for spheroidal luminosity distributions. The $e^{-R^{1/4}}$ law was fully discussed in relation to its dynamical relation to the galactic structure based on the more general profile (Ciotti 1991; Trujillo 2002). 

Fig. \ref{fig-rc-arm} shows the calculated rotation curve for the \dv  bulge model. The result shows a reasonable fit of the inner rotation curve. It reproduces the central steep rise and sharp peak at $R=300$ pc, which is due to  the high density mass concentration in the central region. 

The best-fit total mass of the bulge of our Galaxy is as large as $1.8\times10^{10}\Msun$ and the scale radius is as compact as $R_b=0.5$ kpc with the accuracy of about 5\%. As Fig. \ref{fig-smd} indicates, the surface mass density in the central 0.5 kpc is dominated by the bulge component, and nuclear surface density reaches a value as high as $6.8 \times 10^6 \Msun{\rm pc}^{-2}$. The total projected mass included in the central 500 pc (scale radius) is $9.0 \times 10^9 \Msun$. The total spheroidal mass integrated in a sphere of $R_{\rm b}$ is 0.39 times the total bulge mass, which is for the present case $7.0 \times 10^{9}\Msun$. 

\vskip 5mm \subsubsection{\bf M/L ratio in bulge}

The surface mass distribution in Fig. \ref{fig-smd}  may be compared with the $K$ band surface brightness profile for the inner Galaxy at $| b |<10^\circ$ as presented by Kent et al. (1991, 1992). The $K$ band luminosity profile for the central 1 kpc is expressed as  
\be
\nu_K=1.04 \times 10^6 (r/0.482 {\rm pc})^{-1.85} L_\odot {\rm pc}^{-3}. \label{Kent92}
\ee
This expression approximates the density profile for an isothermal sphere with a constant mass-to-luminosity ratio, for which the power-law index is $-2$. Since the functional form is different, it cannot be compared directly with the \dv profile. However, it may be worthy to compare the integrated luminosity within a radius with the corresponding mass. The integrated luminosity within a sphere of radius 0.5 kpc from Eq. (\ref{Kent92}) is calculated to be  $3.74 \times 10^9 L_\odot$. Thus, a mass-to-luminosity ratio for the bulge within a sphere of the scale radius, $R_{\rm b}=0.5$ kpc, is obtained to be $M/L({\rm bulge ~in~ 0.5 kpc})\simeq 7.1\Msun/L_\odot$.

As Fig. \ref{fig-smd} indicates, the volume density increases rapidly toward the Galactic Center, approaching an infinite value. Note that the SMD at the center is finite as represented by Eq. (\ref{eq-smdb}), whereas the volume density diverges to infinity at the center as in Eq. (\ref{eq-rhob}). The central mass within 1 pc is estimated to be as high as several $10^6 \Msun$.  

\vskip 5mm \subsubsection{\bf Exponential Disk}

The galactic disk is represented by an exponential disk (Freeman 1970). The surface mass density is expressed as
\be 
\Sigma_{\rm d} (R)=\Sigma_{dc} {\rm exp}(-R/R_{\rm d}), 
\label{eq-smdd}
\ee 
where $\Sigma_{dc}$ is the central value, $R_d$ is the scale radius. The total mass of the exponential disk is given by $M_{\rm disk}= 2 \pi \Sigma_{dc} R_{\rm d}^2$.
The rotation curve for a thin exponential disk without perturbation, e.g.  $\Delta=0$, is expressed by (Freeman 1970; Binney and Tremaine 1987).  
\be
V_{\rm d}=\sqrt{R{\partial \Phi \over \partial R}}
=\sqrt{4 \pi G \Sigma_0 R_{\rm d}y^2[I_0(y)K_0(y)-I_1(y)K_1(y)]},
\ee
where $y=R/ (2R_{\rm d}) $, and $I_i$ and $K_i$ are the modified Bessel functions.

If the rotation curve is affected by additional masses $\Delta$ due to arms, rings, bar, and/or interstellar gas, the surface-mass density is represented as
\be 
\Sigma_{\rm d} (R)=\Sigma_{dc} {\rm exp}(-R/R_{\rm d})+\Delta, 
\label{eq-smddelta}
\ee 
The gravitational force $f(R)$ acting on a point at galacto-centric distance $x=R$ is then directly calculated by integrating the $x$ directional component of force due to a mass element $\Sigma_{\rm d}' (x) dx dy$ in the Cartesian coordinates $(x,y)$:
\be 
f(R)=G \int_{-\infty}^\infty \int_{-\infty}^\infty 
{\Sigma_d (x) (R-x) \over s^3} dx dy ,
\ee
where $ s=\sqrt{(R-x)^2+y^2} $ is the distance between the mass element and the point.  The rotation velocity is given by 
\be
V_d(R) = \sqrt{f R}.
\ee

\vskip 5mm \subsubsection{\bf  Dark Halo}

Fitting by a dark halo model within $\sim 20$ kpc is crude not only because of the large scatter and errors of observed rotation velocities in the outer disk, but also for the weaker response of the rotation curve to the halo models. The data may be fitted by an isothermal halo model with the asymptotic velocity of 200 \kms at infinity and a scale radius of $h=10$ kpc. Reasonable fit is also obtained with the same scale radius for the NFW and Burkert profiles, for which the rotation velocity at 20 kpc is approximately 200 \kms. The result is weakly dependent on the adopted model. The dark halo models are described in detail in the next section, the rotation curve is combined with radial velocities of objects surrounding the Galactic halo such as globular clusters, satellite galaxies, and member galaxies in the Local Group. 

\vskip 5mm \subsection{\bf Galactic Mass Parameters} 

Using the rotation curve of the Galaxy as in Fig. \ref{fig-obs} and adopting the classical decomposition method, the parameters are obtained for individual mass components of the bulge, disk and dark halo. The functional form of the bulge was so adopted that the surface mass density is represented by the \dv\ law. The disk was approximated by an exponential disk, and the halo by an isothermal sphere. The observed characteristics are well fitted by superposition of these components. The central steep rise and the high rotation peak at $R= 300$ pc is quite well reproduced by the \dv\ bulge of half-mass scale radius $R_b=0.5$ kpc. The broad maximum at around $R\sim 6$ kpc was fitted by the exponential disk, and the flat outer part by a usual dark halo. 
Table \ref{tab_milkyway} lists the fitting parameters for individual mass components. Since the used data in Fig. \ref{fig-obs} were compiled from different observations, their errors are not uniform, and only eye-estimated values are given, which were evaluated after trial and error of fitting to the observed points. 

\begin{table*}
\begin{center}
\caption{Parameters for Galactic mass components$^*$ } 
\begin{tabular}{llllll}
\hline\hline   
Component & Parameter & Value$^{**}$ \\
\hline
Massive black hole$^\dagger$ & Mass& $M_{\rm BH}=3.7\times 10^6 \Msun$  \\
\hline
Bulge &Mass & 	$M_{\rm b}=1.80\times10^{10}\Msun$  \\
&Half-mass scale radius & $R_b=0.5$ kpc   \\
&SMD at  $R_{\rm b}$ &$\Sigma_{\rm be}=3.2 \times 10^3 \Msun{\rm pc}^{-2}$  \\
&Center SMD & 	$\Sigma_{\rm bc}=6.8\times10^6 \Msun{\rm pc}^{-2}$   \\
&Center volume density & $\rho_{\rm bc}=\infty$  	\\
 \hline  
Disk &Mass & $M_{\rm d}=6.5\times10^{10} $    \\
&Scale radius & $R_{\rm d}=3.5$ kpc   \\
&Center SMD & $\Sigma_{\rm dc}=8.44 \times 10^2 \Msun{\rm pc}^{-2}$  \\
&Center volume density & $\rho_{\rm dc}=8 \Msun{\rm pc}^{-3}$   \\
 \hline 
Bulge and disks &  Total mass & $M_{\rm b+d}=8.3\times10^{10}\Msun$ \\ 
\hline 
Dark halo 
&Mass in $r \le 10$ kpc sphere & $M_{\rm h}(\le 10{\rm kpc})=1.5\times10^{10}\Msun$ \\
(Isothermal sphere)& Mass in $r \le 20$ kpc sphere$^\ddagger$  & $M_{\rm h}(\le 20{\rm kpc})=7.1 \times 10^{10} \Msun$  \\
&Core radius	& $h=12$ kpc    \\
&Central SMD in $|z|< 10$ kpc& $\Sigma_{\rm hc}=4.4 \times 10^2 \Msun{\rm pc}^{-2}$   \\
  &Central volume density & $\rho_{\rm hc}=5.1 \times 10^{-3} \Msun{\rm pc}^{-3}$    \\
  &Circular velocity at infinity  & $V_\infty=200$ \kms  \\ \hline 
Total Galactic Mass 
 & Mass in $r\le 20$ kpc sphere  & $ M_{\rm total}(\le 20 {\rm kpc})=1.5 \times 10^{11} \Msun $  \\
 & Mass in $r\le 100$ kpc NFW sphere  & $ M_{\rm total}(\le 100 {\rm kpc})=3 \times 10^{11} \Msun $  \\
 \hline
\end{tabular} 
\end{center}
$*$ Sofue et al. (2009) \\
$**$ Uncertainty is $\sim 10$\% for bulge and disk, and $\sim 20$ \% for halo. \\ 
$\dagger$ Genzel et al. (2000) \\
$\ddagger$ Mass within 20  kpc is weakly dependent on the halo models, roughly equal to those for NFW and Burkert models. At larger distances beyond 30 kpc, it varies among the halo models (Sofue 2009).
\label{tab_milkyway}
\end{table*} 

\begin{table*}
\caption{Local values near the Sun at $R=R_0 $ (8.0 kpc).} 
\begin{center}
\begin{tabular}{lllll}
\hline\hline  
& Components & Local values   \\
\hline 
Surface Mass Density & Bulge (\dv) & $\Sigma_{\rm b}^\odot =1.48\msqpc$\\
& Disk (exponential; including gas) &  $\Sigma_{\rm d}^\odot =87.5 \msqpc$ \\
& Interstellar gas (HI + H$_2$) & $\Sigma_{\rm gas}^\odot = 5.0 \msqpc$\\
& Bulge+Disk & $\Sigma_{\rm bd}^\odot = 89 \msqpc$\\
& Dark halo (isothermal, $|z|<10$ kpc)& $\Sigma_{\rm halo}^\odot= 3.2 \times 10^2 \msqpc$\\
& Total (bulge+disk+halo)& $\Sigma_{\rm total}^\odot= 4.2 \times 10^2 \msqpc$ \\ 
\hline
Volume Mass Density & Bulge  & $\rho_{\rm b}^\odot=1.3 \times 10^{-4}\mcupc$\\
& Disk$^\dagger$ for $z_0=144$ pc & $\rho_{\rm d}^\odot=0.30 \mcupc~(=\Sigma_{\rm d}^\odot/2 z_0)$ \\ 
& Disk for $z_0=247$ pc & $\rho_{\rm d}^\odot=0.18 \mcupc$ \\ 
& Disk Oort's value & $\rho_{\rm d}^\odot=0.15 \mcupc$ \\ 
& Interstellar gas (included in disk) & $ \rho_{\rm gas}^\odot=0.05 \mcupc$ \\ 
& Bulge+Disk   & $\rho_{\rm bd}^\odot=0.18-0.3 \mcupc$\\
& Dark halo    & $\rho_{\rm halo}^\odot=3.5 \times 10^{-3} \mcupc$\\
& Total (bulge+disk+halo) & $\rho_{\rm total}^\odot=0.2 - 0.3 \mcupc$ \\
\hline 
Total Mass in & Bulge  &  $M_{\rm b}^\odot=1.75\times10^{10}\Msun$ \\
solar sphere$^\ddagger$ ($r\le R_0$) & Disk   & $M_{\rm d}^\odot=  4.33\times10^{10}\Msun$ \\
& Bulge+Disk &  $M_{\rm bd}^\odot=6.08\times10^{10}\Msun$ \\
& Dark halo &  $M_{\rm dh}^\odot=8.7\times10^9\Msun$ \\
& Total (bulge+disk+halo) & $M_{\rm total}^\odot=7.3\times10^{10}\Msun$ \\
& Representative mass $M_0=R_0 V_0^2/G $ &  $M_0=7.44\times10^{10}\Msun$ \\
\hline
\end{tabular} \\
\end{center}
$\dagger$ For the scale height, $z_0=247$ pc (Kent et al. 1991) and 144 pc (Kong and Zhu 2008) are adopted.\\
$\ddagger$ The "Solar sphere" is a sphere of radius $R_0=8$ kpc centered on the Galactic Center.
\label{tab_local}
\end{table*}

The local values of the surface mass and volume densities in the solar vicinity  calculated for these parameters are also shown in table \ref{tab_local}. The  volume density of the disk has been calculated by $\rho_{\rm d}=\Sigma_{\rm d} /(2 z_0)$ with $z_0$ being the scale height at $R=R_0$, when the disk scale profile is approximated by $\rho_{\rm d}(R_0, z)=\rho_{\rm d0}(R_0) {\rm sech} (z/z_0)$. For the local galactic disk, two values are adopted: $z_0=144\pm 10$ pc for late type stars based on the Hipparcos star catalogue (Kong and Zhu 2008) and 247 pc  from Kent et al. (1991). The local volume density by the bulge is four orders of magnitudes smaller than the disk component, and the halo density is two orders of magnitudes smaller. However, the surface mass densities as projected on the Galactic plane are not negligible. The bulge contributes to 1.6\% of the disk value, or the stars in the direction of the galactic pole would include about 2\% bulge stars, given the  \dv\ density profile. The surface mass density of the dark halo integrated at heights of $-10<z<10$ kpc exceeds the disk value by several times.

\vskip 5mm \subsection{\bf  Miyamoto-Nagai Potential}

Although the deconvolution using the \dv  and exponential disk well represents the observations, the mass models are not necessarily self-consistent in the sense that the model mass profiles are the solutions of the Poisson equation. One of the convenient methods to represent the Galaxy's mass distribution by a self-consistent dynamical solution is to use superposition of multiple Plummer-type potentials.  
 
The mass distribution $\rho(R, z)$ and the gravitational potential $\Phi(R,z)$ are related by the Poisson's equation: 
\begin{equation}
 \Delta \Phi= 4 \pi \rho (R, z),
 \label{eqLaplace}
\end{equation}
Let us recall that the potential for a point mass is given by 
\begin{equation}
 \Phi = {-GM \over r} = {-GM \over \sqrt{R^2+z^2}},
\end{equation}
with $r=\sqrt{R^2+z^2}$ being the distance from the center.
An extended spherical mass is often described by a Plummer's law: 
\begin{equation}
 \Phi =  {-GM \over \sqrt{r^2+b^2}} = {-GM \over \sqrt{R^2+z^2+b^2}}.
\label{eqPlum}
\end{equation}
Here, $b$ is a constant representing the scale radius of the sphere.

The most convenient Plummer-type formula, which describes the potential and realistic mass distribution in the Galaxy has, been obtained by Miyamoto and Nagai (1975). The potential is a modified one from Eq. (\ref{eqPlum}) for an axisymmetric spheroid. A galaxy is represented by superposition of several mass components in the same functional form as
\begin{equation}
 \Phi = \sum_{i=1}^n \Phi_i= \sum_{i=1}^n {-GM_i \over \sqrt{R^2+(a_i+\sqrt{z^2+b_i^2})^2}},
 \label{eqMN}
\end{equation}
where, $a_i$ and $b_i$ are constants representing the scale radius and scale height of the $i$-th spheroidal component. The rotation velocity in the galactic plane at $z=0$ is given by
\begin{equation}
 V_{\rm rot}(R)  = \sqrt{\sum_{i=1}^n R {\partial  \Phi_i \over \partial R}}
 =R\sqrt{\sum_{i=1}^n {GM_i \over [R^2+(a_i+b_i)^2]^{3/2}}}.
\label{eqVrot}
\end{equation}
The mass distribution is calculated from the Poisson's equation Eq. \ref{eqLaplace}:
\be
\rho(R, z)  ={1 \over 4 \pi} 
\sum_{i=1}^n M_i {a_iR^2+[a_i+3(z^2+b_i^2)^{1/2}][a_i+(z^2+b_i^2)^{1/2}]^2
\over
\{R^2+[a_i+(z^2+b_i^2)^{1/2}]^2\}^{5/2}(z^2+b_i^2)^{3/2}\}}.
\label{eqMNden}
\ee 

\begin{figure} 
\begin{center}
\includegraphics[width=8cm]{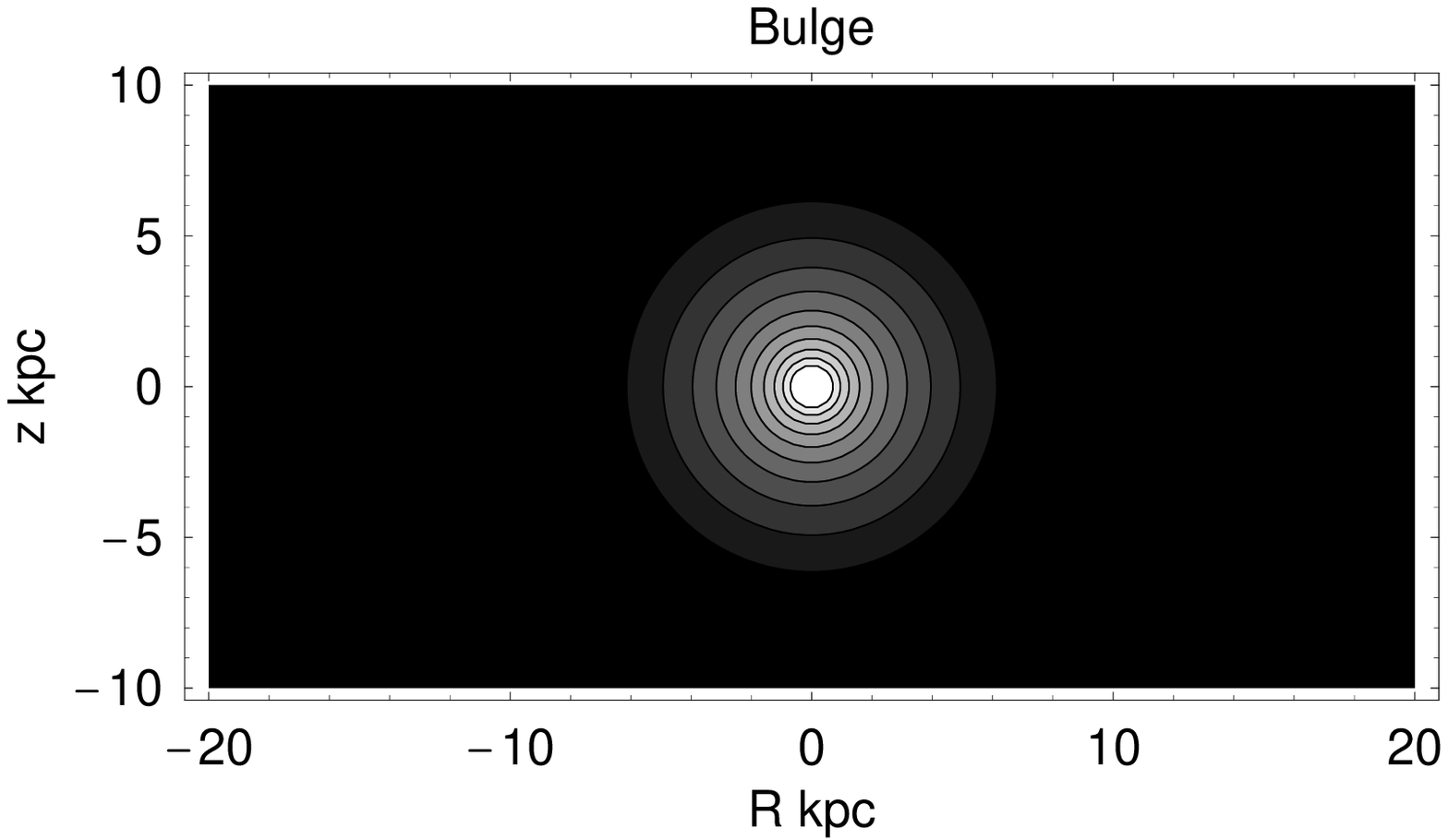} \\
\includegraphics[width=8cm]{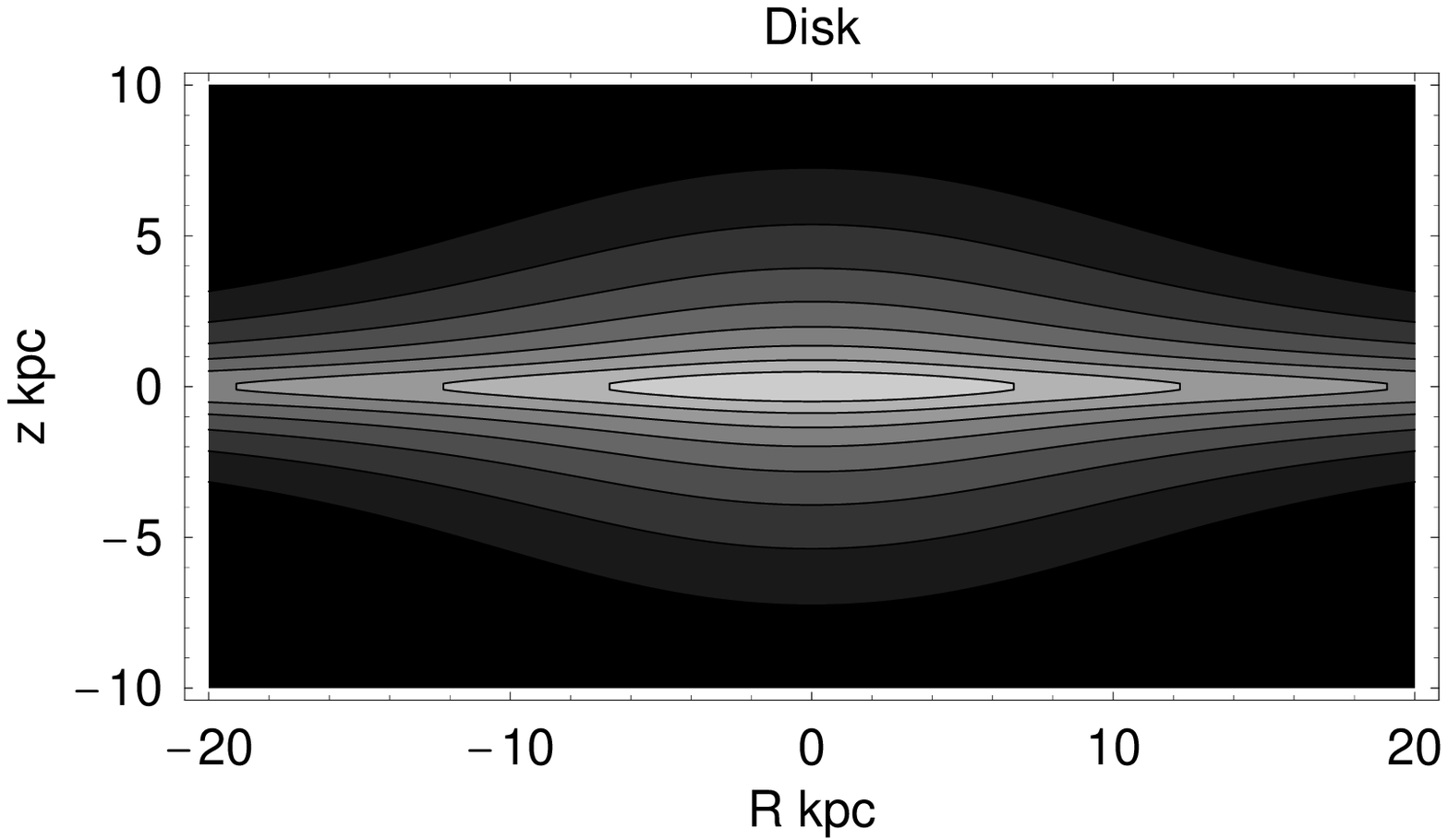} \\
\includegraphics[width=8cm]{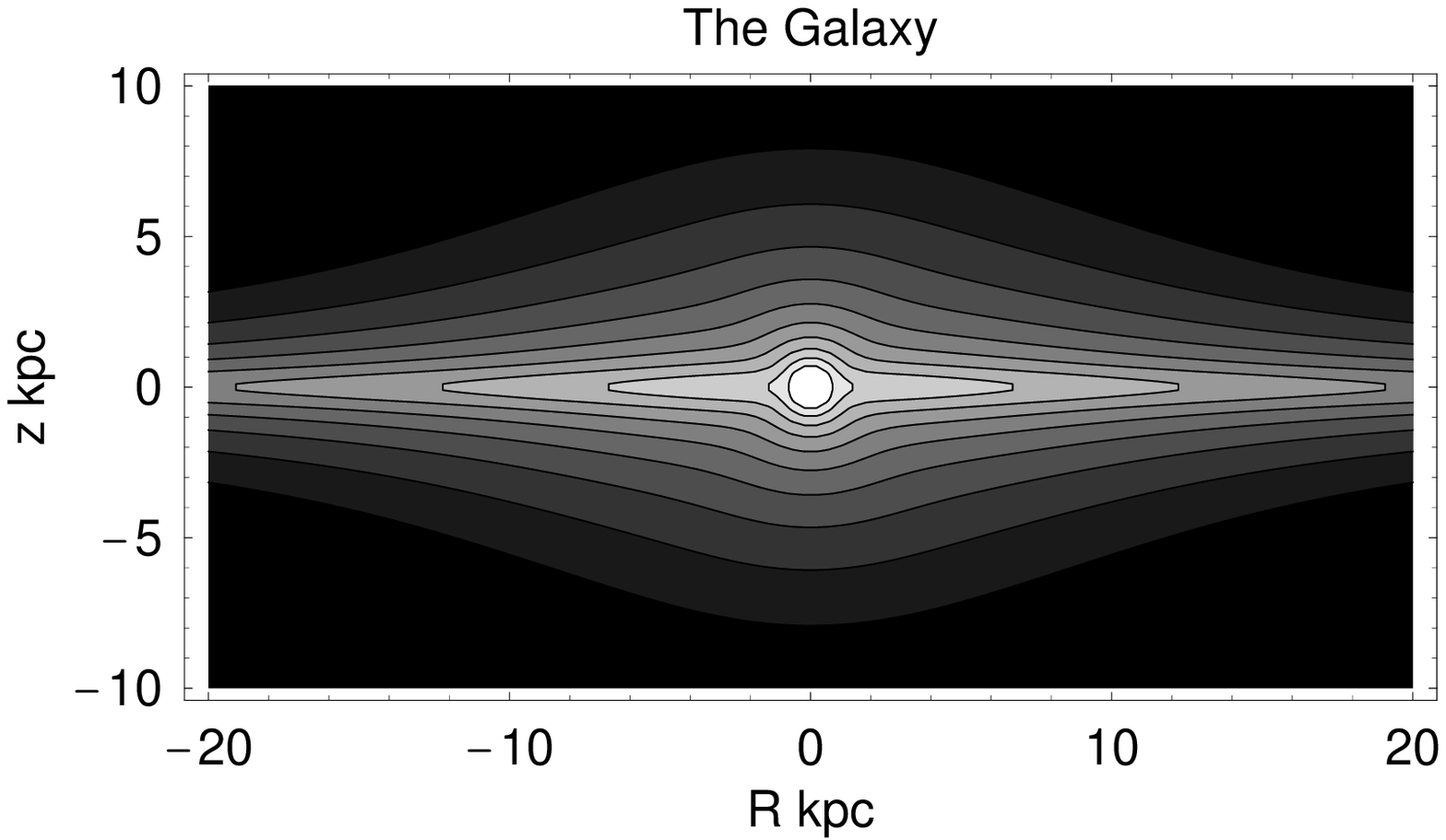} \\
\includegraphics[width=7cm]{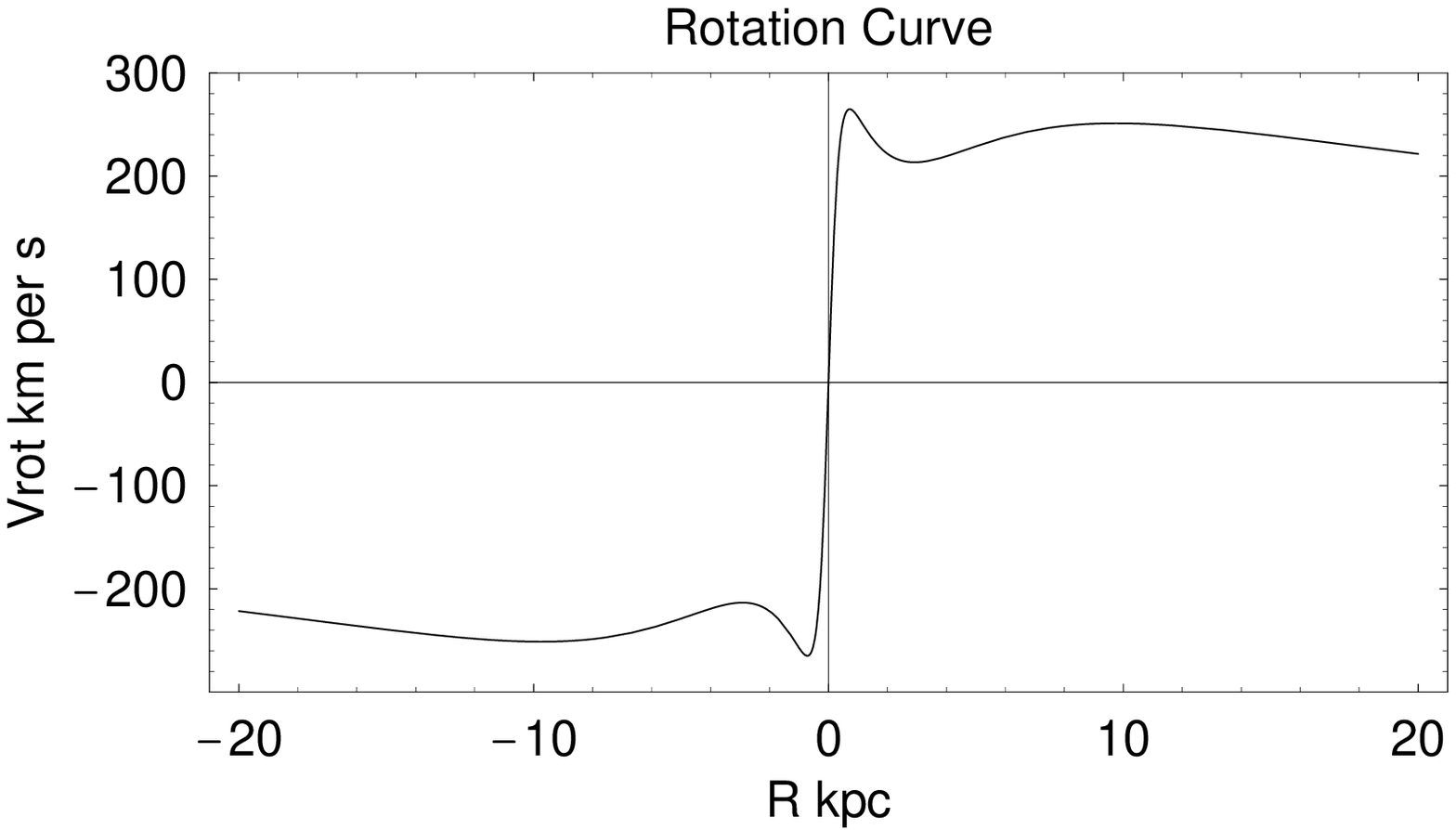} 
\end{center}
\caption{Meridional distributions of the volume density in the bulge, disk and their superposition to represent the Galaxy calculated for the Miyamoto-Nagai potential (1975) with parameters in table \ref{tabMN}. Contours and grey scales are logarithmic. The corresponding rotation curve is shown at bottom.  }
\label{figMN}
\end{figure}

Figure \ref{figMN} shows the meridional distribution of volume mass-density  calculated for a model Galaxy composed of two components of a bulge and disk with the parameters as given in table \ref{tabMN} (Miyamoto and Nagai 1975). This model approximately reproduces the rotation curve and the Oort's (1965) value of the local mass density of $0.15 \Msun {\rm pc}^{-3}$ at $R=10$ kpc. Note that the dark halo was not taken into account, but instead the mass  ($2.5 \times 10^{11}\Msun$)  and scale radius (7.5 kpc) were taken to be larger than the present-day values of $\sim 10^{11} \Msun$ and $\sim 3.5 $ kpc, in order to mimic the flat part of the rotation curve. Today, as discussed in the next section, the outer flat rotation is well understood as due to the dark halo. Nevertheless, the MN potential is often used, for its analytical form, to represent a galaxy by modifying the parameters and adding the halo and central components.

\begin{table*}[htbp]
\bc
\begin{tabular}{llll}
\hline
   Component  & $M_i~(\Msun)$  & $a_i$ (kpc) & $b_i$ (kpc) \\
\hline
   Bulge & $2.05\times 10^{10}$ & 0.0 & 0.495 \\
   Disk  & $2.547\times10^{11}$ & 7.258 & 0.520 \\ 
\hline
\end{tabular}
\ec
\caption{Parameters determining the Miyamoto-Nagai Potential of the Galaxy. }
\label{tabMN}
\end{table*}

\vskip 5mm \subsection{\bf Direct Method}

In the above methods, several mass components are assumed a priori, each of which had representative functional form, and therefore, the results depend on the assumed profiles. In order to avoid this dependence on the adopted functions, a direct method to obtain the mass distribution without assuming the functional form can be applied, where the mass distribution is calculated directly using the observed rotation velocity data. Since the rotation curve is restricted to give force in the galactic plane, the method cannot give the three dimensional information. Hence, a 'true' mass profile in a real disk galaxy is assumed to lie between two extreme profiles, which are either spherical or axisymmetric flat-disk.  

\vskip 5mm \subsubsection{\bf Spherical Mass Distribution}

The mass $M(R)$ of a spherical body inside radius $R$ is given by
\begin{equation}
M(R)=\frac{R {V(R)}^{2}}{G},
\label{masssphere}
\end{equation}
where $V(R)$ is the rotation velocity at $r$. Then the SMD
${\Sigma}_{S}(R)$ at $R$ is calculated by, 
\be
\Sigma_{\rm S}(R) = 2 \int\limits_0^{\infty} \rho (r) dz = \frac{1}{2 \pi} \int\limits_R^{\infty} \frac{1}{r \sqrt{r^2-R^2}} \frac{dM(r)}{dr}dr .
\label{smdsphere}
\ee 
The volume mass density $\rho(R)$ is given by
\begin{equation}
\rho(R) =\frac{1}{4 \pi r^2} \frac{dM(r)}{dr}.
\label{rhosphere}
\end{equation}
 
For a given rotation curve $V(R)$, Eq. (\ref{smdsphere}) can be computed numerically. In a galaxy, this gives a good approximation for the central region where the spheroidal component dominates. On the other hand, the equation gives underestimated mass density near the outer edge at  $R \sim R_{\rm max}$ because of the edge effect due to the finite radius of data points. The edge effect is negligible in the usual disk regions.

\vskip 5mm \subsubsection{\bf Flat-Disk Mass Distribution}

The surface mass density (SMD) for a thin disk, ${\Sigma}_{\rm D}(R)$, can be obtained by solving the Poisson's equation on the assumption that the mass is distributed in a flat disk with negligible thickness. It is given by
\begin{equation}
{\Sigma}_{\rm D}(R) =\frac{1}{{\pi}^2 G} \left[ \frac{1}{R} \int\limits_0^R 
{\left(\frac{dV^2}{dr} \right)}_x K \left(\frac{x}{R}\right)dx + 
\int\limits_R^{\infty} {\left(\frac{dV^2}{dr} \right)}_x K \left
(\frac{R}{x}\right) \frac{dx}{x} \right],
\label{smdflat}
\end{equation}
where $K$ is the complete elliptic integral and becomes very large when $x\simeq R$ (Binney \& Tremaine 1987). For the calculation, it must be taken into account that Eq. (\ref{smdflat}) is subject to the boundary condition, $V(0)=V(\infty)=0$. Also it is assumed that $V(0)=0$ at the center. Since a central black hole of a mass on the order of  $10^{6}-10^{7}M_{\odot}$ dominates the RC only within a few pc, it does not influence the galactic scale SMD profile. 

When calculating the first term on the right hand side of Eq. (\ref{smdflat}) for the central region, it happens that there exist only a few data points, where the reliability of the calculated $V(R)$ is lower than the outer region. In addition, the upper limit of the integration of the second term is $R_{\rm max}$ instead of infinity. Since the rotation curves are nearly flat or declining outward from $R = R_{\rm max}$, the second term becomes negative. Thus, the values are usually slightly overestimated for ${\Sigma}_{\rm D}(R)$ at $R\simeq R_{\rm max}$.

\vskip 5mm \subsubsection{\bf Verification using the Miyamoto-Nagai Potential}

\begin{figure} 
\bc
\includegraphics[width=6cm]{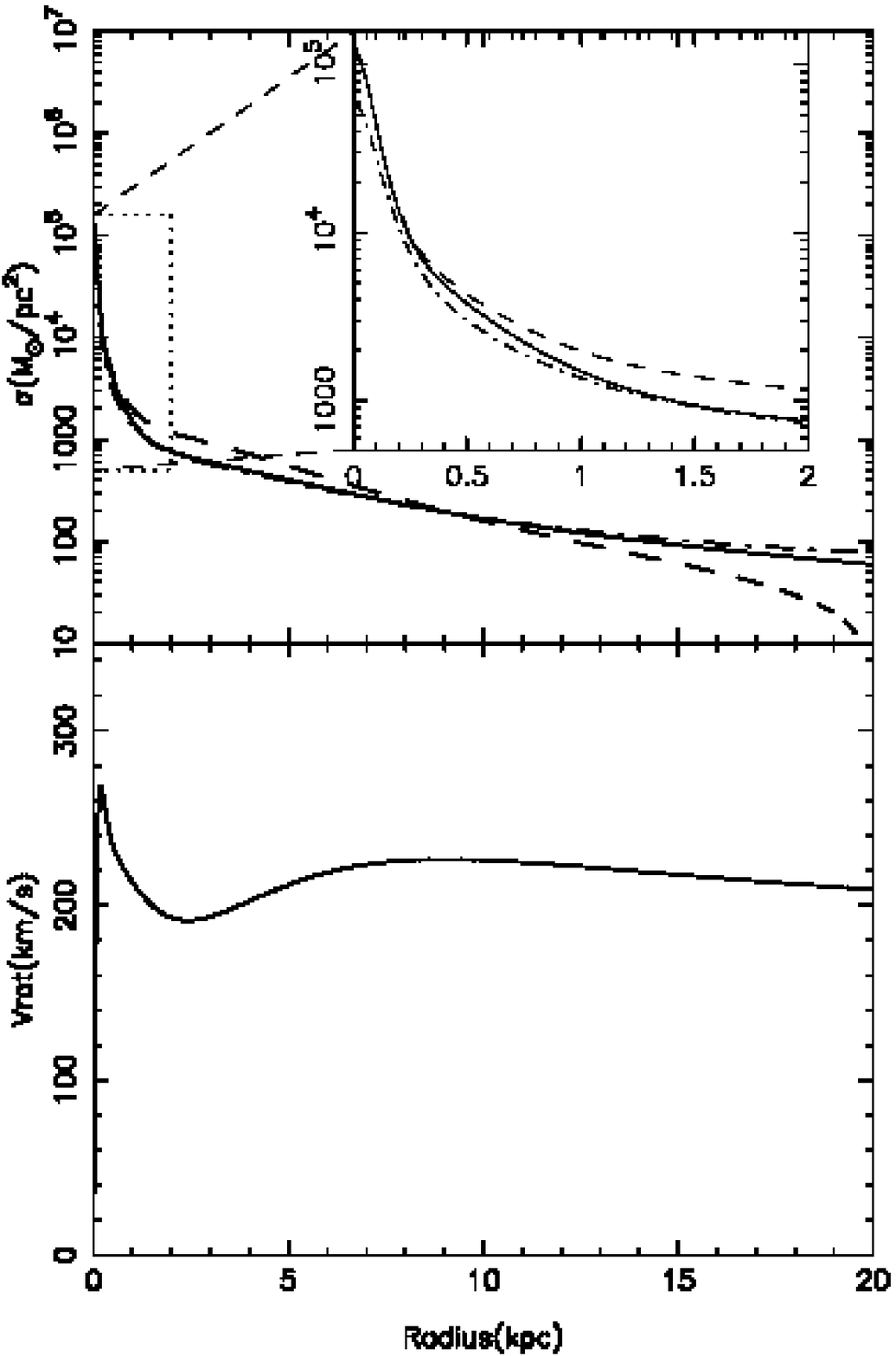}  \\ 
\ec
\caption{Comparison of the mass deconvolution (top panel) using spherical and flat-disk models applied to the analytical rotation curve for a Miyamoto-Nagai potential (bottom panel). (Takamiya and Sofue 1999) .}
\label{fig-mass-deconv-test}  
\end{figure}  

It is interesting to examine how the results for the spherical and flat-disk assumptions, as computed from equations \ref{smdsphere} and \ref{smdflat}, respectively, differ from each other as well as from the `true' SMD for the Miyamoto-Nagai (1975) potential. Given a set of parameters, the `true' SMD as well as the rotation curve can be obtained. Using this rotation curve,  the SMDs both for a spherical and flat-disk are calculated using the methods as described in the previous sections. 

Fig. \ref{fig-mass-deconv-test} shows the 'true' SMD and computed SMDs from the rotation curve for  spherical and flat-disk assumptions calculated. Here, the rotation curve only up to $R= 20$ kpc is used. The figure demonstrates that the spherical case well reproduces the true SMD for the inner region. This is reasonable, because the spherical component is dominant within the bulge. On the other hand, the flat-disk case better reproduces the true SMD in the disk region, which is also reasonable. Near the outer edge, the flat-disk case better traces the true SMD, while the spherical case is affected by the edge effect significantly. 

It must be stressed that the results from the two extreme assumptions, spherical and flat-disk, differ at most by a factor of two, and do not differ by more than a factor of 1.5 from the true SMD in most regions except for the edge. It may be thus safely assumed that the true SMD is in between the two extreme cases. It should be remember that the SMD is better represented by a spherical case for the inner region, while  a flat-disk case is better for the disk and outer part. 

\vskip 5mm \subsection{\bf Direct Mass Distribution in the Galaxy}

\begin{figure*} 
\bc 
\includegraphics[width=9cm]{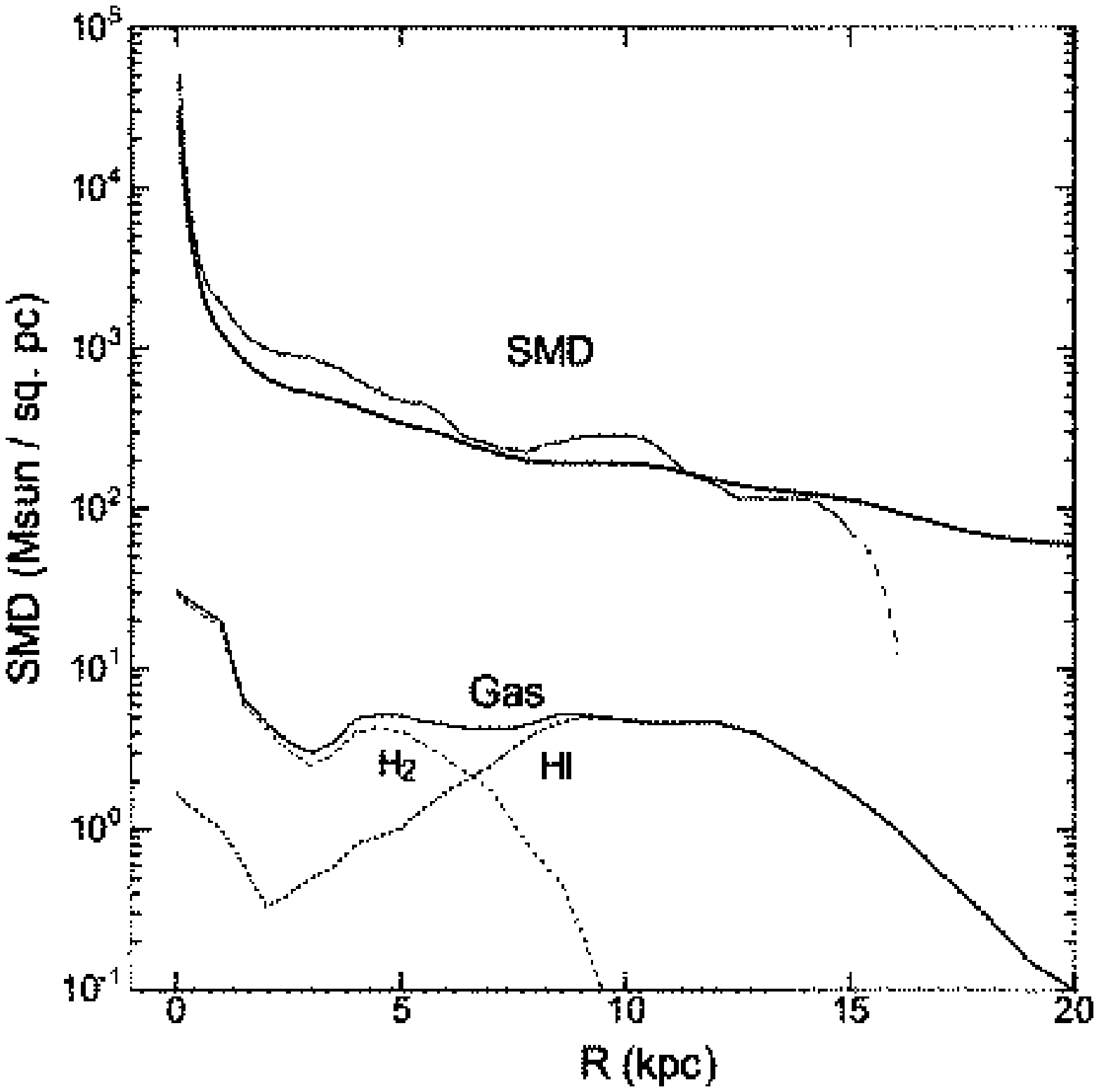}  
\ec
\caption{Radial distribution of the surface-mass density (SMD) in the Galaxy directly calculated from the rotation curve on the disk assumption (thick line). The thin line is the result for spherical assumption which better represents the innermost bulge component. The outer truncation (dashed part) is due to edge effect. Lower thin line is SMD of interstellar gas made by annulus-averaging the face-on projected distribution in Fig. \ref{HIH2face} (Nakanishi and Sofue 2003, 2006). HI and H$_2$ gas SMDs are also shown separately by dotted lines.}
\label{fig-smd-mw}   
\label{HIH2smd} 
\end{figure*}

Figure \ref{fig-smd-mw}  shows the obtained direct SMD distribution in the Galaxy. There is remarkable similarity between this result and the total SMD obtained by deconvolution of the rotation curve into the bulge, disk and halo components, as indicated by the upper dashed line in Fig. \ref{fig-smd}.

The mass is strongly concentrated toward the nucleus, and the bulge component dominates in the central region. The calculated SMD reaches a value as high as $\sim 10^5 \Msun~{\rm pc}^2$ at radius of a few tens pc. Higher density concentration has been observed from high-resolution infrared photometry and spectroscopy, indicating SMD as high as $\sim 10^6 \Msun~{\rm pc}^3$ within a few pc (Genzel et al. 1996). These values may be compared with the central value of the bulge's SMD as fitted by the \dv  profile, $\Sigma_{\rm bc}=6.8\times 10^6\Msun~{\rm pc}^2$, from Eq. \ref{eq-smdb} (table \ref{tab_milkyway}).

The galactic disk appears as the straight-line part at $R \sim 3 $ to  8 kpc on this semi-logarithmic plot ($R$ vs log $\Sigma$), indicating the exponential character. Even in these radii, as well as in the solar vicinity at $R\sim 8$ kpc, the dynamical surface mass (not volume density) is dominated by the dark matter, because the SMD is the projection of huge extent of the dark halo. There is slight difference between SMDs from the deconvolution method and direct method: for example at $R\sim 8$ kpc, the SMD is $\sim 300 \Msun~{\rm pc}^{-2}$ in Fig.\ref{fig-smd}, whereas it is $\sim 200-250 \Msun~{\rm pc}^{-2}$ in Fig.\ref{fig-smd-mw} . This discrepancy is due to the difference caused by the limited areas of integration from the finite data for the infinitely extended dark halo as well as due to difference in the adopted methods.

The outer disk, as indicated by the flat-disk model (full line in Fig. \ref{fig-smd-mw}), is followed by an outskirt with a more slowly declining density profile, gradually detaching from the straight line part of the disk. This outskirt indicates the dark halo, which extends to much larger radii, as is discussed in the next section.

\vskip 5mm \subsection{\bf Direct Mass Distributions in Spiral Galaxies}

\begin{figure} 
\bc
\includegraphics[width=8cm]{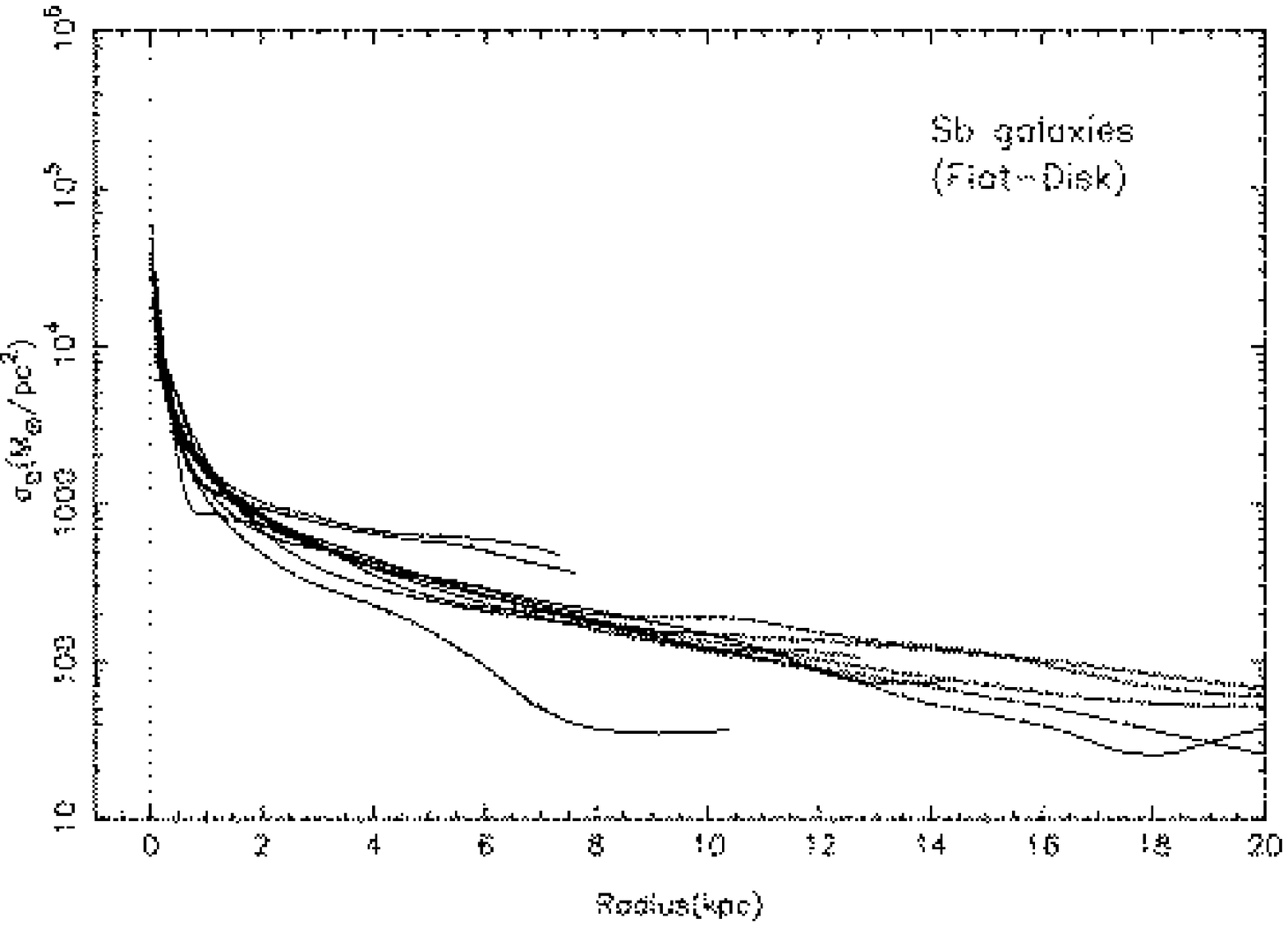}  \\
\includegraphics[width=8cm]{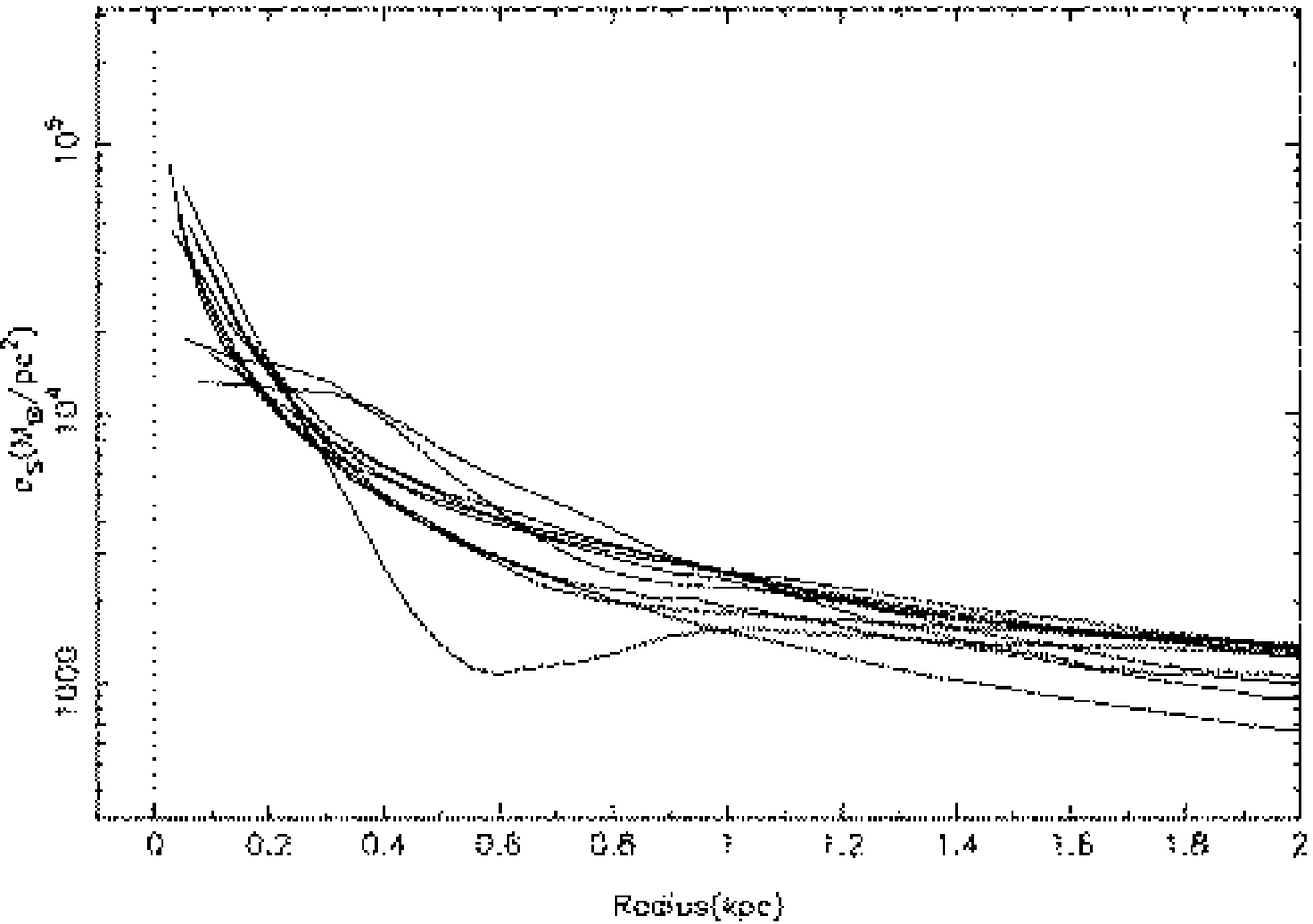}  
\ec
\caption{ [Top]: Mass distributions in Sb galaxies using flat-disk. [Bottom]: Same but for the central regions using spherical model (Takamiya and Sofue 2000).}
\label{fig-smd-Sb}  
\end{figure}  

The mass distributions in spiral galaxies can be also directly computed from observed rotation curves in the same way as for the Milky Way. Figure \ref{fig-smd-Sb} shows the SMD distributions of Sb galaxies obtained for the rotation curves in Fig. \ref{rc-all} using Eq. \ref{smdflat}. The central regions are enlarged in the bottom panel, where Eq. (\ref{smdsphere}) is adopted for a spherical model, which better represents the inner spheroidal components. The calculated SMD profiles for the Sb galaxies are similar to that of the Milky Way. It is also known that the profiles for Sa to Sc galaxies are very similar to each other, except for the absolute values (Takamiya and Sofue 2000). It is stressed that the dynamical structure represented by the density profile is very similar to each other among spiral galaxies. The SMD profiles have a universal characteristics as shown in these figures: high central concentration, exponential disk (straight line on the semi-logarithmic plot), and outskirt due to the dark halo.

\vskip 5mm \subsection{\bf Distribution of Interstellar Gas in the Galaxy}

The rotation curve of the Milky Way Galaxy is useful not only for deriving the mass distribution, but also for mapping the interstellar gas.
Given a rotation curve $V(R)$, radial velocity $\vr$ of any object near the galactic plane at $b \sim 0\Deg$ is uniquely calculated for its distance and longitude $(l, r)$.
Inversely, given the radial velocity and galactic longitude ($\vr, l$) of an object, its distance from the Sun, and therefore, its position in the galactic disk is determined. Thus, the distribution of objects and gases can be obtained by measuring radial velocities from spectroscopic observations such as of recombination lines (HII regions), $\lambda$21-cm line emission (HI clouds and diffuse gas),  molecular lines (CO lines), and/or maser lines (e.g. SiO).

\begin{figure} 
\begin{center}
\includegraphics[width=6cm]{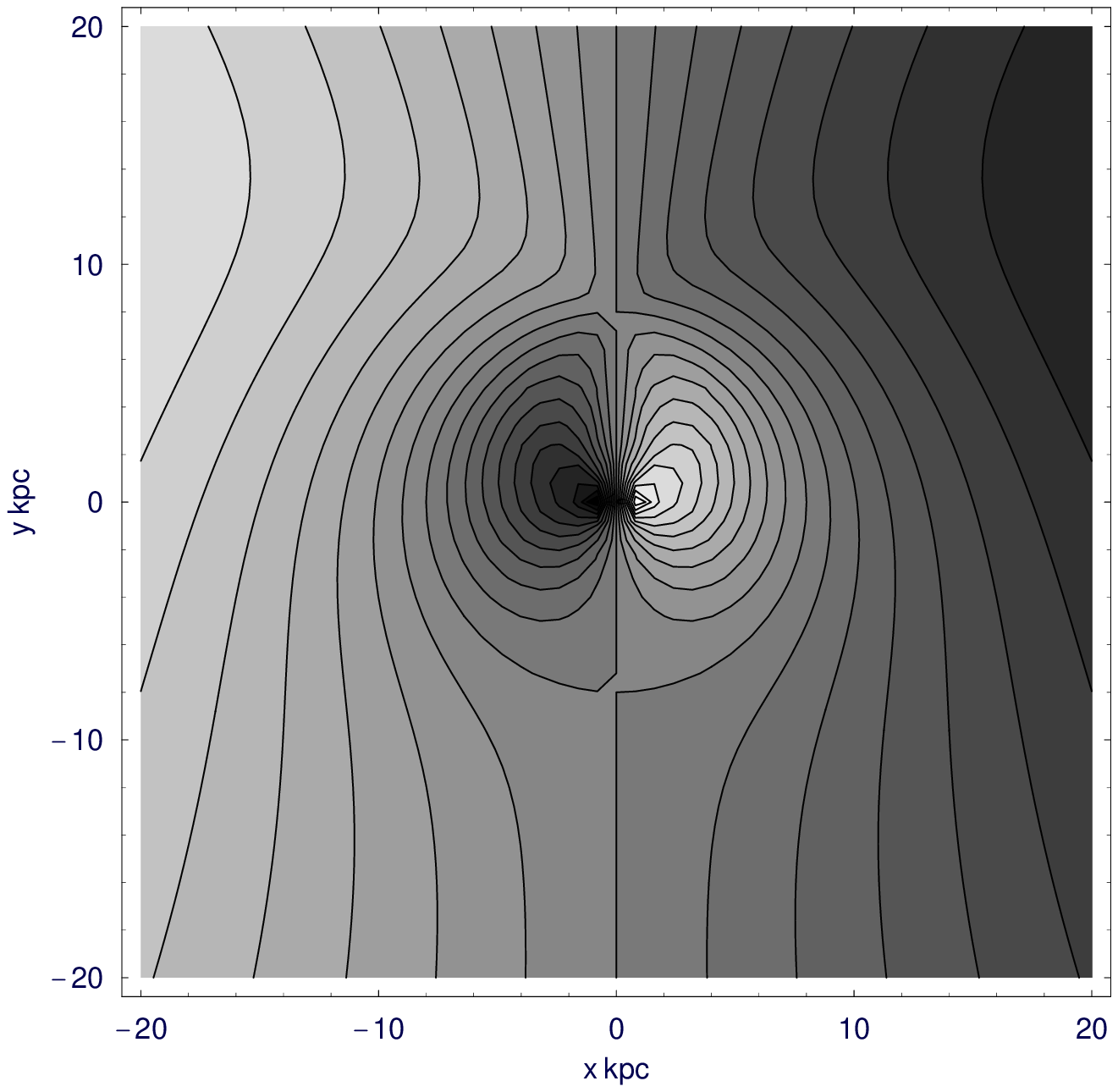} 
\end{center}
\caption{Radial-velocity field in the galactic plane.}
\label{vfield} 
\begin{center}
\includegraphics[width=7cm]{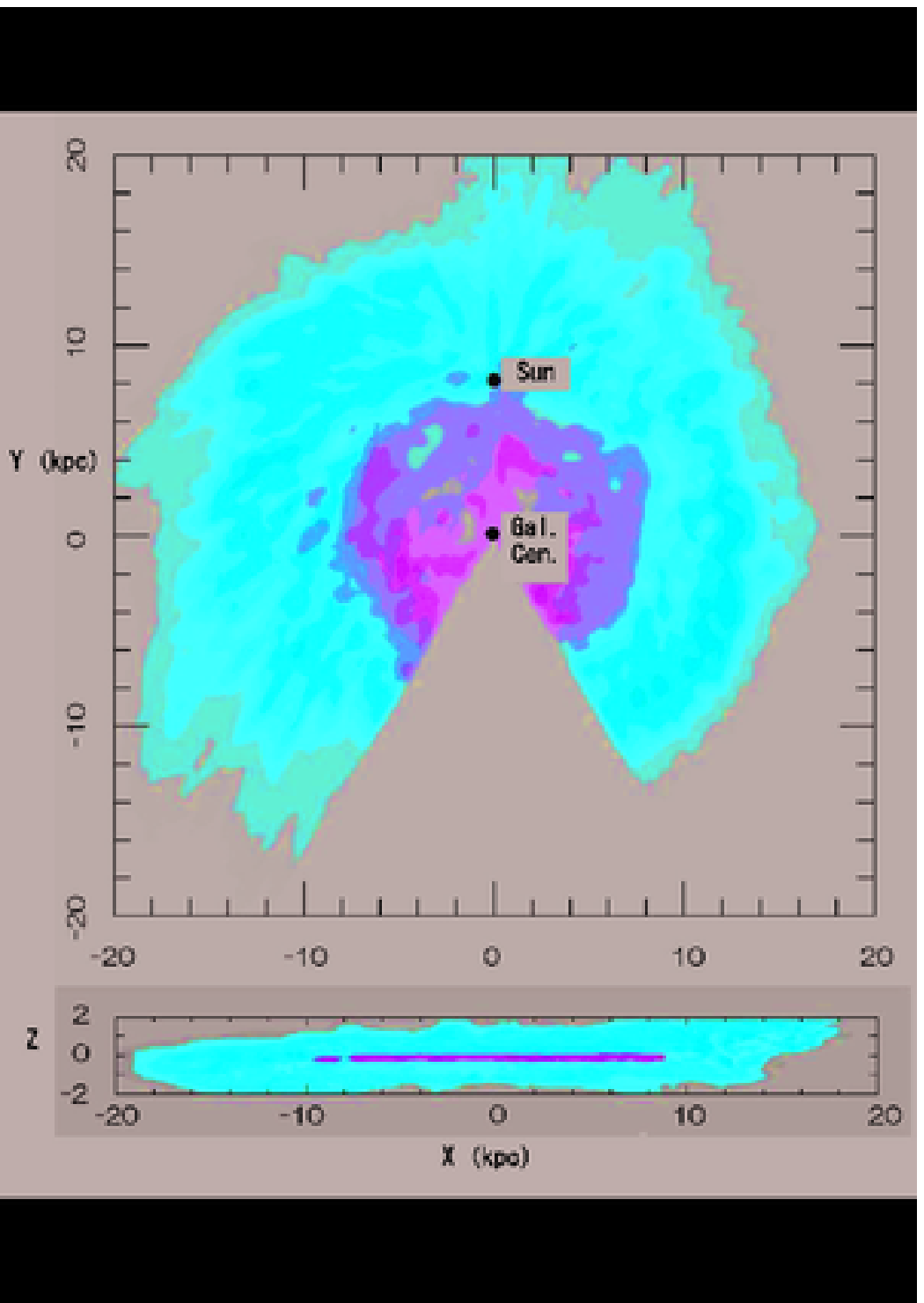} 
\end{center}
\caption{Distribution of surface density of interstellar gas in the Galaxy obtained from spectral data of the HI 21-cm and CO-2.6 mm line emissions on an assumption of circular rotation of the gas (Nakanishi and Sofue 2003, 2006).}
\label{HIH2face}
\end{figure}

The radial velocity of an object in the galactic plane is calculated by
\begin{equation}
 \vr = \Rsun(\w - \wsun)\sin~ l = \left\{{\Rsun \over R} V(R) -\Vsun \right\} \sin~ l.
\label{eq-vr}
\end{equation}
Using this equation, a radial-velocity diagram is obtained for the Galaxy as shown in Fig. \ref{vfield}.  

The radial-velocity field is used to determine the position (distance $r$) of an object in the Galaxy for which the coordinates and radial velocity, $(l, b; \vr)$, are known. The volume density of interstellar gas at the position is obtained from the line intensity. This procedure is called the  velocity-to-space transformation (VST) and is useful to produce a face-on map of the density distribution. Positions of objects outside the solar circle are uniquely determined by this method. However, inside the solar circle the solution for the distance is two-fold, appearing either at near or at far side of the tangent point. For solving this problem, additional information's such as apparent diameters of clouds and thickness of the HI and molecular disks are required.

 With the VST method, the HI and H$_2$ gas maps can be obtained by the following procedure. The column density of gas is related to the line intensities as
\be
 N_{\rm HI, H_2}~[{\rm H, H_2~cm^{-2}}] = C_{\rm HI, H_2} \int T_{\rm HI, CO}(v) dv~{\rm[K~km~s^{-1}]},
\label{eq-NHIH2} 
\ee
where $T_{\rm HI}$ and $T_{\rm CO}$ are the brightness temperatures of the HI and CO lines, and $C_{\rm HI}=1.82\times10^{18}$ [H atoms cm$^{-2}$] and $C_{\rm H_2}\sim 2\times10^{20}$ f(R, Z) [H$_2$ molecules cm$^{-2}$] are the conversion factors from the line intensities to column densities of HI and H$_2$ gases, respectively. Here, $f(R, Z)$ is a correction factor ranging from 0.1 to 2 for the galacto-centric distance $R$ and metallicity $Z$ in the galaxy with the solar-vicinity value of unity (Arimoto et al. 1996). 
The local density of the gas at the position that corresponds to the radial velocity $v$ is observed is obtained by
\be
 n_{\rm HI,H_2} = {dN_{\rm HI, H_2} \over dr}={dN_{\rm HI, H_2} \over dv}{dv \over dr}
 =C_{\rm HI, H_2} T_{\rm HI,CO}(v)dv/dr.
 \label{eq-denHIH2}
 \ee 

Combining the LV diagrams as shown in Fig. \ref{pvdHI} and the velocity field, a ``face-on" distribution is obtained of the density of interstellar gas in the Galaxy.  Fig. \ref{HIH2face} shows a face-on view of the Galactic disk as seen in the HI and molecular line emissions (Nakanishi and Sofue 2003, 1006).  Figure \ref{HIH2smd} shows azimuthally averaged radial profiles of the interstellar gas density. 
The gaseous mass density is far less compared to the dynamical mass densities by stars and dark matter by an order of magnitude. The interstellar gas density at $R\sim 8$ kpc is $\sim 5.0 \msqpc$, which shares only several percents of disk mass density of $\sim 87.5 \msqpc$ (Fig. \ref{fig-smd}, \ref{HIH2smd}, table \ref{tab_local}). 
   
\section{DARK HALO}
 
\vskip 5mm \subsection{\bf Dark Halo in the Milky Way}

The mass of the Galaxy inside the solar circle is $\approx 10^{11}\Msun$. The mass  interior to the distance of the Large Magellanic Cloud at 50 kpc may grow to $6 \times 10^{11}\Msun$ (Wilkinson \& Evans 1999). Interior to 200 kpc, the mass would be at least $2 \times 10^{12}\Msun$ (Peebles 1995). 
The Milky Way and M31 are the major members of the Local Group, around which many satellites and dwarf galaxies are orbiting(Sawa and Fujimoto 2005). Masses and extents of dark halos around these two giant galaxies are crucial for understanding the dynamics and structure of the Local Group. Outer rotation curve of the Galaxy is a key to put constraints on the dark halo structure.  
Although the outer rotation curve is reasonably fitted by an isothermal model, the halo model might not be unique, and is difficult to clearly discriminate the isothermal model (Begeman et al. 1991) from NWF (Navarro et al. 1996) and Burkert models (Burkert 1995), because of the large scatter of data as well as for the limited radius within which the rotation curve is observed. 

Kinematics of satellites and member galaxies in the Local Group is a crew to estimate the dark halo, which is considered to be extended far outside the galactic disk and in the intracluster space (Kahn and Woltjer 1959;  Li and White 2008, van der Marel and Guhathakurta 2008). Sawa and Fujimoto 2005) have computed the past probable orbits of the major members of the Local Group under the condition that their positions and radial velocities are satisfied at the present time.  
In this section, the behaviors of rotation curves calculated for different dark halo models are examined, and are compared with a "pseudo rotation curve" of the Local Group, which combines the rotation curve of the Galaxy and radial velocities of the members of the Local Group.

\vskip 5mm \subsubsection{\bf Isothermal halo model}

The simplest interpretation of the flat rotation curve observed in the outer Galaxy, and outer rotation curves in many spiral galaxies is to adopt the semi-isothermal spherical distribution for the dark halo (Kent 1986; Begeman et al. 1991). In the isothermal model, the density profile is written as
\be
\rho_{\rm iso} (R)={\rho_{\rm iso} ^0 \over  1+ (R/h)^2}, 
\label{eq_iso} 
\ee
where $\rho_{\rm hc}$ and $h=R_{\rm h}$ are constants giving the central mass density and scale radius of the halo, respectively. This profile gives finite mass density at the center, but yields a flat rotation curve at large radius. The circular velocity is given by 
\be
V_h(R)=V_\infty \sqrt{1-\left(h \over R \right) {\rm tan}^{-1}\left(R \over h \right) },  
\ee
where $V_\infty$ is a constant giving the flat rotation velocity at infinity. The constants are related to each other as
\be
V_\infty=\sqrt{ 4 \pi G \rho_{\rm iso}^0 h^2},
\ee
or the central density is written as 
\be
\rho_{\rm iso}^0=0.740 \left(V_\infty \over {200 {\rm km ~s^{-1}}} \right) \left(h \over {1 {\rm kpc}} \right)^{-2}   \Msun {\rm pc}^{-3}.
\ee
The enclosed mass within radius $R$ is given by
\be
M(R)=4 \pi \int_0^R \rho_i(r) r^2 dr
\label{eq-enclosedmass}
\ee
with $i=$ iso (NFW or Bur for the other two models as discussed below).

At small radius, $R\ll h$, the density becomes nearly constant equal to $\rho_{\rm iso}^0$ and the enclosed mass increases steeply as $M(R) \propto R^3$. At large radius of $R\gg h$ the density decreases as $\rho_{\rm iso} \propto R^{-2}$ and the enclosed mass tends to increase linearly with radius as $M(R) \propto R$.

\vskip 5mm \subsubsection{\bf  NFW and Burkert Profiles}

Based on numerical simulations of the formation of galaxies in the cold-dark matter scenario in the expanding universe, several model profiles have been found to fit better the calculated results. The most well known model is the NFW model proposed by Navarro, Frenk and White (1996), and Burkert (1995) has modified this model. The NFW and Burkert density profiles are written, respectively, as 
\be 
\rho_{\rm NFW} (R)={\rho_{\rm NFW} ^0 \over (R/h)[1+(R/h)]^2} ,
\label{eq-nfw}
\ee
and
\be \rho_{\rm Bur} (R)={\rho_{\rm Bur} ^0 \over [1+(R/h)][1+(R/h)^2)]} .
\label{eq-bur}
\ee 
The circular rotation velocity is calculated by
\be
V_h (R)=\sqrt{GM_{\rm h} (R)\over R},
\ee
where $M_h$ is the enclosed mass within $h$ as calculated by Eq. (\ref{eq-enclosedmass}).

\begin{figure} 
\bc
\includegraphics[width=7cm]{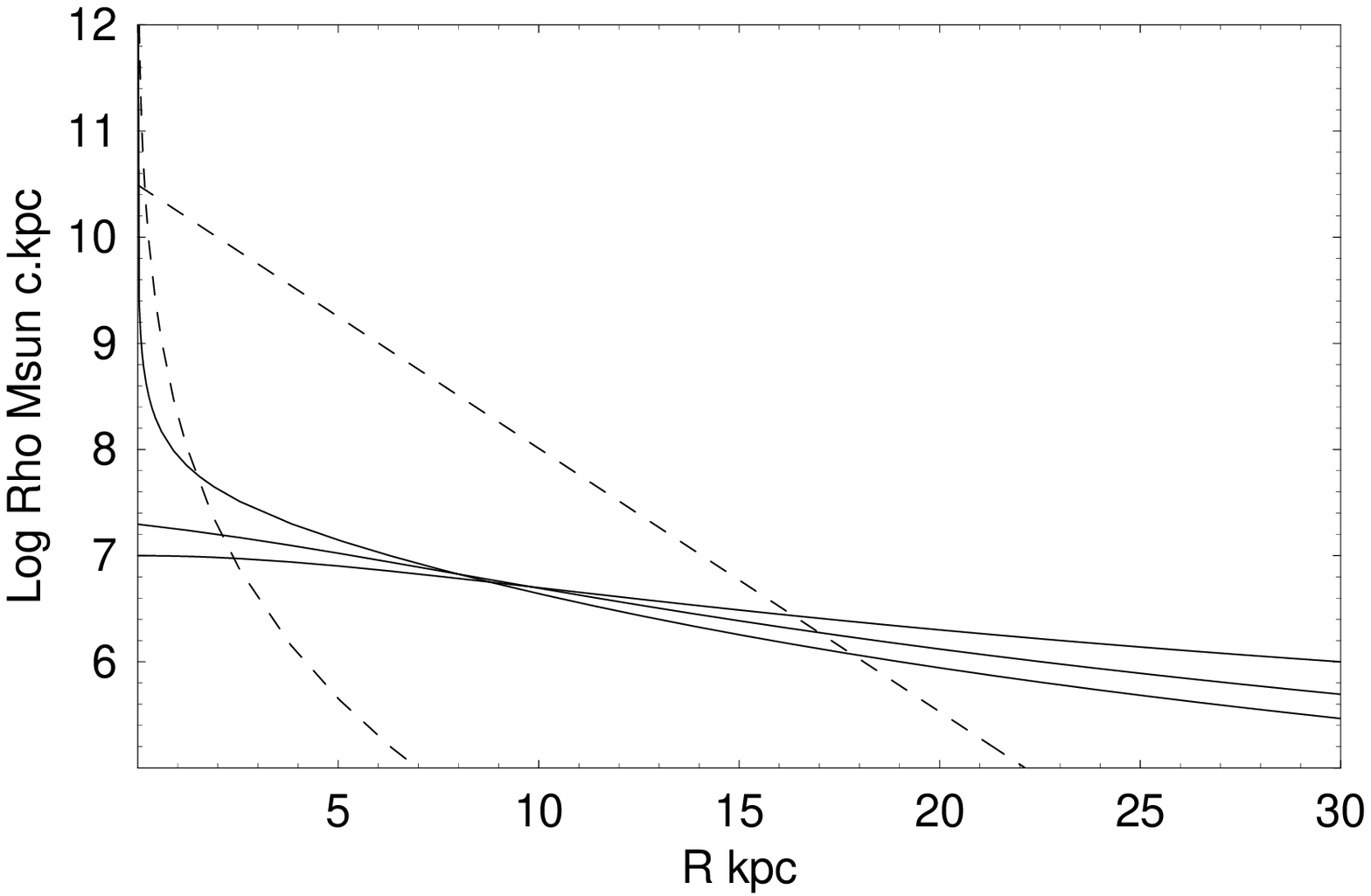} \\   
\includegraphics[width=7cm]{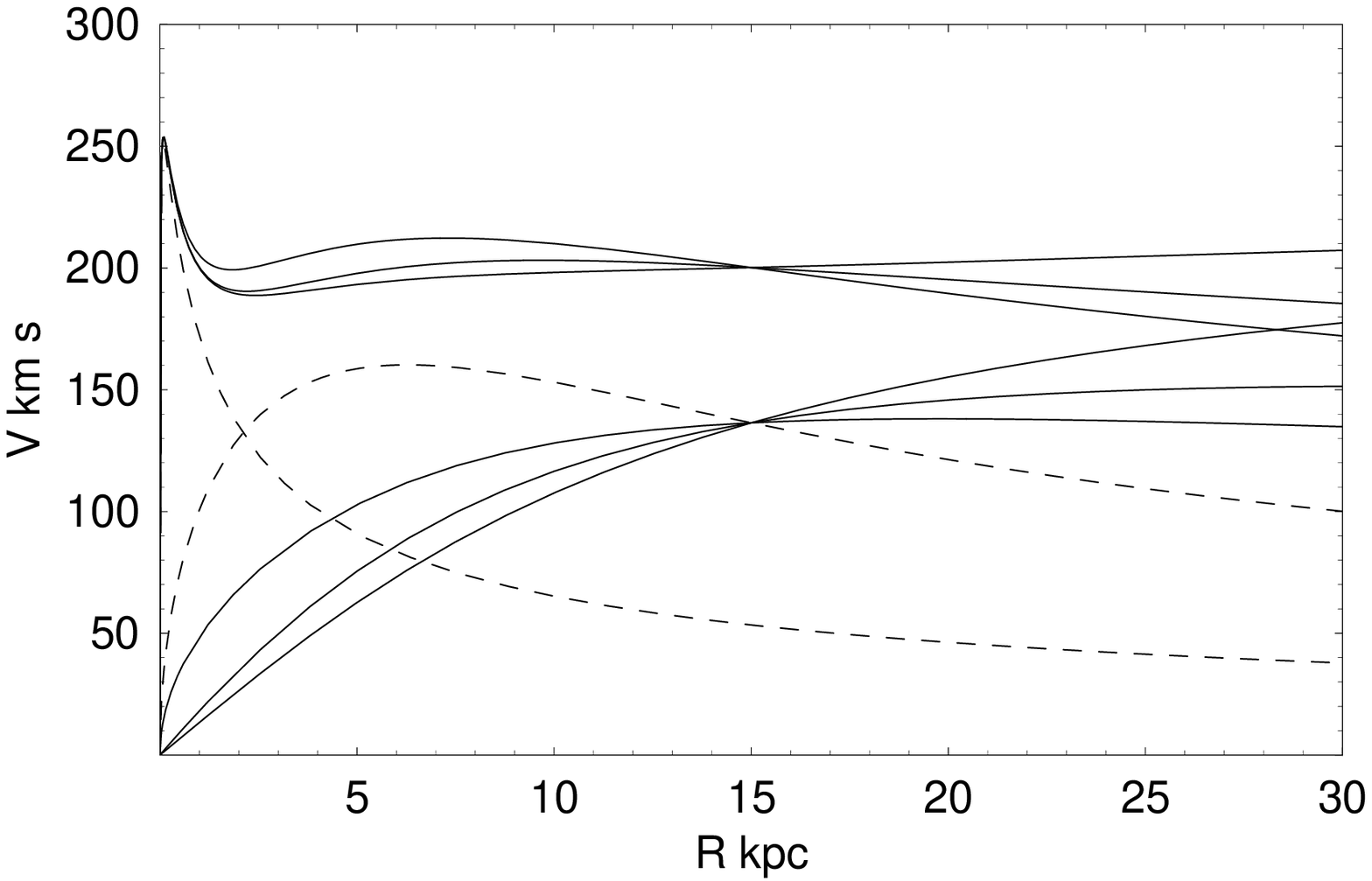} \\
\includegraphics[width=8cm]{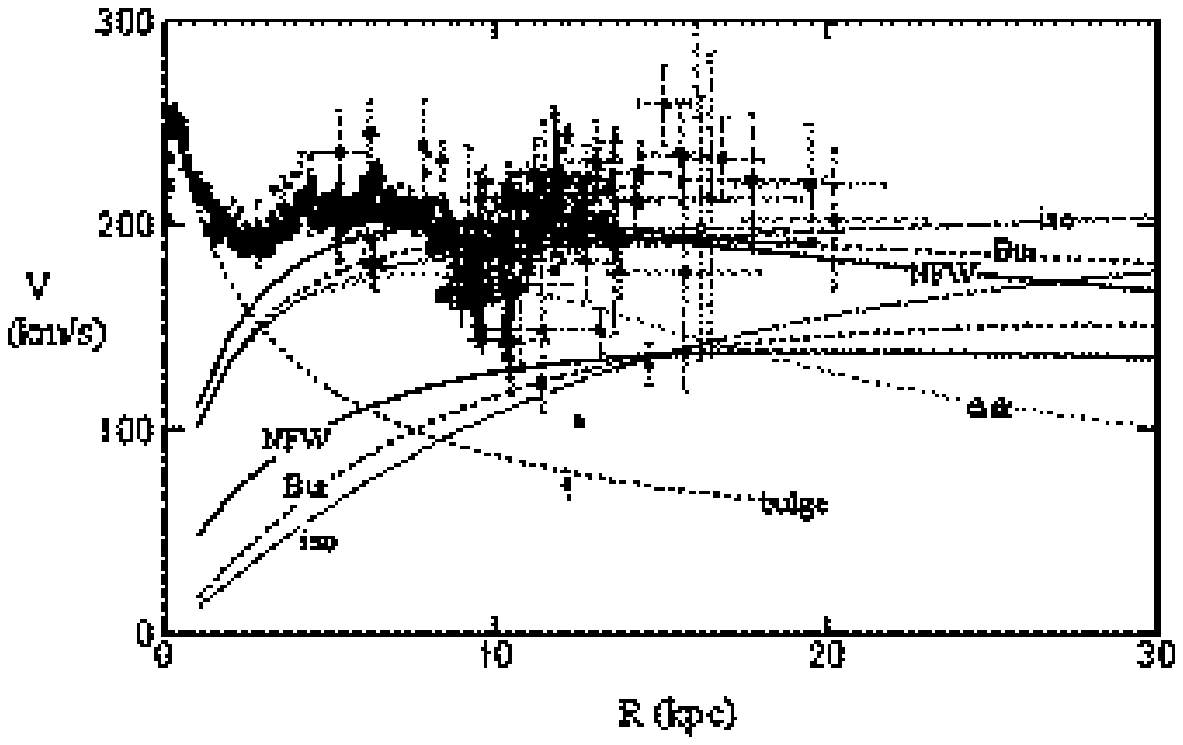} \\ 
\ec
\caption{[Top] Schematic volume density profiles for dark halo models - isothermal, Burkert and NKF models (full lines from top to bottom at $R=30$ kpc) - compared with the disk and bulge (dashed lines). 
[middle] Rotation curves.
[bottom] Rotation curves compared with observations. }
\label{fig-dh}  
\bc
\vskip 3mm \includegraphics[width=6.7cm]{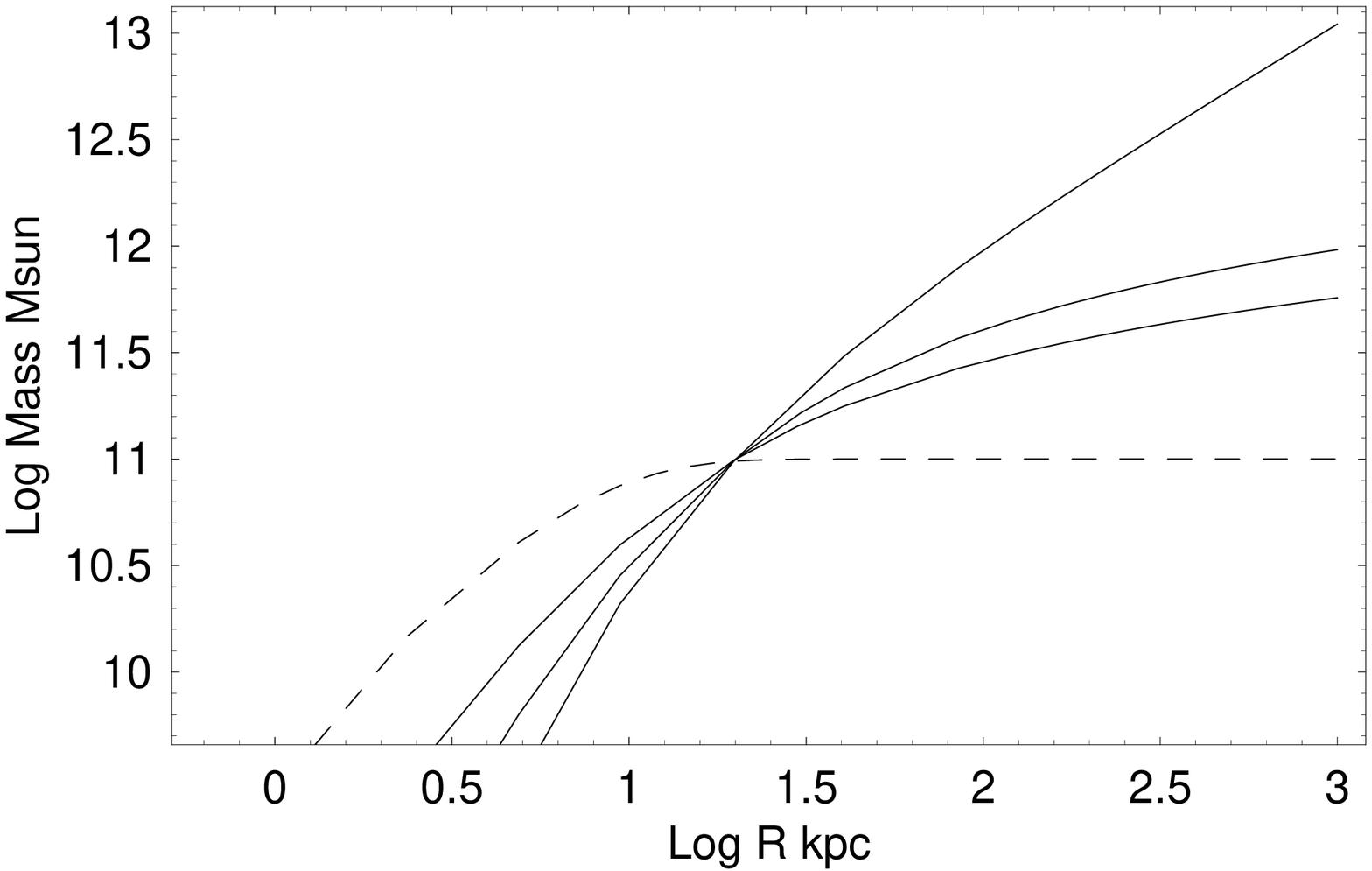} \\ \ec
\caption{Enclosed masses of dark halos up to 1 Mpc (isothermal, Burkert and NFW from top to bottom at 1 Mpc), compared with the disk mass (dashed line) that approximately manifests the galactic luminous mass.}
\label{fig-dhMpc}  
\end{figure}

At small radius with $R \ll h$, the NFW density profile behaves as $\rho_{\rm NFW}\propto 1/R$, yielding an infinitely increasing density toward the center, and the enclosed mass behaves as $M(R) \propto R^2$. On the other hand, the Burkert profile tends to constant density $\rho_{\rm Bur}^0$, yielding steeply increasing enclosed mass as $M(R) \propto R^3$, similarly to the isothermal profile. At large radius with $R \gg h$, both the NFW and Burkert profiles have densities $\rho_{\rm NFW,~Bur} \propto R^{-3}$, and they yield milder logarithmic increase of mass as $M(R) \propto {\rm ln}~ R$ (Fig. \ref{fig-dhMpc} )

The scale radius of a dark halo model is usually supposed to be between 3.5 to 10 kpc. Here, a value of $h=10$ kpc is adopted commonly for the three models.
Figure \ref{fig-dh} shows density distributions for the three different models of the dark halo as well those for the disk and bulge. Corresponding rotation curves are then compared with the observations. Fig. \ref{fig-dh} also shows the masses enclosed within a sphere of radius $R$. In these figures, the total masses of the disk and halo are taken to be the same at  $R=15$ kpc, so that the rotation velocity is nearly flat at $R= 10$ to 20 kpc and fit the observations. Fig. \ref{fig-dhMpc} calculates the enclosed masses up to radius of 1 Mpc for the three models. This figure demonstrates that the isothermal model predicts rapidly increasing mass merging with the neighboring galaxies' halos, whereas the NFW and Burkert models predict a rather isolated system with mild (logarithmic) increase of enclosed mass for the Milky Way.
 
\vskip 5mm \subsubsection{\bf Dark halo contribution in the inner galaxy}

In the inner Galaxy at $R<\sim 10$ kpc, the rotation velocity is predominantly determined by the bulge and disk contributions. Among the three dark halo profiles, only the NFW model predicts a nuclear cusp in which the dark matter density increases toward the nucleus inversely proportional to the radius. However, the bulge density is kept sufficiently large compared to the dark halo density in the Galactic Center. The Burkert and isothermal halo models indicate a mild density plateau at the center, which is also much smaller than the disk and bulge densities. 

Accordingly, in any models, the contribution from the dark halo to the rotation velocity is negligible in the inner Galaxy. Thus, it is hard to discriminate the dark halo models by comparing with the observations in the inner Galaxy, although the data are most accurately obtained there. The rotation velocity becomes sensitive to the dark halo models only beyond $\sim 20$ kpc.  

\vskip 5mm \subsection{\bf  Pseudo Rotation Curve of the Local Group}

In order to discuss the mass distribution in the dark halo, kinematics of objects surrounding the Galactic disk is essential (e.g., Kahn and Woltjer 1959; Sawa and Fujimoto 2005). A method for deriving dark halo density profiles by using radial velocities of the Local Group of galaxies as an extension of the rotation curve of the Galaxy is described below. 

\begin{figure}
\bc
\includegraphics[width=8cm]{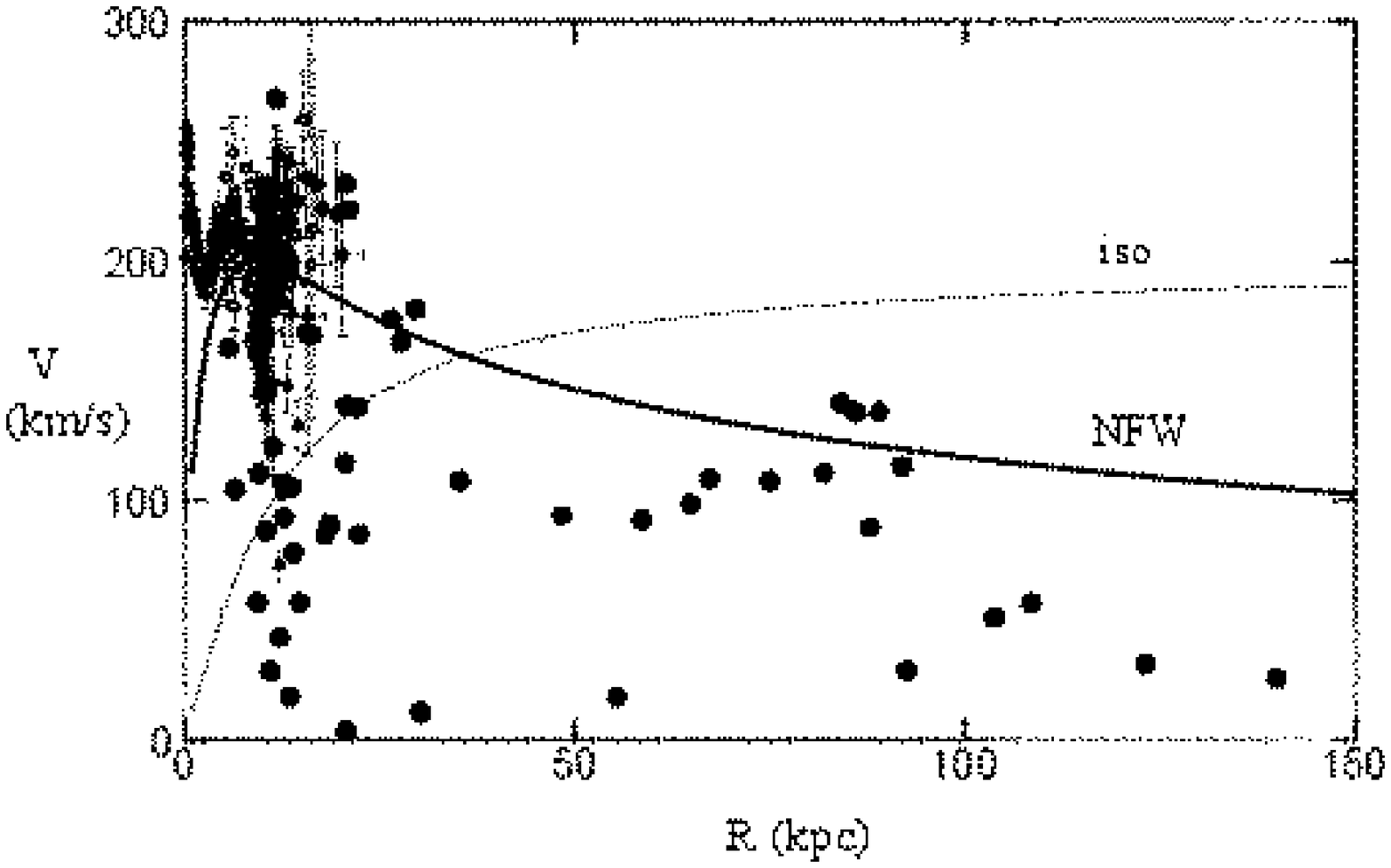} \\ 
\includegraphics[width=8cm]{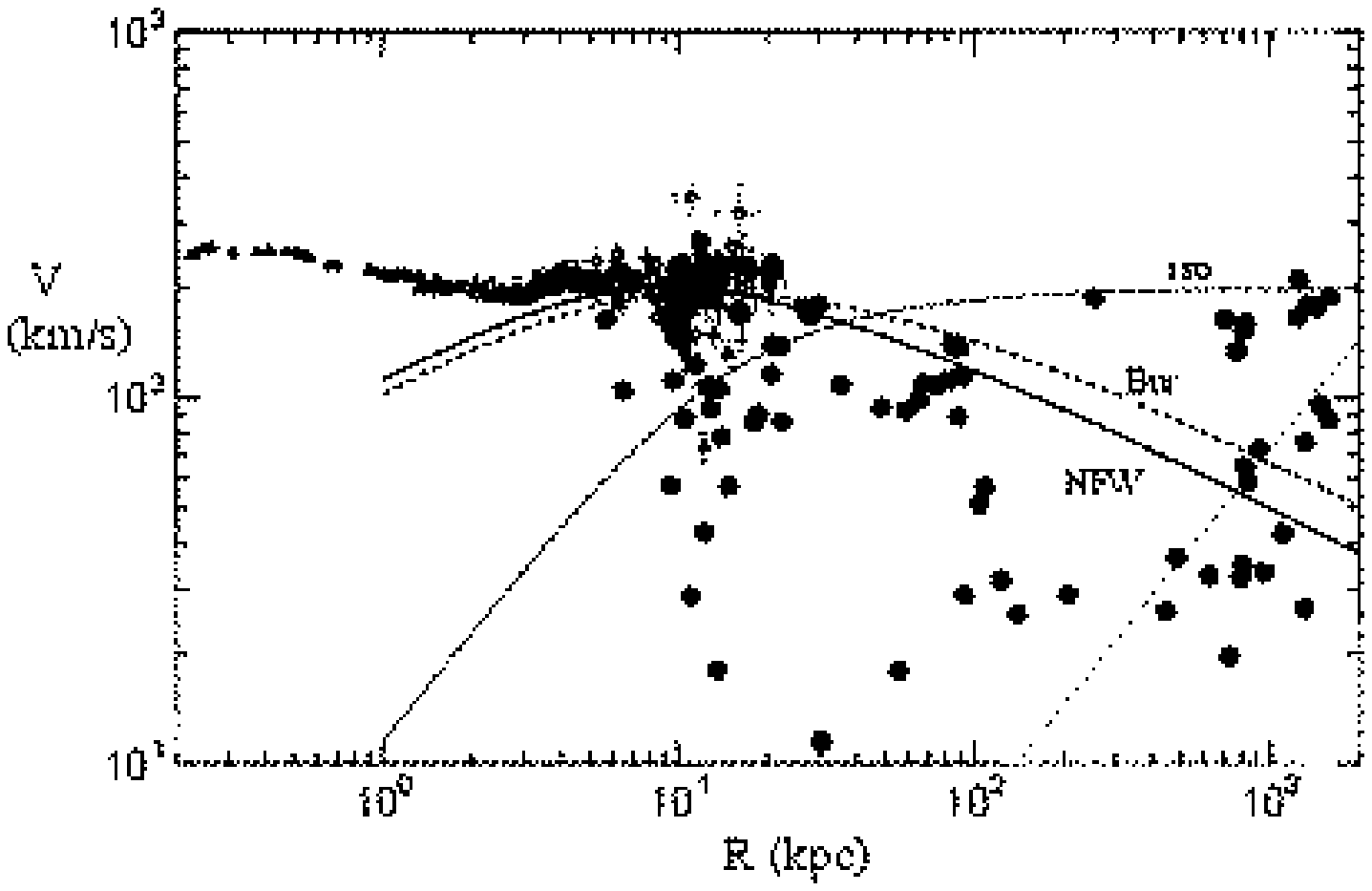} 
\ec
\caption{Pseudo rotation curves within 150 kpc (top) and 1.5 Mpc in logarithmic distance scale (bottom). Data beyond the galactic disk are absolute values of Galacto-centric radial velocities of globular clusters, satellite galaxies, and Local Group members. The big circle at 770 kpc denotes M31. The thick line shows the rotation curve for the NFW halo model, which traces well the upper boundary of the plot at $R\le 150$ kpc. The asymptotic horizontal thin line represents the isothermal model. Hubble flow for $H_0=72$ \kms Mpc$^{-1}$ is shown by a dashed line near the right-bottom corner.}
\label{fig-150}  
\label{fig-vgc}    
\end{figure}

In Fig. \ref{fig-vgc}, absolute values of galacto-centric radial velocities $V_r=|V_{\rm GC}|$ of outer globular clusters and member galaxies of the Local Group are plotted against their galacto-centric distances $R_{\rm GC}$ (Sofue 2009, and the literature cited therein for the data. In the figure, calculated rotation curves for the NFW model is shown by the thick line, and individual components are shown by dashed lines. The isothermal halo model is shown by the upper thin line, which is horizontal at large radius. The upper boundaries of the plot are well fitted by the rotation curve for the NFW model. Fig. \ref{fig-vgc} shows the same for entire region of the Local Group up to 1 Mpc.   The galacto-centric distance $R_{\rm GC}$ was calculated from galactic coordinates and helio-centric distance, and $V_r$ was calculated from observed helio-centric radial velocity by correcting for the solar rotation of $V_0=200$ \kms at $R=R_0$. 

In these figures the  calculated rotation curves for the three dark halo models are superposed. The dark halo models are so chosen that they are smoothly connected to the inner composite rotation curve for the best-fit bulge and disk. The scale radius of the dark halo is taken to be $h=10$ kpc. The total mass of the dark halo within the critical radius $R_{\rm c}=15$ kpc in each model is taken to be equal to the disk mass in the same radius. 

As equations (\ref{eq-nfw}) and (\ref{eq-bur}) indicate, the NFW and Burkert profiles are similar to each other at radii sufficiently greater than the scale radius at $R\gg h$ kpc. Accordingly, the rotation curves corresponding to these two models are similar at $R>30$ kpc.

If the area at $R<\sim 150$ kpc in Fig. \ref{fig-vgc} are carefully inspected, the NFW traces the upper envelope of plot better than isothermal profile. On the other hand, the isothermal model can better trace the upper boundary of the entire observations in the Local Group up to $\sim 1.5$ Mpc. M31, one of the major two massive galaxies in the Local Group, lies on the isothermal curve. 
These facts imply that the mass concentration around the Galaxy within $\sim 150$ kpc is isolated from the larger-scale mass distribution controlling the whole Local Group members. Hence, if it is gravitationally bound to the Local Group, there exists a much larger amount of dark mass than the dark halo of the Galaxy, filling the entire Local Group.

\vskip 5mm \subsection{\bf Mass of the Galaxy embedded in the Dark Halo}

The upper boundary of the pseudo rotation curve within $\sim 150$ kpc radius is well fitted by the NFW and Burkert models, as shown in Fig. \ref{fig-150}. In these models, the total mass of the Galaxy involved within a radius 770 kpc, the distance to M31, is $8.7\times 10^{11}\Msun$, and the mass within 385 kpc, a half way to M31, is $4.4 \times 10^{11}\Msun$, respectively. These values may be adjustable within a range less than $10^{12}\Msun$ by tuning the parameters $h$ and $R_{\rm c}$, so that the flat galactic rotation at $R\sim 10-20$ kpc can be approximated. 

In either case, the Galaxy's mass is by an order of magnitude smaller than the total mass required for gravitational binding of the Local Group. If the two galaxies, M31 and the Galaxy, are assumed to be gravitationally bound to the Local Group, the total mass is estimated to be on the order of $M_{\rm tot} \sim V^2 R/G \sim 4.8 \times 10^{12}\Msun$, as estimated from the mutual velocity of the two galaxies of $V \sim \sqrt{3} V_{\rm GC}$ with $V_{\rm GC}=-134$ \kms and the radius of the orbit is $R=385$ kpc. Here, the factor $\sqrt{3}$ was adopted as a correction for the freedom of motion. Thus, the Local Group may be considered to contain a dark matter core of mass comparable to this binding mass, $\sim 4.8 \times 10^{12}\Msun$. The mean mass density required to stabilize the Local Group is, thus, estimated to be 
$\rho_{\rm LG} \sim 2 \times 10^{-5}\Msun{\rm pc}^{-3}$.   

Now, the boundary of the Galaxy is defined as the radius, at which the density of the galactic dark halo becomes equal to the dark matter density of the Local Group. The thus defined radius is $R_{\rm G}\sim 100$ kpc for the NFW model. The Galaxy's mass within this boundary is $M_{\rm G: NFW}= 3\meleven$ (Fig. \ref{fig-dh}).
 
The high-velocity ends of the pseudo rotation curve for the entire Local Group up to $\sim 1.5$ Mpc in Fig. \ref{fig-vgc} are well represented by an isothermal dark halo model. The estimated total mass is $M \sim 4.8 \times 10^{12}\Msun$  for the terminal flat velocity of $V=134 \sqrt{3} \sim 230$ \kms. This may be compared with the estimates by Li and White (2008) ($5.3\times 10^{12}\Msun$) and van der Marel and Guhathakurta (2008) ($5.6\times10^{12}\Msun$).

The mean dark matter density of the Universe is on the order of 
\be
\rho_{\rm uni} \sim {{(H_0 R)^2  R \over G} {1\over {4\pi \over 3} R^3}} = {3H_0^2 \over {4\pi G}},
\ee
which is approximately $\sim 2\times 10^{-29}{\rm g ~cm^{-3}} \sim 1.2\times10^{-6}\Msun {\rm pc}^{-3} $ for $H_0=72$ \kms kpc$^{-1}$. The total mass of this "uniform" dark matter is $\sim 1.2 \meleven$ inside a sphere of radius 385 kpc, and $\sim 1 \mtwelve$ in 770 kpc, which is small enough compared to the dynamical mass of the Local Group.

The ratio of the baryonic mass (luminous mass) of a galaxy to the dark matter mass gives important information for the formation scenario of galaxies. The upper limit to the baryonic mass of the Galaxy is approximately represented by the disk plus bulge mass, which was estimated to be $M_{\rm baryon}\simeq 0.83\times 10^{11}\Msun$ by fitting of the inner rotation curve using de Vaucouleurs bulge and exponential disk. This yields an upper limit to the baryon-to-dark matter mass ratio of the Galaxy within radius $R$ to be $\Gamma (R) \simeq 0.38$ for $R=100$ kpc, and $\Gamma \simeq 0.23$ for $R=385$ kpc, where $\Gamma$ is defined by
\be
\Gamma(R)={M_{\rm baryon}(R)\over M_{\rm dark ~matter} (R)}={M_{\rm baryon}(R) \over \left\{M_{\rm total} (R)-M_{\rm baryon}(R)\right\}}.
\ee
 These values are on the same order of the cosmological baryon-to-dark matter ratio of 0.20 (=4.6\%/23\%) from WMAP (Spergel et al. 2003). Thus, the baryon-to-dark matter ratio within the Galaxy's boundary is close to the cosmological value.

The baryonic mass of galaxies in the Local Group may be further compared with the total Local Group mass. The total luminous mass is represented by those of the Galaxy and M31, which is about twice the Galaxy's baryonic mass, because the contribution from the other small and dwarf galaxies are negligible. Then, the upper limit to the baryon-to-dark matter mass ratio in the whole Local Group is estimated to be $\Gamma \sim 0.036$, which is only 0.2 times the cosmic value. It is expected that approximately 80\% of baryonic matter of the Local Group exists in the intergalactic space without being captured to the present galaxies.

\vskip 5mm \subsection{\bf The Galaxy, M31 and Local Group}

\begin{figure} 
\bc
\includegraphics[width=8cm]{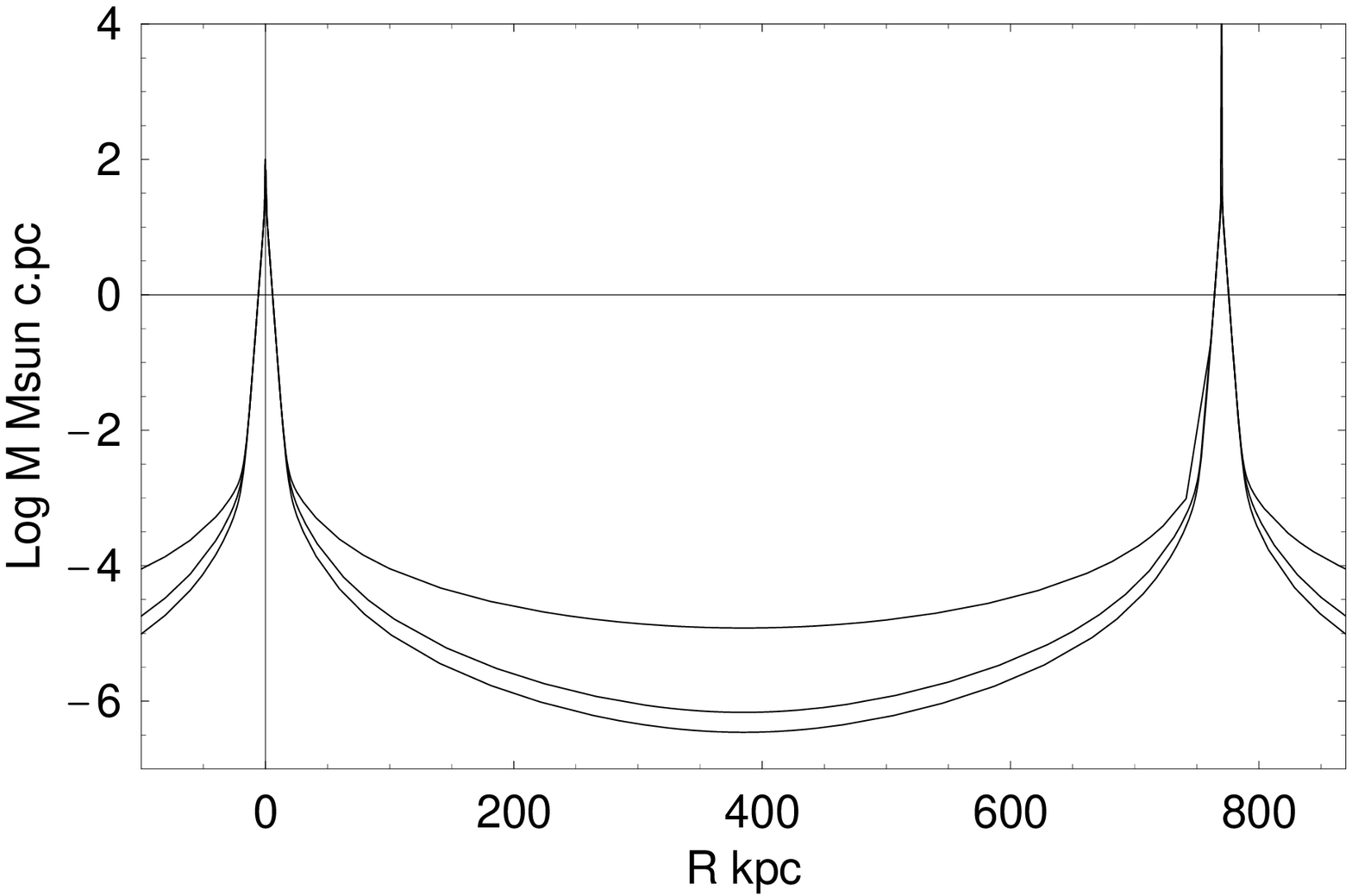} \\  
\includegraphics[width=8cm]{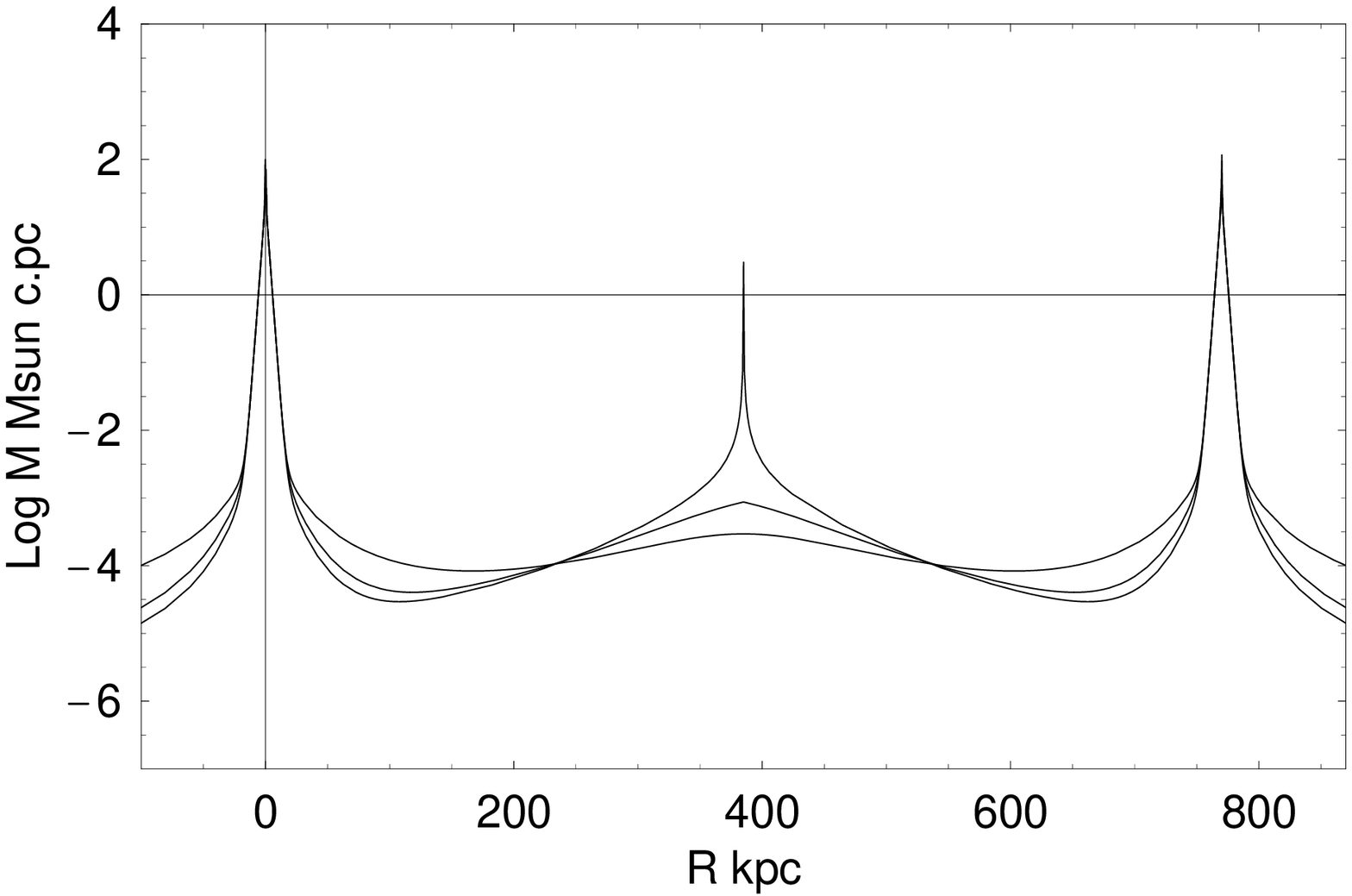} 
\ec
\caption{
[Top] Density profiles between the Milky Way and M31 for three dark halo models --- isothermal, Burkert and NFW --- from top to bottom. The isothermal halo (upper curve) gravitationally binds the two galaxies, while the other two potentials cannot bind the two galaxies.
[Bottom] If the Galaxy and M31 are located on both sides of a dark matter core of the Local Group, the dark matter can bind the two galaxies. 
}
\label{fig-mwm31}   
\end{figure}

 HI observations by Carignan et al. (2006) showed that M31's rotation curve is flat till a radius 35 kpc. They estimate the  mass of M31 within 35 kpc to be $3.4\meleven$. This is comparable to the total mass of the Milky Way Galaxy within 35 kpc of $3.2\meleven$. The Galaxy and M31 may be assumed to have similar mass profiles.  Fig. \ref{fig-mwm31} shows calculated density profiles along a line crossing the centers of the Galaxy and M31, where the two galaxies are separated by 770 kpc. 
 
If the halo profile is either Burkert or NFW, which is favored from the observations inside $R\sim 150$ kpc, the two galaxies are not massive enough to bind each other. In order to stabilize the Local Group, an extended massive component of intracluster dark matter of several$\mtwelve$ is required. The bottom panel in Fig. \ref{fig-mwm31} illustrates a density profile for such a case that there exists a dark matter core  at the midpoint between the Galaxy and M31. For each model, the core radius is  $h=100$ kpc and the integrated mass within a sphere of radius 385 kpc is adjusted to $5 \mtwelve$, a sufficient mass to bind the Local Group.

Less massive galaxies are floating in the dark matter ocean between the two big continents, M31 and the Galaxy. 
It is known that rotation curves of dwarfs and latest type galaxies increase monotonically toward their edges, as is typically observed for M33 (Corbelli and Salucci 2000). M33 is speculated to border M31 by their dark halos, and the rotation velocity increases until it reaches the velocity dispersion in the Local Group of $\sim 200$ \kms. 

A question is encountered about "temperatures" of dark matter. Inside small galaxies, rotation velocities are as small as $\sim 100$ \kms, much less than those in giant spirals and intracluster space, and therefore, the potential is shallow. Accordingly, the dark matter is cooler, so that it is gravitationally bound to the shallow potential. On the other hand, the halo continuously merges with halos of bigger galaxies and/or intracluster matter with higher velocities of $\sim 200$ \kms. Thus, the temperature of dark matter increases from inside to outside of a dwarf galaxy. 
This temperature transfer of dark matter occurs also from inside the outskirts of the Galaxy at  $R\sim 100-150$ kpc, where the rotation velocity is $\sim 100$ \kms according to the well fitted NFW model, to outside where the Local Group dark matter with velocity dispersion of $\sim 200$ \kms is dominant.

It is considered that there are three components of dark matter. First, the galactic dark matter having the NFW profile, which defines the mass distribution in a galaxy controlling the outer rotation curve; second, extended dark matter filling the entire Local Group having a velocity dispersion as high as $\sim 200$ \kms, which gravitationally stabilize the Local Group; and third, uniform dark matter having much higher velocities originating from super galactic structures. The third component, however, does not significantly affect the structure and dynamics of the present Local Group. It is therefore speculated that at any place in the Galaxy, there are three different components of dark matter having different velocities, or different temperatures. They may behave almost independently from each other, but are interacting by their gravity.  

\vskip 5mm \subsection{\bf Dark Halos in Galaxies}

By the 1970's, it had been recognized that rotation curves of spiral galaxies are flat to distances as large as $R \sim 30-50$ kpc from the nuclei by spectroscopy of HII regions (Rubin \& Ford 1970; Rubin et al. 1982, 1985) and HI line emission (Roberts \& Rots 1973). The flat rotation curves suggested that the dynamical mass continued to rise to the last measured regions of the galaxies. Theoretically, dark matter halos were postulated to exist in any spiral galaxies in order for the disk to be stable from gravitational fragmentation (Ostriker \& Peebles 1973). By analyzing motions of satellite and companion galaxies, Einasto et al. (1974) had shown that spiral galaxies are surrounded by massive dark halos. 

Deeper and higher-resolution HI observations with synthesis telescopes reveal that for the majority of spiral galaxies, rotation curves remain flat beyond the optical disks (Bosma 1981a, b;   van Albada et al. 1985; Begeman 1989). The largest HI disk has been known for Sc galaxy UGC 2885 with HI radius of 83 kpc for $H_0=72$ \kmsmpc\, where the rotation curve is still flat (Roelfsema \& Allen 1985). 

\vskip 5mm \subsection{\bf Mass-to-Luminosity Ratio} 

The best indicator of dark matter in a galaxy is the difference between the galaxy mass predicted by the luminosity and the mass predicted by the rotation velocities. This difference is usually indicated by the ratio of the dynamical mass to luminosity, which is called the mass-to-luminosity ratio ($M/L$). The M/L ratio is  a clue to the distribution of visible and dark mass. 
Most investigations have assumed that the luminous part of a galaxy consists of bulge and  disk, each with a constant mass-to-luminosity ratio. The ``maximum-disk'' assumption, that the disk component corresponding the rotation curve is dominated totally by baryonic mass, is often adopted to derive $M/L$ ratios in the individual components (Kent 1986,  1992; Takamiya and Sofue 2000).  

\begin{figure} 
\bc
\includegraphics[width=6cm]{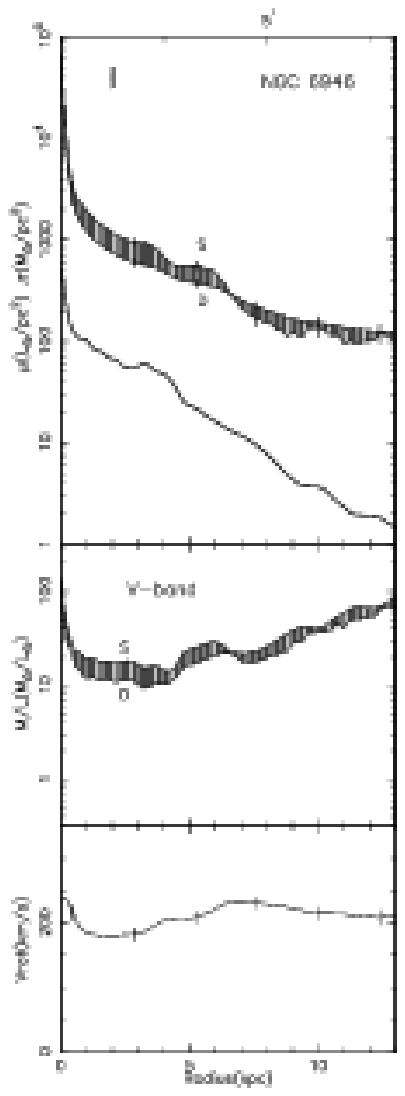}  \\ 
\ec
\caption{[Top] Surface mass density distribution of the spiral galaxy NGC 6946 calculated from the rotation curve by spherical and flat disk assumptions. The surface luminosity distribution is shown by the thick line. [Middle]: M/L ratio. [Bottom]: Rotation curve (Takamiya and Sofue 2000).}
\label{ML_ngc6946}  
\end{figure}

\begin{figure} 
\bc
\includegraphics[width=7.6cm]{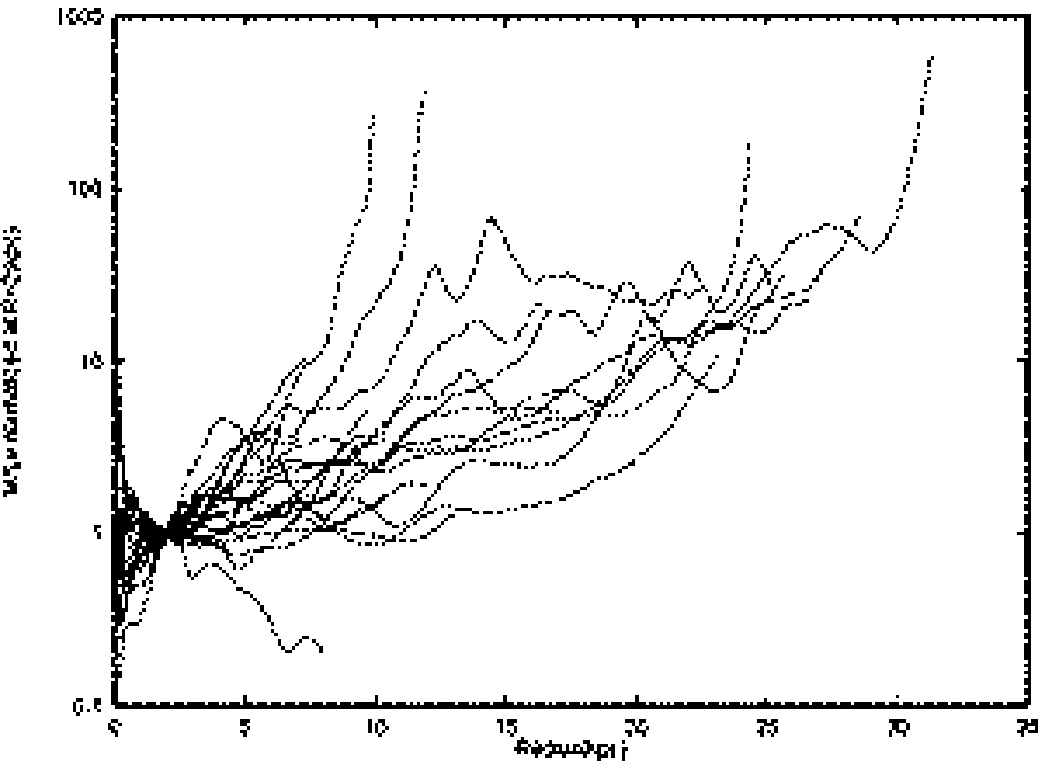}  \\ 
\hskip -1mm \includegraphics[width=8cm]{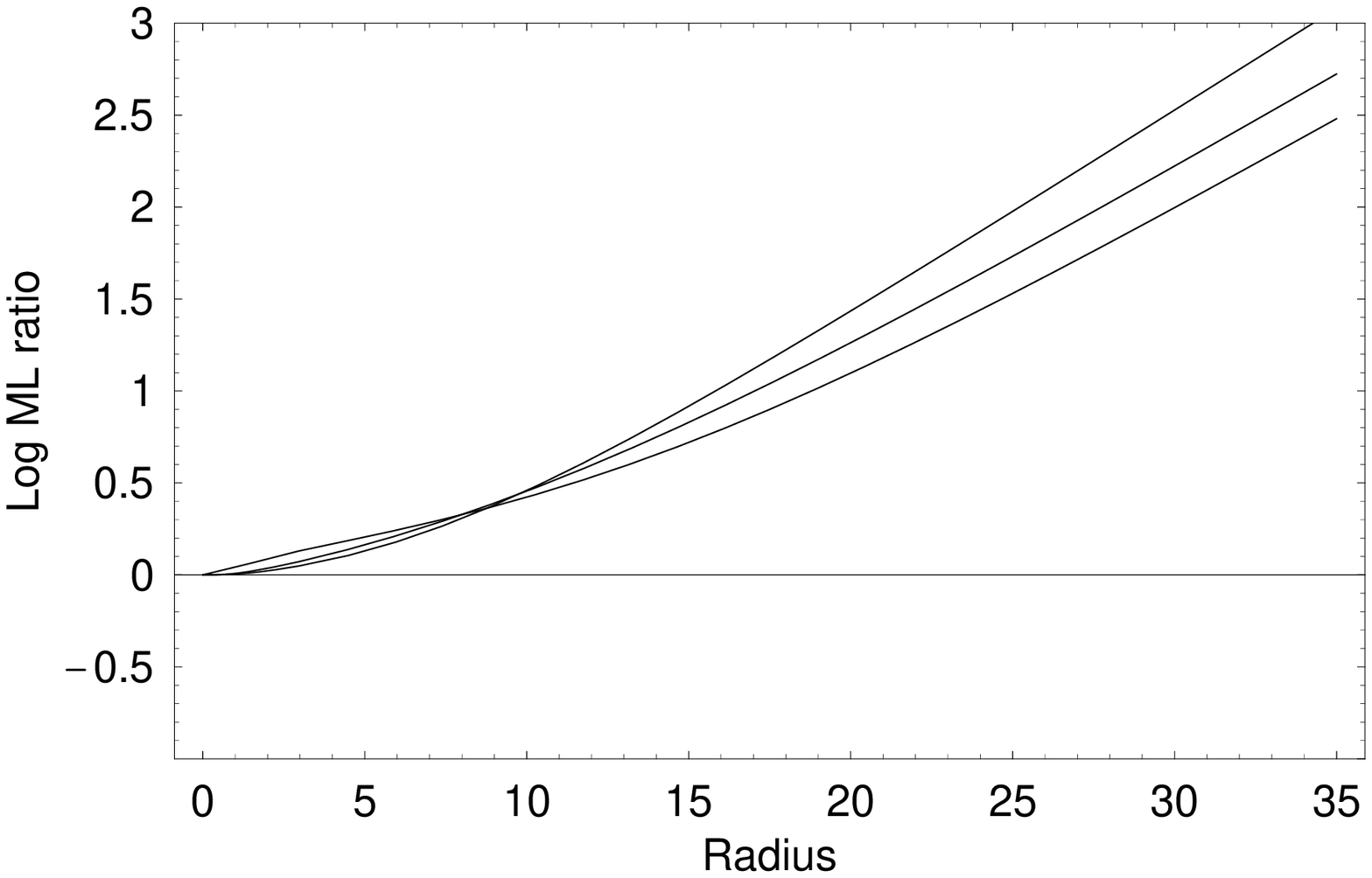}  \
\ec
\caption{[Top] M/L ratios for spiral galaxies normalized at $R=2$ kpc (Takamiya and Sofue 2000). [Bottom] M/L ratio approximately calculated for the dark halo models in the Galaxy. From upper to lower lines: isothermal, NFW and Burkert models, respectively.}
\label{fig-MLratio}  
\end{figure}  

The distribution of M/L ratio in a galaxy is obtained by measuring the  surface-mass density (SMD) and surface luminosity profiles (Forbes 1992; Takamiya and Sofue 2000). The observations have indicated that the M/L ratio is highly variable within individual galaxies, and it increases rapidly beyond the disk toward the furthest measured points. Fig. \ref{ML_ngc6946} shows the mass distribution in the typical Sc galaxy NGC 6946. The mass-to-luminosity (M/L) ratio for the same galaxy is also shown in the same figure. The surface mass density distribution was directly derived from the rotation curve both by spherical and flat disk assumptions. The surface luminosity distribution is shown by the thick line (Takamiya and Sofue 2000). Fig. \ref{fig-MLratio} shows the mass-to-luminosity ratios for various spiral galaxies normalized at $R=2$ kpc. The M/L ratio generally increases toward the outer edges. In the figure these M/L ratios are compared with that calculated for the dark halo models of the Milky Way under the maximum-bulge and disk assumption.
 
\section{SMALLER MASS STRUCTURES}

\vskip 5mm \subsection{\bf Spiral arms and rotation dips} 

Although rotation curves of galaxies are generally smooth and flat, wavy ripples are often superposed as shown in Fig. \ref{fig-obs} and \ref{rc-all}.  The rotation maximum and wavy ripples in the Galaxy's rotation curve around $\sim 5 $ kpc are significantly deviated from the curve for exponential disk. The deep minima at $3$ and 9 kpc are another prominent perturbations on the rotation curve. 

In addition to the smooth and axisymmetric dynamical components by the bulge, disk and dark halo, which are the fundamental structure, the effects by smaller-scale structures such as spiral arms and bar should be considered. If the amplitude of the ripple is not large, they may be treated as a linear perturbation on the main structure, and can be represented by adding a small perturbation term to the main components as shown by $\Delta$ in Eq. \ref{eq-smddelta}. If the ripple amplitude is large and comparable to the rotation velocities, more sophisticated modeling is needed for non-linear perturbation. There have been a number of numerical simulations to discuss the effect of a bar and spiral arms on the kinematics of interstellar gas and stars (e.g., Athanassoula 1992; Wada and Habe 1995). However, fitting to the observations by such non-linear numerical simulation is not straightforward, and the results do not necessarily give unique solutions about the mass and shape of the perturbed structure. Here, only a brief discussion is given of qualitative properties noticed on the rotation curve, and is interpreted by simple analyses in a linear approximation. 

The Galaxy has prominent gaseous spiral arms as shown in Fig. \ref{HIH2face}. The surface mass fraction of gas reaches several percent in the inner disk. In the outer disk, it reaches a few tens of percents, but the mass of the dark halo much exceeds the disk. Hence, the contribution from the gaseous mass to the circular velocity is negligible in any region of the Galaxy. This means that the observed ripples of $\sim \pm 10$ \kms on the rotation curve cannot be attributed to the gaseous arms alone. Some underlying more massive structures may be considered. If logarithmic spiral arms of density perturbation of a few tens of percent superposed on the disk are assumed to be superposed, the ripples on the rotation curve can be reproduced as shown in Fig. \ref{fig-rc-arm} (Sofue et al. 2009). 
In order to obtain a better fit to the observations including the 9 kpc dip, a ring or a density wave of radius 11 kpc with amplitude as high as $\sim 0.3$ times the underlying disk density is required.  It is possible that the dip is related to a density wave corresponding to the Perseus Arm.

\vskip 5mm \subsection{\bf Bar} 

The effect of a bar on the kinematics is significant in the inner galaxy (Athanassoula 1992), and observations have indeed revealed a bar in our Galaxy (Blitz et al. 1993; Weiland et al. 1994; Freudenreich 1998).  
Nakai (1992) noticed asymmetry in the distribution and kinematics of the CO-line emission at $360^\circ \le l \le 30^\circ$ with respect to the Galactic Center (Fig.\ref{pvdCO}), and argued that it may be attributed to non-circular motion by a bar. Similar bar-driven non-circular motions are also observed in molecular disks of many spiral galaxies (Sakamoto et al. 1999; Kenney et al. 1992). 
 
  The rotation dip near 3 kpc  may be discussed in the scheme of non-linear response of gas to the barred potential. According to near-infrared observations  (Freudenreich 1998), the bar length is $\sim  1.7$ kpc and the tilt angle of $13^\circ$ from the Sun-Galactic center line. If the bar's amplitude is taken to be on the order of about $\delta{\rm bar}\simeq 20$\% of the background smooth component, the 3 kpc dip in the observed rotation curve is well explained as due to non-circular motion by the bar's potential (Fig. \ref{fig-rc-arm}). The excess mass by the bar may be estimated to be on the order of $M_{\rm bar}\sim \delta M_{\rm disk+bulge}(R \le 2 kpc) \sim  10^{10}\Msun$. 

In more inner region, observations show that the majority ($\sim 95$\%) of gas in the Galactic Center at $R\le \sim 300$ pc is rotating in a ring showing rigid-body features o n the PV diagram (Bally et al. 1987; Oka et al. 1998; Sofue 1995). A few percent of the gas exhibits forbidden velocities, which may be either due to non-circular motion in an oval potential (Binney et al. 1991) or to expanding motion. The fact that the majority of the gas is regularly rotating indicates that the gas is on circular orbits.

\vskip 5mm \subsection{\bf  Massive central component and Black Hole}

Central rotation curves and mass distributions have been produced for a number of galaxies by a systematic compilation of PV diagrams in the CO and \ha\ lines (Sofue 1996; Sofue et al. 1999; Bertola et al. 1998). 
When they are observed with sufficiently high resolutions, innermost velocities  of galaxies start already from high values at the nucleus, indicating the existence of high density cores in their nuclei.  High central density is not a characteristic only for massive galaxies. The nearby Sc galaxy M33, having a small bulge, exhibits central velocities at about V=100 \kms\ (Rubin and Graham 1987) which do not decrease to zero at the origin. Thus, in spiral galaxies, the contribution from a peaked central mass is remarkable, exceeding the density contribution from the disk and bulge. 

For many spirals, including the Milky Way, the innermost region exhibits rapid rotation velocities, offering evidence for massive nuclear black holes.
Consequently, orbital velocities in the center decrease rapidly from a velocity close to the speed of light down to 100 to 200 \kms of the bulge and disk rotation velocities (Fig. \ref{rc-mw-log}).
It has been established that the spheroidal component, e.g. bulge, is deeply coupled with the central black hole, as inferred from the tight correlation between the black hole mass and bulge luminosity (Magorrian et al. 1998, Kormendy 2011).    

Infrared observations of proper motions of circum nuclear stars revealed that the Galactic Center nests a black hole with mass of $3.7 \times 10^6\Msun$ coinciding with Sgr A$^*$ (Genzel et al. 1997, 2000; Ghez 1998, 2005).
In the nearby galaxy, NGC 4258, water maser lines at 22 GHz are observed  from a disk of radius 0.1 pc in Keplerian rotation, which is attributed to a massive black hole of a mass of $3.9 \times 10^{7}\Msun$ (Nakai et al. 1993;   Miyoshi et al. 1995; Herrnstein et al. 1999).  VLBI observations of the water maser lines have revealed a rapidly rotating nuclear torus of sub parsec scales in several nearby active galactic nuclei. There have been an increasing number of evidences for massive black holes in galactic nuclei (Melia 2011).   

The existence of massive objects in the nuclei is without doubt related to nuclear activity such as via rapid gas inflow in the steep gravitational potential. However, the activity is not always activated by the dynamical structure. It is well established that the rotation curves, and therefore, the fundamental mass distributions in the central regions are very similar to each other. Showing the same dynamical structure, some galaxies exhibit nuclear activity, while the others do not. 

In fact, high-accuracy central rotation curves for starburst galaxies, Seyferts , LINERs, and galaxies with nuclear jets do not show any particular peculiarity in their mass distributions. The typical starburst galaxy M82 shows normal central rotation, except that its outer rotation is declining due to the truncation of dark halo. Even such a very active galaxy like NGC 5128 (Cen  A) shows a normal rotation curve (van Gorkom et al. 1990). The radio lobe galaxy NGC 3079 has both strong nuclear activity and usual rotation properties. 

It is likely that nuclear activity is triggered by local and temporary causes around central massive cores and black holes, while the underlying dynamical structure is stable and universal in any galaxies. Possible mechanism would be intermittent inflow of gas from a circum nuclear torus, to which the disk gas is continuously accumulating by secular angular momentum loss by a bar or spiral arms. Because the accreting gas would be clumpy, the torus will  be distorted from axisymmetric potential, triggering non-circular streaming flow to cause shocks and angular momentum exchange among the clumps. Some clumps will lose angular momentum, and are accreted to the nucleus.  
 
\section{SUMMARY}

The methods to derive rotation curves of the Galaxy and spiral galaxies were described in this chapter, and explanations were given of the methods to calculate the mass distribution using the rotation curve. Observed rotation curves for the Milky Way as well as for nearby spiral galaxies were presented, and were used to derive the mass distributions.
The rotation curve is the most fundamental tool to discuss the dynamics and mass distribution in the Galaxy and in disk galaxies, and there are various methods to derive rotation curves. A unified rotation curve of the Milky Way was obtained by integrating the current various observations as well those for nearby spiral galaxies (Fig. \ref{fig-obs}, \ref{rc-all}, \ref{rc-type}).

The mass distribution is determined both by dynamical and photometric methods. In this chapter, the dynamical method was described in detail. Compared to the statistical method, which assumes such parameters as the mass-to-luminosity ratio, the dynamical method was shown to be more direct to measure the mass including the dark matter and black holes.
The dynamical method may be further categorized into two ways: one is the decomposition method, by which the rotation curve is fitted by a calculated one as summation of several mass components  (Fig. \ref{fig-smd},  \ref{figMN}). The dynamical parameters of individual components are determined during the fitting process. This method is convenient to discuss separately the basic galactic structures, which are usually the central massive object, bulge, disk and dark halo (table \ref{tab_milkyway}). Each of these components may have its own evolutional and dynamical properties. Inconvenience of this method was that one has to assume a priori the functional forms for the mass components, which may not necessarily be selected uniquely by observations. Therefore, the results depend on the adopted models, functional forms, as well as on one's consideration on the galactic structure. 

Another method is the 
t method, in which the mass distribution is directly calculated from the data of rotation velocity (Fig. \ref{fig-smd-mw}, \ref{fig-smd-Sb}). This method does not employ any galactic models or functional forms, but straightly compute the mass distribution. The results can be compared with surface photometry to obtain the distribution of mass-to-luminosity ratio by simply dividing the surface-mass density by surface luminosity (Fig. \ref{ML_ngc6946}). The thus computed M/L ratio is found to generally increase monotonically toward the edges of spiral galaxies, indicating the direct evidence for dark matter halo.
As discussed in the previous sections, these two dynamical methods, decomposition and direct methods, are complimentary, and the derived mass distributions are consistent with each other.

It was shown that the rotation characteristics of spiral galaxies are similar to each other, and so are the mass distributions. The dynamical back bones are, thus, universal from Sa to Sc galaxies, and the structures are almost identical among the galaxies in so far as the dynamical mass is concerned. The Milky Way exhibits equally the most typical universal characteristics. 

The dark halo of the Galaxy was  discussed in detail. It was shown that the NFW and Burkert profiles better represent the observations of the outermost rotation characteristics up to $R\sim 150$ kpc compared to the isothermal profile. It was also shown that the dark halo extends far out to the intracluster space almost a half way to M31, where the dark matter properly possessed by the Local Group dominates in order for the group to be gravitationally bound.

\end{document}